\documentclass[a4paper,12pt]{article}
\usepackage{cite}
\usepackage{epsfig}
\usepackage{amssymb,amsmath}
\usepackage{graphicx}
\usepackage{float}
\usepackage{comment}
\usepackage{epstopdf}

\pdfoutput=1

\newlength{\dinwidth}
\newlength{\dinmargin}
\newcommand{\re}{{\mathrm{Re}}}
\newcommand{\im}{{\mathrm{Im}}}
\newcommand{\cF}{{\cal{F}}}

\newcommand{\cR}{\mathcal{R}}
\newcommand{\eqn}[1]{(\ref{#1})}
\newcommand{\bel}[1]{\be\label{#1}}
\newcommand{\e}{\mathrm{e}}

\newcommand{\lrder}{\stackrel{\leftrightarrow}{\partial}}

\newcommand{\athdm}[0]{A2HDM}

\newcommand{\ta}[0]{\tilde \alpha}

\def\be{\begin{equation}}
\def\ee{\end{equation}}
\def\beqn{\begin{eqnarray}}
\def\eeqn{\end{eqnarray}}
\def\no{\nonumber}
\def\ba{\begin{array}{c}}
\def\bat{\begin{array}{cc}}
\def\ea{\end{array}}
\def\bi{\begin{itemize}}
\def\ei{\end{itemize}}

\def\cL{{\cal L}}

\def\be{\begin{equation}}
\def\ee{\end{equation}}
\def\sm{\text{SM}}
\usepackage{color}

\definecolor{DarkGreen}{rgb}{0.3,0.5,0.35}
\definecolor{Violet}{rgb}{0.5,0,1}
\definecolor{Brown}{rgb}{0.5,0.25,0}

\begin{document}

\title{
\begin{flushright}\vbox{\normalsize \mbox{}\vskip -6cm
FTUV/13-0217 \\[-3pt] IFIC/12-89
}
\end{flushright}\vskip 45pt
{\bf LHC constraints on two-Higgs doublet models}}
\bigskip

\author{Alejandro Celis, Victor Ilisie and Antonio Pich\\[15pt]
{\small IFIC, Universitat de Val\`encia -- CSIC, Apt. Correus 22085, E-46071 Val\`encia, Spain}}

\date{}
\maketitle
\bigskip \bigskip

\begin{abstract}
\noindent
A new Higgs-like boson with mass around 126 GeV has recently been discovered at the LHC.
The available data on this new particle is analyzed within the context of two-Higgs doublet models
without tree-level flavour-changing neutral currents.
Keeping the generic Yukawa structure of the Aligned Two-Higgs Doublet Model framework, we study the implications of the LHC data on the allowed scalar spectrum. We analyze both the CP-violating and CP-conserving cases, and
a few particular limits with a reduced number of free parameters, such as the
usual models based on discrete ${\cal Z}_2$ symmetries.
\end{abstract}

\newpage

\section{Introduction}
\label{sec:intro}

The ATLAS and CMS collaborations have recently announced the discovery of a new neutral boson,
with a measured mass of $125.2\pm 0.3 \pm 0.6$ GeV \cite{:2012gk} and $125.8\pm 0.4 \pm 0.4$ GeV  \cite{:2012gu}, respectively.
The LHC data is compatible with the expected production and decay of the Standard Model (SM) Higgs boson, the most significant decay modes
being $H\to \gamma\gamma$ and $H\to ZZ^{(*)}\to \ell^+\ell^-$. The excess of events observed by ATLAS (CMS) has a (local) statistical significance of
$6.1\,\sigma$ ($6.9\,\sigma$). Although the spin of the new particle has not been measured yet, the observed diphoton decay channel shows clearly
that it is a boson with $J\not= 1$, making very plausible the scalar hypothesis.     Preliminary analyses of $H \rightarrow ZZ \rightarrow 4 \ell$~\cite{Chatrchyan:2012dg,ATLAS:2012ac} and $H \rightarrow \gamma \gamma $~\cite{ATLAS:2012ad,Chatrchyan:2012twa} events suggest indeed the assignment $J^P = 0^+$, though more statistics is still needed to give a definite answer.

Additional (but less significant) evidence has been reported by the CDF and D\O~collaborations \cite{Aaltonen:2012qt}, which observe
an excess of events in the mass range between 120 and 135 GeV
(the largest local significance is $3.3\,\sigma$). The excess seems consistent with a SM Higgs produced in association with a $W^{\pm}$ or
$Z$ boson and decaying to a bottom-antibottom quark pair.

While more experimental analyses are needed to assess the actual nature of this boson, the present data give already very important
clues, constraining its couplings in a quite significant way. The stringent exclusion limits set previously on a broad range of masses provide also
complementary information which is very useful to establish allowed domains for alternative new-physics scenarios. A SM Higgs boson has been
already excluded at 95\% CL in the mass ranges
0--122.5  
and 127--600 GeV
\cite{ATLAS:2012ad,Barate:2003sz,Aaltonen:2010yv,:2012zzl,Aaltonen:2012if,:2012tf,:2012an,Chatrchyan:2012tx}.

The new boson appears to couple to the known gauge bosons ($W^\pm$, $Z$, $\gamma$, $g$) with the strength expected for the SM Higgs~\cite{Espinosa:2012im,Klute:2012pu,Carmi:2012yp,Azatov:2012bz,Azatov:2012rd,Giardino:2012dp,Corbett:2012dm,Masso:2012eq,Ellis:2012hz,Cheung:2013kla},
although a slight excess of events in the $2\gamma$ decay channel, compared with the SM expectation, is observed by ATLAS and CMS~\cite{:2012gk,:2012gu}.  Moreover, its
fermionic couplings seem compatible with a linear dependence with the fermion mass, scaled by the electroweak scale $v\approx 246$~GeV \cite{Ellis:2012hz}.
Thus, it has the properties expected for a Higgs-like particle, related with the spontaneous breaking of the electroweak symmetry.
An obvious question to address is whether it corresponds to the unique Higgs boson incorporated in the SM, or it is just the first signal of a much richer scalar sector.

The simplest modification of the SM Higgs mechanism consists in incorporating additional
scalar doublets, respecting the custodial symmetry, which can easily satisfy the electroweak precision tests. This leads to a rich spectrum
of neutral and charged scalars, providing a broad range of dynamical possibilities with very interesting phenomenological implications.
The minimal extension of the scalar sector with only one additional doublet contains
five physical scalars: two charged fields $H^{\pm}$ and three neutral ones $h$, $H$ and $A$; thus, there are three possible candidates for the recently discovered neutral boson. If the scalar potential preserves the CP symmetry,
$h$ and $H$ are CP-even, while $A$ is CP-odd; in this case there are no $AW^+W^-$
and $AZZ$ couplings at tree level, which makes the $A$ possibility quite unlikely.

Generic multi-Higgs doublet models give rise to unwanted flavour-changing neutral current~(FCNC) interactions through non-diagonal couplings of neutral scalars to fermions.
The tree-level FCNCs can be eliminated requiring the alignment in flavour space of the Yukawa matrices coupling to a given right-handed fermion~\cite{Pich:2010ic}. The
Aligned Two-Higgs Doublet Model (A2HDM)~\cite{Pich:2009sp} results in a very specific structure, with all fermion-scalar interactions being proportional to the corresponding fermion masses. This leads to a rich and viable  phenomenology~\cite{Pich:2009sp,Pich:2010ic,Jung:2010ik,Jung:2010ab,Jung:2012vu,Celis:2012dk} with an interesting hierarchy of FCNC effects, suppressing them in light-quark systems while allowing potentially relevant signals in heavy-quark transitions.
The A2HDM constitutes a very general framework which includes, for particular values of its parameters, all previously considered two-Higgs doublet models (2HDMs) without FCNCs~\cite{Branco:2011iw,Gunion:1989we}, and incorporates in addition new sources of CP violation.

In the following, we will analyze the recent discovery of a Higgs-like object within the A2HDM. We will study the different possible interpretations of the new boson, the corresponding experimental constraints on its couplings, and the implications for the remaining scalar spectrum. Previous analyses
\cite{Craig:2012vn,Ferreira:2011aa,Ferreira:2012my,Barroso:2012wz,Burdman:2011ki,Arhrib:2011wc,Arhrib:2012ia,Gabrielli:2012yz,Blum:2012kn,Belanger:2012gc,Chen:2013kt}
have only considered more specific scenarios based on discrete ${\mathcal Z}_2$ symmetries \cite{Glashow:1976nt}, {\it i.e.}, the so called 2HDMs of types I \cite{Haber:1978jt,Hall:1981bc}, II \cite{Hall:1981bc,Donoghue:1978cj}, X (leptophilic or lepton specific), Y (flipped)  \cite{Barger:1989fj,Grossman:1994jb,Akeroyd:1994ga,Aoki:2009ha} and inert \cite{Deshpande:1977rw}. The more general A2HDM framework opens a wide range of additional possibilities, which we will try to characterize keeping in mind the high-statistics data samples that the LHC is expected to deliver in the future, at higher energies. Two very recent works have already employed the A2HDM, in the limit of CP conservation, to analyze the Higgs data \cite{Altmannshofer:2012ar,Bai:2012ex}. Another previous work has considered the CP-conserving A2HDM with a custodial symmetry imposed on the Higgs potential~\cite{Cervero:2012cx}. We will compare our results in that limit and will also explore the consequences of allowing CP-violating phases, either in the scalar potential (mixing of the three neutral scalars) or in the Yukawa couplings.
While parts of our analysis remain valid in more general 2HDM settings, the flavour constrains would necessary be different in models with tree-level FCNCs \cite{Branco:1996bq,Botella:2009pq,Botella:2011ne} and, therefore, the appropriate modifications should be taken into account.

Our paper is organized as follows: In section~\ref{sec:A2HDM}, we describe the theoretical framework adopted in our analysis, indicating the relevant couplings of the A2HDM scalars.  In section~3 we define the Higgs signal strengths, which are used to make contact with the experimental measurements.  Section~\ref{sec:fits} presents our results and shows the scalar parameter ranges needed to explain the present data. Our conclusions are given in section~\ref{sec:conclusion}. The appendices include a compilation of useful formulae as well as the statistical treatment and data used in this work.

\section{The Aligned Two-Higgs-Doublet Model}
\label{sec:A2HDM}

The 2HDM extends the SM with a second scalar doublet of hypercharge $Y=\frac{1}{2}$.
The neutral components of the scalar doublets $\phi_a(x)$ ($a=1,2$) acquire vacuum expectation values that are, in general, complex: $\langle 0|\phi_a^T(x)|0\rangle =\frac{1}{\sqrt{2}}\, (0,v_a\, \e^{i\theta_a})$. Through an appropriate $U(1)_Y$ transformation we can enforce $\theta_1=0$, since only the relative phase $\theta \equiv \theta_2 - \theta_1$ is observable.
It is convenient to perform a global SU(2) transformation in the scalar space $(\phi_1,\phi_2)$ and work in the so-called Higgs basis
$(\Phi_1,\Phi_2)$, where only one doublet acquires a vacuum expectation value:
\begin{equation}
\left( \begin{array}{c} \Phi_1 \\ -\Phi_2  \end{array} \right) \; \equiv \;
\left[\bat \cos{\beta} & \sin{\beta}\\ \sin{\beta} & -\cos{\beta}\ea\right]\;
\left( \begin{array}{c} \phi_1 \\ e^{-i\theta}\phi_2  \end{array} \right) \; ,
\end{equation}
with $\tan{\beta}= v_2/v_1$.
In this basis, the two doublets are parametrized as
\begin{equation}  \label{Higgsbasis}
\Phi_1=\left[ \begin{array}{c} G^+ \\ \frac{1}{\sqrt{2}}\, (v+S_1+iG^0) \end{array} \right] \; ,
\qquad\qquad\qquad
\Phi_2 = \left[ \begin{array}{c} H^+ \\ \frac{1}{\sqrt{2}}\, (S_2+iS_3)   \end{array}\right] \; ,
\end{equation}
where $G^\pm$ and $G^0$ denote the Goldstone fields
and $\langle 0|H^+| 0\rangle=\langle 0|G^+|0\rangle=\langle 0|G^0|0 \rangle=\langle 0|S_i|0\rangle=0$.
Thus, $\Phi_1$ plays the role of the SM scalar doublet with
$v\equiv \sqrt{v_1^2+v_2^2}\simeq (\sqrt{2}\, G_F)^{-1/2} = 246~\mathrm{GeV}$.

The physical scalar spectrum contains five degrees of freedom: the two charged fields $H^\pm(x)$
and three neutral scalars $\varphi_i^0(x)=\{h(x),H(x),A(x)\}$, which are related with the $S_i$ fields
through an orthogonal transformation $\varphi^0_i(x)=\mathcal{R}_{ij} S_j(x)$.
The form of the $\mathcal{R}$ matrix is fixed by the scalar potential, which determines the neutral scalar mass matrix
and the corresponding mass eigenstates. A detailed discussion is given in appendix \ref{app:potential}. In general, the CP-odd component $S_3$ mixes with the CP-even fields
$S_{1,2}$ and the resulting mass eigenstates do not have a definite CP quantum number.
If the scalar potential is CP symmetric this admixture disappears; in this particular case, $A(x) = S_3(x)$
and\footnote{
In the usually adopted notation $\tilde\alpha =\alpha-\beta$, where $\alpha$ is the rotation angle expressing the
two mass eigenstates $h$ and $H$ in terms of the CP-even neutral fields of the original scalar basis $\phi_1(x)$ and $\phi_2(x)$.
Since the choice of initial basis is arbitrary, the parameters $\alpha$ and $\beta$ are in general unphysical;
their values can be changed at will through SU(2) rotations. These angles only become meaningful in particular models where a specific basis is singled out
(through a symmetry for instance).
}
%
\bel{eq:CPC_mixing}
\left(\ba h\\ H\ea\right)\; = \;
\left[\bat \cos{\tilde\alpha} & \sin{\tilde\alpha} \\ -\sin{\tilde\alpha} & \cos{\tilde\alpha}\ea\right]\;
\left(\ba S_1\\ S_2\ea\right) \, .
\ee
Performing a phase redefinition of the neutral CP-even fields, we can fix the sign of $\sin{\ta}$.  In this work we adopt the conventions\ $M_h \le M_H$\ and\
$ 0 \leq \ta \leq \pi$, so that $\sin{\ta}$ is positive.

\subsection{Yukawa Alignment}
\label{sec:A2HDM-alignment}

The most generic Yukawa Lagrangian with the SM fermionic content gives rise to FCNCs because the fermionic couplings of the two scalar doublets cannot be simultaneously diagonalized in flavour space. The non-diagonal neutral couplings can be eliminated by requiring the alignment in flavour space of the Yukawa matrices~\cite{Pich:2009sp}; {\it i.e.}, the two Yukawa matrices coupling to a given type of right-handed fermions are assumed to be proportional to each other and can, therefore, be diagonalized simultaneously. The three proportionality parameters $\varsigma_f$~($f=u,d,l$) are arbitrary complex numbers and introduce new sources of CP violation.

In terms of the fermion mass-eigenstate fields, the Yukawa interactions of the A2HDM read~\cite{Pich:2009sp}
\beqn\label{lagrangian}
 \mathcal L_Y & = &  - \frac{\sqrt{2}}{v}\; H^+ \left\{ \bar{u} \left[ \varsigma_d\, V M_d \mathcal P_R - \varsigma_u\, M_u^\dagger V \mathcal P_L \right]  d\, + \, \varsigma_l\, \bar{\nu} M_l \mathcal P_R l \right\}
\nonumber \\
& & -\,\frac{1}{v}\; \sum_{\varphi^0_i, f}\, y^{\varphi^0_i}_f\, \varphi^0_i  \; \left[\bar{f}\,  M_f \mathcal P_R  f\right]
\;  + \;\mathrm{h.c.} \, ,
\eeqn
where $\mathcal P_{R,L}\equiv \frac{1\pm \gamma_5}{2}$ are the right-handed and left-handed chirality projectors,
$M_f$ the diagonal fermion mass matrices
and the  couplings of the neutral scalar fields are given by:
\begin{equation}    \label{yukascal}
y_{d,l}^{\varphi^0_i} = \cR_{i1} + (\cR_{i2} + i\,\cR_{i3})\,\varsigma_{d,l}  \, ,
\qquad\qquad
y_u^{\varphi^0_i} = \cR_{i1} + (\cR_{i2} -i\,\cR_{i3}) \,\varsigma_{u}^* \, .
\end{equation}
As in the SM, all scalar-fermion couplings are proportional to the corresponding fermion masses.   This linear dependence on the fermion mass is characteristic of the A2HDM framework and does not hold in non-aligned 2HDMs with FCNCs.
The only source of flavour-changing interactions is the Cabibbo-Kobayashi-Maskawa~(CKM) quark mixing matrix $V$~\cite{CKM}.
All possible freedom allowed by the alignment conditions is determined by the three family-universal complex parameters $\varsigma_f$, which provide new sources of CP violation without tree-level FCNCs~\cite{Pich:2009sp}.
The usual models with natural flavour conservation, based on discrete ${\cal Z}_2$ symmetries, are recovered for particular (real) values of the couplings $\varsigma_f$,
as indicated in Table~\ref{tab:models}.

\begin{table}\begin{center}
\caption{\it \small CP-conserving 2HDMs based on discrete $\mathcal Z_2$ symmetries.}
\vspace{0.2cm}
\begin{tabular}{|c|c|c|c|}
\hline
Model & $\varsigma_d$ & $\varsigma_u$ & $\varsigma_l$  \\
\hline
Type I  & $\cot{\beta}$ &$\cot{\beta}$ & $\cot{\beta}$ \\
Type II & $-\tan{\beta}$ & $\cot{\beta}$ & $-\tan{\beta}$ \\
Type X  & $\cot{\beta}$ & $\cot{\beta}$ & $-\tan{\beta}$ \\
Type Y  & $-\tan{\beta}$ & $\cot{\beta}$ & $\cot{\beta}$ \\
Inert  & 0 & 0 & 0 \\
\hline
\end{tabular}
\label{tab:models}
\end{center}\end{table}

Quantum corrections induce a misalignment of the Yukawa matrices, generating small FCNC effects suppressed by the corresponding loop factors \cite{Pich:2009sp,Pich:2010ic,Jung:2010ik,Ferreira:2010xe,Braeuninger:2010td}. However, the flavour symmetries of the A2HDM tightly constraint the possible FCNC structures, keeping their effects well below the present experimental bounds
\cite{Pich:2009sp,Pich:2010ic,Jung:2010ik,Jung:2010ab,Jung:2012vu,Celis:2012dk}.\footnote{
The only FCNC structures induced at one loop take the form \cite{Pich:2010ic,Jung:2010ik}:
\begin{eqnarray}
\mathcal L_{\mathrm{FCNC}} & =& \frac{C(\mu)}{4\pi^2 v^3}\; (1+\varsigma_u^*\varsigma_d^{\phantom{*}})\;
\sum_i\, \varphi^0_i(x)\;\left\{
(\cR_{i2} + i\,\cR_{i3})\, (\varsigma_d^{\phantom{*}}-\varsigma_u^{\phantom{*}})\;
\left[\bar d_L\, V^\dagger M_u^{\phantom{\dagger}} M_u^\dagger\, V M_d^{\phantom{\dagger}}\, d_R\right]-\right.\label{eq:FCNCop}
\\
&&\hskip 4.3cm \left.
-\, (\cR_{i2} - i\,\cR_{i3})\, (\varsigma_d^*-\varsigma_u^*)\;
\left[\bar u_L\, V M_d^{\phantom{\dagger}} M_d^\dagger\, V^\dagger M_u^{\phantom{\dagger}}\, u_R\right] \right\}
\; +\; \mathrm{h.c.}\nonumber
\end{eqnarray}
with $C(\mu)=C(\mu_0)-\log{(\mu/\mu_0)}$.
These FCNC effects vanish identically in the $\mathcal{Z}_2$ models where the alignment condition is protected by a discrete symmetry. In the most general case, assuming the alignment to be exact at some scale $\mu_0$, i.e. $C(\mu_0)=0$, a non-zero value for the FCNC coupling is generated when running to a different scale. However, the numerical effect
is suppressed by $m_{q}m_{q'}^2/v^3$ and quark-mixing factors, avoiding the stringent experimental constraints for light-quark systems. Explicit examples of symmetry-protected underlying theories leading to a low-energy A2HDM structure have been discussed in Refs.~\cite{Serodio:2011hg,Varzielas:2011jr,Cree:2011uy}.
}

The orthogonality of the rotation matrix $\mathcal{R}$, implies the following relations among the Yukawa couplings of the three neutral scalars:
\beqn \label{sumrules}
\sum_{i=1}^{3} \, (y_{f}^{\varphi_i^0})^2\, =\, 1 \, ,
\qquad\qquad
\sum_{i=1}^{3} \, |y_{f}^{\varphi_i^0}|^2 &\! =\! & 1 + 2\, |\varsigma_f |^2   \, ,
\qquad\qquad
\sum_{i=1}^{3} \,  y_{f}^{\varphi_i^0} \,\cR_{i1}\, =\, 1 \, ,
\nonumber\\[5pt]
\sum_{i=1}^{3} \,  y_{d,l}^{\varphi_i^0} \, \cR_{i2}\, =\, \varsigma_{d,l}  \, ,
\qquad\quad && \qquad\quad
\sum_{i=1}^{3} \,  y_{u}^{\varphi_i^0} \, \cR_{i2}\, =\, \varsigma_u^*  \, ,
\nonumber\\[5pt]
\sum_{i=1}^{3} \,  y_{d,l}^{\varphi_i^0}\, \cR_{i3} = i \, \varsigma_{d,l}  \, ,
\qquad\quad && \qquad\quad
\sum_{i=1}^{3} \,  y_{u}^{\varphi_i^0}\, \cR_{i3} = -i \, \varsigma_{u}^*  \, .
\eeqn

\subsection{Bosonic Couplings}
\label{bosonic_cpl}

The full set of interactions among the gauge and scalar bosons is given in appendix~\ref{app:scalar-gauge}.
The relevant vertices for our analysis are the ones coupling a single neutral scalar with a pair of gauge bosons.
As shown in Eq.~\eqn{eq:Lphv2}, they are identical to their SM counterpart, with the field $S_1$ taking the role of the SM Higgs.
Therefore ($VV= W^+W^-, ZZ$),
\be
g_{\varphi^0_i VV}\; =\; \mathcal{R}_{i1}\; g_{h VV}^{\mathrm{SM}}\, ,
\ee
which implies
\be   \label{sumrule}
g_{hVV}^2 + g_{HVV}^2 + g_{AVV}^2\; = \; \left(g_{hVV}^{\mathrm{SM}}\right)^2\, .
\ee
The strength of the SM Higgs interaction is shared by the three 2HDM neutral bosons.  In the CP-conserving limit, the CP-odd field decouples while the strength of the $h$ and $H$
interactions is governed by the corresponding $\cos{\tilde\alpha}$ and $\sin{\tilde\alpha}$ factors.
Thus, a general feature of 2HDMs is that, at tree level, the couplings of the neutral scalars to vector bosons cannot be enhanced over the SM value and obey the custodial symmetry relation $g_{\varphi_i^0ZZ} = g_{\varphi_i^0WW}$.   Observing a scalar boson with a somewhat enhanced coupling to vector bosons or a deviation from custodial symmetry~\cite{Farina:2012ea} would therefore be in clear contradiction with the predictions of this class of models.
The relations \eqref{sumrules} and \eqref{sumrule} establish a connection between the couplings of the observed 126~GeV resonance and searches for other neutral and charged scalars within the A2HDM.

In order to compute the two-photon decay widths of the neutral scalars, one also needs their couplings to a pair
of charged scalars, generated through the scalar potential discussed in appendix~\ref{app:potential}. Since
these couplings depend on still unknown parameters, we will parametrize the corresponding interaction as
\be
\cL_{\varphi^0 H^+H^-}\; =\; - v \;\sum_{\varphi^0_i}\, \lambda_{\varphi^0_i H^+H^-}\;\, \varphi^0_i\, H^+H^-\, .
\ee
Explicit expressions for the cubic couplings $\lambda_{\varphi^0_i H^+H^-}$, in terms of the Higgs potential parameters, can be found in appendix~\ref{app:potential}. If CP is assumed to be an exact symmetry, $\lambda_{A H^+H^-}=0$.

\section{Higgs Signal Strengths}
\label{sec:SignalStrengths}

The experimental data on Higgs searches is given in terms of the so-called signal strengths, measuring the observable cross sections in units of the  corresponding SM expectations. At the LHC, the relevant production mechanisms for a SM-like Higgs particle are gluon fusion
($ g g \rightarrow H$), vector boson fusion ($ q q^{\prime} \rightarrow q q^{\prime} VV\rightarrow q q^{\prime} H$), associated production with a vector boson ($q \bar q^{\prime} \rightarrow W H/Z H $) and the associated production with a $t \bar t$ pair  ($q \bar q/ gg \rightarrow t \bar t  H$).    The Higgs decay channels explored so far are $\gamma \gamma$, $Z Z^{(*)}$, $W W^{(*)}$, $b \bar b$ and $\tau^+ \tau^-$.

In order to fit the experimental measurements, we consider the ratios :
\begin{align}\label{eq:ratios}
\notag \\[-15pt]
\mu_{\gamma\gamma}^{\varphi^0_i} &\equiv
\frac{\sigma(pp\to \varphi^0_i)\, \text{Br} (\varphi^0_i\to \gamma\gamma)}{\sigma(pp\to h)_{\mathrm{SM}}\, \text{Br} (h\to \gamma\gamma)_{\mathrm{SM}}}\, ,
\qquad &
\mu_{\gamma\gamma jj}^{\varphi^0_i} &\equiv
\frac{\sigma(pp\to jj\varphi^0_i)\, \text{Br} (\varphi^0_i\to \gamma\gamma)}{\sigma(pp\to jjh)_{\mathrm{SM}}\, \text{Br} (h\to \gamma\gamma)_{\mathrm{SM}}}\, ,
\notag \\[7pt]
\mu_{VV}^{\varphi^0_i} &\equiv
\frac{\sigma(pp\to \varphi^0_i)\, \text{Br} (\varphi^0_i\to VV)}{\sigma(pp\to h)_{\mathrm{SM}}\, \text{Br} (h\to VV)_{\mathrm{SM}}}\, ,
&
\mu_{WWjj}^{\varphi^0_i} &\equiv
\frac{\sigma(pp\to jj\varphi^0_i)\, \text{Br}(\varphi^0_i\to WW)}{\sigma(pp\to jjh)_{\mathrm{SM}}\, \text{Br}(h\to WW)_{\mathrm{SM}}}\, ,
\notag \\[7pt]
\mu_{\tau\tau}^{\varphi^0_i} &\equiv
\frac{\sigma(pp\to \varphi^0_i)\, \text{Br} (\varphi^0_i\to \tau\tau)}{\sigma(pp\to h)_{\mathrm{SM}}\, \text{Br} (h\to \tau\tau)_{\mathrm{SM}}}\, ,
&
\mu_{bbV}^{\varphi^0_i} &\equiv
\frac{\sigma(pp\to V\varphi^0_i)\,\text{Br}(\varphi^0_i\to b\bar{b})}{\sigma(pp\to Vh)_{\mathrm{SM}}\,\text{Br}(h\to b\bar{b})_{\mathrm{SM}}}\, ,
\\[-15pt]\notag
\end{align}
where $V={W, \!\ Z}$ and $j$ stands for jet.
QCD corrections cancel to a large extend in these ratios, provided that a single production mechanism dominates. This certainly applies
to $\mu_{\gamma\gamma}^{\varphi^0_i}$, $\mu_{VV}^{\varphi^0_i}$ and $\mu_{\tau\tau}^{\varphi^0_i}$ which are governed by the
dominant production channel through gluon fusion.
The same would be true for $\mu_{WWjj}^{\varphi^0_i}$ and $\mu_{\gamma\gamma jj}^{\varphi^0_i}$ (gauge-boson fusion), and
$\mu_{bbV}^{\varphi^0_i}$ (associated production), assuming that there is no contamination from other channels. It is convenient to express the ratio of the branching fractions as:
\be \label{Brratios}
\frac{\text{Br}(\varphi^0_i \to X )}{\text{Br}(h\to X)_{\mathrm{SM}}}\; =\;
\dfrac{1}{\rho(\varphi^0_i)}\;  \frac{\Gamma(\varphi^0_i \to X )}{\Gamma(h\to X)_{\mathrm{SM}}}
\;\, ,
\ee
where $\rho(\varphi^0_i)$ measures the total decay width of the scalar $\varphi_i^0$ in units of the SM Higgs width,
\be
\Gamma (\varphi^0_i) = \rho(\varphi^0_i)  \, \Gamma_{\mathrm{SM}} (h)\, .
\ee
Particularizing to the A2HDM and assuming only one dominant production channel in each case,\footnote{
The contamination of the different Higgs production mechanisms in
$h \rightarrow \gamma \gamma(jj)$
is discussed in appendix~\ref{statistical}.}
one finds:
\begin{align}
  \mu_{bbV}^{\varphi^0_i}\; &=\; (\mathcal{R}_{i1})^2\, \Bigl[ \re(y_d^{\varphi^0_i})^2 + \im(y_d^{\varphi^0_i})^2 \beta_{b}^{-2} \Bigr]  \, \rho(\varphi^0_i)^{-1}  ,
\qquad &
\mu_{WWjj}^{\varphi^0_i}\; &=\; (\mathcal{R}_{i1})^4  \; \rho(\varphi^0_i)^{-1} ,
\notag \\
\mu_{\tau\tau}^{\varphi^0_i}\; &=\; C_{gg}^{\varphi^0_i}\,  \Bigl[ \re(y_l^{\varphi^0_i})^2 + \im(y_l^{\varphi^0_i})^2 \beta_{\tau}^{-2} \Bigr]  \, \rho(\varphi^0_i)^{-1} ,   &
\mu_{VV}^{\varphi^0_i}\; &=\; C_{gg}^{\varphi^0_i}\; (\mathcal{R}_{i1})^2 \; \rho(\varphi^0_i)^{-1} ,
\notag \\
\mu_{\gamma\gamma}^{\varphi^0_i}\; &=\; C_{gg}^{\varphi^0_i}  \; C_{\gamma\gamma}^{\varphi^0_i} \; \rho(\varphi^0_i)^{-1}  , &
 \mu_{\gamma\gamma jj}^{\varphi^0_i}\; &=\; (\mathcal{R}_{i1})^2  \; C_{\gamma\gamma}^{\varphi^0_i}  \; \rho(\varphi^0_i)^{-1}  ,\qquad
\label{ratios2}
\end{align}
where $\beta_{f} = (1-4m_f^2/M_{\varphi_i^0}^2)^{1/2}$. The one-loop functions are given by
\be  \label{gluonfusion}
C_{gg}^{\varphi^0_i}\; =\; \frac{\sigma(gg\to\varphi^0_i)}{\sigma(gg\to h)_{\mathrm{SM}}}\; =\;
\frac{\Big|\sum_q \re (y_q^{\varphi^0_i})  \!\ \mathcal{F}(x_q)\Big|^2 +\, \Big|\sum_q \im(y_q^{\varphi^0_i})  \!\  \mathcal{K}(x_q)\Big|^2}{\Big|\sum_q  \!\ \mathcal{F}(x_q)\Big|^2}
\ee
and
\begin{align}\label{gammascaling}
C_{\gamma\gamma}^{\varphi^0_i} \; &=\; \frac{\Gamma(\varphi^0_i\to\gamma\gamma)}{\Gamma(h\to\gamma\gamma)_{\mathrm{SM}}}
\\[5pt] & =\;
\frac{\Big|\sum_f \re(y_f^{\varphi^0_i})  \!\  N_C^f  \!\ Q_f^2   \!\ \mathcal{F}(x_f) + \mathcal{G}(x_W) \mathcal{R}_{i1} + \mathcal{C}_{H^\pm}^{\varphi^0_i} \Big|^2 +\,
\Big|\sum_f \im(y_f^{\varphi^0_i})  \!\  N_C^f  \!\ Q_f^2   \!\  \mathcal{K}(x_f)\Big|^2}
{\Big|\sum_f   \!\  N_C^f  \!\ Q_f^2   \!\ \mathcal{F}(x_f) + \mathcal{G}(x_W)  \Big|^2} \, , \qquad
\notag\end{align}
with $N_C^f$ and $Q_f$ the number of colours and the electric charge of the fermion $f$, $x_f = 4 m_f^2/M_{\varphi_i^0}^2$
and $x_W = 4 M_W^2/M_{\varphi_i^0}^2$.
Notice that the ratios \eqn{eq:ratios} are defined for $M_{\varphi^0_i} = M_{h_{\mathrm{SM}}}$.
The two separate terms in the numerators of Eqs.~(\ref{gluonfusion}) and (\ref{gammascaling}) correspond to the CP-even and CP-odd structures $\varphi^0_i X_{\mu\nu} X^{\mu\nu}$ and $\varphi^0_i X_{\mu\nu} \widetilde X^{\mu\nu}$, with $X_{\mu\nu} = G_{\mu\nu}$ ($F_{\mu\nu}$) in the gluon (photon) case and $\widetilde X^{\mu\nu} = \epsilon^{\mu\nu\sigma\rho} X_{\sigma\rho}$.
The functions $\mathcal{F}(x_f)$, $\mathcal{K}(x_f)$ and $\mathcal{G}(x_W)$ contain the triangular 1-loop contributions from fermions and $W^\pm$ bosons.
We will neglect the masses of the first two fermion generations. Since $\mathcal{F}(x_f)$ and $\mathcal{K}(x_f)$ vanish for massless fermions, we only need to consider the top, bottom and tau contributions; the last two are negligible in the SM, but in the A2HDM could be enhanced by the
alignment factors $\varsigma_d$ and $\varsigma_{l}$. In $C_{\gamma\gamma}^{\varphi^0_i}$ we have also considered the contribution from a charged-scalar loop parametrized by
\be
\mathcal{C}_{H^\pm}^{\varphi^0_i}\; =\; \frac{v^2}{2M_{H^\pm}^2}\; \lambda_{\varphi^0_i H^+H^-}\; \mathcal{A}(x_{H^\pm})\, ,
\ee
with $x_{H^\pm} = 4 M_{H^\pm}^2/M_{\varphi_i^0}^2$.
The explicit expressions of the different loop functions are:
\begin{align}  \label{functions}
\mathcal{F}(x)\; &=\; \frac{x}{2} [4+(x-1)f(x)]\, , \qquad &
\mathcal{G}(x)\; &=\; -2 -3x + \Big(\frac{3}{2}x - \frac{3}{4}x^2\Big)f(x)\, ,
\notag \\
\mathcal{A}(x)\; &=\; -x - \frac{x^2}{4}f(x)\, ,  &
 \mathcal{K}(x) \; &= \;  - \frac{x}{2} f(x)   \,,
\end{align}
with
\begin{equation}
f(x)\; =\; \begin{cases} -4\arcsin^2(1/\sqrt{x})\, , \quad & x\geqslant1 \\[3pt] \Big[\ln\Big( \frac{1+\sqrt{1-x}}{1-\sqrt{1-x}}\Big)- i\pi \Big]^2\, , & x<1 \end{cases} \, .
\end{equation}

\section{Phenomenological Analysis}
\label{sec:fits}

We are interested in analyzing the current LHC and Tevatron data within the A2HDM. The experimental information on the new neutral boson is certainly in early stages; some decay channels have very big uncertainties while some others have not even been seen yet. Nevertheless, while more precise information on all possible production and decay channels is necessary in order to make a detailed study, present data already allow us to extract significant constraints on the parameter space of the model.

The deviations from the SM expectations originate from several sources. The three neutral scalars of the A2HDM have couplings to the gauge bosons which are different (smaller in absolute value) than the ones of the SM Higgs: in SM units they are given by ${\cal R}_{i1}$. The Yukawa couplings get also multiplied by the factors $y_{f}^{\varphi_i^0}$, which are functions of ${\cal R}_{ij}$ and the parameters $\varsigma_f$. Moreover, the presence of a charged scalar manifests in one additional one-loop contribution to the $\varphi_i^0\to 2\gamma$ decay amplitudes, parametrized through the constants $\mathcal{C}_{H^\pm}^{\varphi^0_i}$. In the limit of CP conservation, there are two clear candidates for the new scalar, the CP-even fields $h$ and $H$  (we will nevertheless analyze later the unlikely $A$ possibility). The A2HDM allows in addition for physical CP-violating phases, both in the scalar potential and the Yukawa couplings,  generating mixings among the three neutral scalars and CP-odd contributions to the Higgs-like signal strength parameters.  Being quadratic in the CP-violating parameters, this last type of corrections could be expected to be small. However, the current bounds on the A2HDM couplings still allow for sizeable effects
\cite{Pich:2009sp,Pich:2010ic,Jung:2010ik,Jung:2010ab,Jung:2012vu,Celis:2012dk}.

Sensitivity to the top-quark Yukawa coupling and to a lesser extent to the bottom coupling appears through the one-loop production mechanism of gluon fusion and in the $\gamma \gamma$ decay channel. Neutral scalar production via $pp \rightarrow t \varphi_i^0 j (b)$ could provide complementary information on the top Yukawa coupling when more data becomes available~\cite{Farina:2012xp,Biswas:2012bd}.  The most important constraints on the bottom Yukawa coupling come indirectly from the total decay width, which is in general dominated by $\varphi_i^0\to b\bar b$, and the measurement of scalar production with an associated vector boson ($q\bar q^{\prime}\to \varphi_i^0 V \rightarrow (b \bar b) V$). Neutral boson production via top-quark fusion with subsequent decay into a pair of $b$ quarks, $q\bar q/gg\to t\bar t \varphi_i^0 \rightarrow t\bar t (\bar b b)$, in which the bottom and top Yukawa couplings appear at tree level will also play an important role; the current experimental sensitivities in this channel are still low~\cite{CMS:2012ywa,ATLAS:2012cpa}. The $\tau$ Yukawa coupling is directly tested through $\varphi^0_i\to\tau^+ \tau^-$, the most accessible production mechanisms at the LHC being in this case vector-boson fusion, associated production with a vector boson and gluon fusion.

For a given choice of neutral scalar-field candidate $\varphi_i^0$ and its couplings, we define the $\chi^2$ function as
\begin{align}
\chi^2(\varphi_i^0)\; =\; \sum_k\; \frac{\left(\mu_k^{\varphi_i^0}  - \hat{\mu}_k\right)^2}{\sigma_k^2}\, ,
\end{align}
where $k$ runs over the different production/decay channels considered,
$\hat{\mu}_k$ and $\sigma_k$ are the measured Higgs signal strengths and their one-sigma errors, respectively, and $\mu_k^{\varphi_i^0}$ the corresponding theoretical predictions in terms of the A2HDM parameters, as given in Eqs.~(\ref{eq:ratios}) and (\ref{ratios2}). Scanning over the allowed parameter space, we then look for those sets of couplings minimizing the $\chi^2$ and their corresponding uncertainties. The details about the statistical treatment and data used in this work are presented in appendix \ref{statistical}.

We will first analyze the CP-conserving limit in section~\ref{sec:real}, where we will also study some particular scenarios often adopted in previous works.
In section~\ref{sec:complex} we will discuss the most general case, without making any assumption about the scalar potential, and analyze the present constraints on the complex Yukawa couplings of the assumed 126 GeV scalar boson.

\subsection{The A2HDM in the CP-conserving limit}
\label{sec:real}

Assuming that the Lagrangian preserves the CP symmetry, the two CP-even neutral scalars $h$ and $H$ couple to the gauge bosons with reduced couplings ${\cal R}_{11} = \cos \tilde \alpha$ and ${\cal R}_{21} = - \sin  \tilde \alpha$, respectively, and their Yukawa couplings are real:
\begin{equation}  \label{hyHYukawas}
y_{f}^h \; =\; \cos\tilde{\alpha} + \varsigma_f \sin\tilde\alpha \, ,
\qquad\qquad
y_{f}^H \; =\; -\sin\tilde{\alpha} + \varsigma_f \cos\tilde\alpha \, .
\end{equation}
The CP-odd boson $A$ does not couple at tree-level to $W^+W^-$ and $ZZ$ (${\cal R}_{31} = 0$), while its fermionic couplings are purely imaginary (pseudoscalar interaction):
\be \label{AYukawas}
y_{d,l}^A \; =\; i\,\varsigma_{d,l}\, ,
\qquad\qquad
y_u^A \; =\; -i\,\varsigma_u \, .
\ee

\subsubsection{A light CP-even Higgs at 126 GeV}
\label{sec:light}

We will first focus in the most plausible possibility that the lightest scalar $h$ corresponds to the observed neutral boson with $M_h= 126$~GeV. The alternative choice of the heavier field $H$ can be easily recovered through an appropriate change of the mixing angle, $\tilde \alpha \rightarrow \tilde \alpha - \pi/2 $, and will be further discussed in section~\ref{heavy}. We will also consider later, in section~\ref{CPodd}, the more exotic case of a CP-odd Higgs $A$.
In this first analysis we assume that the charged scalar is either very heavy or its coupling to the neutral Higgs is very small, so that its contribution $\mathcal{C}^h_{H^\pm}$ to the $h\to\gamma \gamma$ decay width  is negligible. We also assume that the bounds from flavour physics are naturally evaded, as it is the case at large values of the charged scalar mass. The $H^\pm$ contribution to the diphoton decay width as well as the flavour constraints will be considered later in section~\ref{sec:charged}.

The minimization of $\chi^2(h)$ leads to two different solutions, differing in the sign of the top Yukawa coupling. The central values of the corresponding A2HDM parameters and their statistical one-sigma errors obtained from the global fit are:
\begin{align}    \label{fitreal1}
\cos \ta &= 0.99^{+0.01}_{-0.06}\, ,  &   y_u^h &= 0.8^{+0.1}_{-0.2}\, ,  & \left| y_d^h\right|  &= 0.7\pm 0.3\, ,  &  \left| y_l^h\right|  =  0.8\pm 0.5  \,,
\end{align}
and
\begin{align}    \label{fitreal2}
\cos \ta &= 0.99^{+0.01}_{-0.04}\, ,  &  y_u^h &= -0.8^{+0.1}_{-0.3}\, ,  & \left| y_d^h\right|  &= 1.1\pm 0.3\, ,  & \left| y_l^h\right|  =  0.9 \pm 0.5  \, .
\end{align}
In both cases, the gauge coupling $g_{hVV}$ is very close to the SM one.
Changing simultaneously the signs of $\cos \tilde \alpha$ and $y_f^h$ leads obviously to identical Higgs signal strengths and, therefore, to two equivalent solutions.

In the first solution the $W^{\pm}$ and top-quark loops contribute with different signs to the $h\to \gamma\gamma$ amplitude, giving a destructive interference as in the SM. The needed enhancement of the $2\gamma$ branching ratio is obtained through a smaller total decay width, $\rho(h)\approx 0.6$.
This pushes upward the ratios $\mu^h_{\gamma\gamma}$ and $\mu^h_{\gamma\gamma jj}$, allowing to explain part of the excess experimentally observed in these two channels. However, the gluon-fusion production channel has a smaller cross section than in the SM. The combined effect results in a small increase of the $\gamma\gamma$ channel, $\mu^h_{\gamma\gamma}\approx 1.1$, while a much larger enhancement remains in the $\gamma\gamma jj$ case, $\mu^h_{\gamma\gamma jj}\approx 1.5$.

The second solution corresponds to a top-quark contribution to $h\to \gamma\gamma$ with the opposite sign, so that it interferes constructively with the $W^{\pm}$ amplitude. This allows one to explain the $2\gamma$ excess without hardly modifying the total decay rate, $\rho(h)\approx 1.1$ and providing a slightly better fit.

In both solutions there is a sign degeneracy in the bottom and tau Yukawa couplings. Although the tree-level decays $h\to \bar b b$ and $h\to \tau^+ \tau^-$ are insensitive to these signs, the loop-induced processes $gg\rightarrow h$ and $h \rightarrow \gamma \gamma$ receive contributions from the bottom and tau (only the $\gamma\gamma$ process) Yukawas, which interfere with the leading top and $W^\pm$ (in the $\gamma\gamma$ decay) amplitudes as shown in \eqref{gluonfusion} and \eqref{gammascaling}. In the SM the bottom and tau contributions are negligible, but their effect could be relevant in the A2HDM if the top Yukawa coupling is considerably suppressed or if the parameters $\varsigma_{d,l}$ are large. However, this is not the case for the fitted Yukawa values in (\ref{fitreal1}) and (\ref{fitreal2}), which are of $\mathcal{O}(1)$ for both solutions, leaving the sign of the bottom and tau Yukawas undetermined.   The relevance of the $\tau^+ \tau^-$ and $\bar b b$ channels to determine possible deviations from the SM and within the different $\mathcal{Z}_2$ versions of the 2HDM, which could be pointing to a more general Yukawa structure as provided by the A2HDM, has been emphasized recently in Ref.~\cite{Altmannshofer:2012ar}.

\begin{figure}[tb]
\centering
\includegraphics[width=7cm,height=7cm]{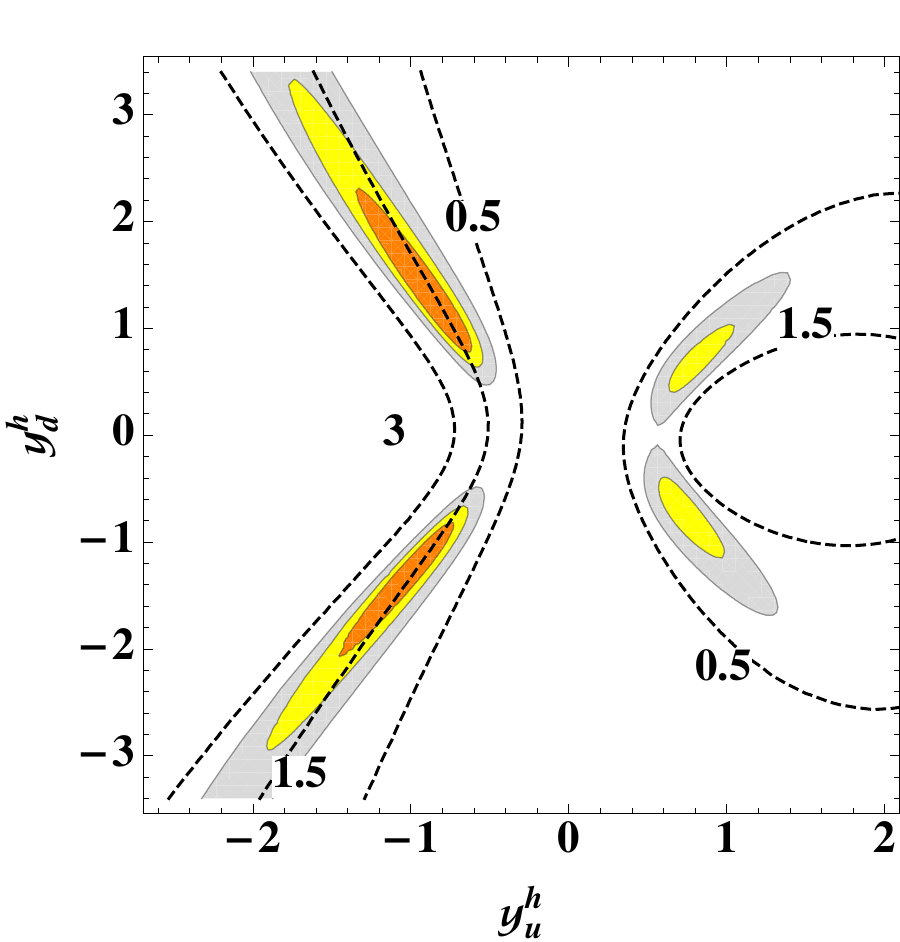}
~
\includegraphics[width=7cm,height=7cm]{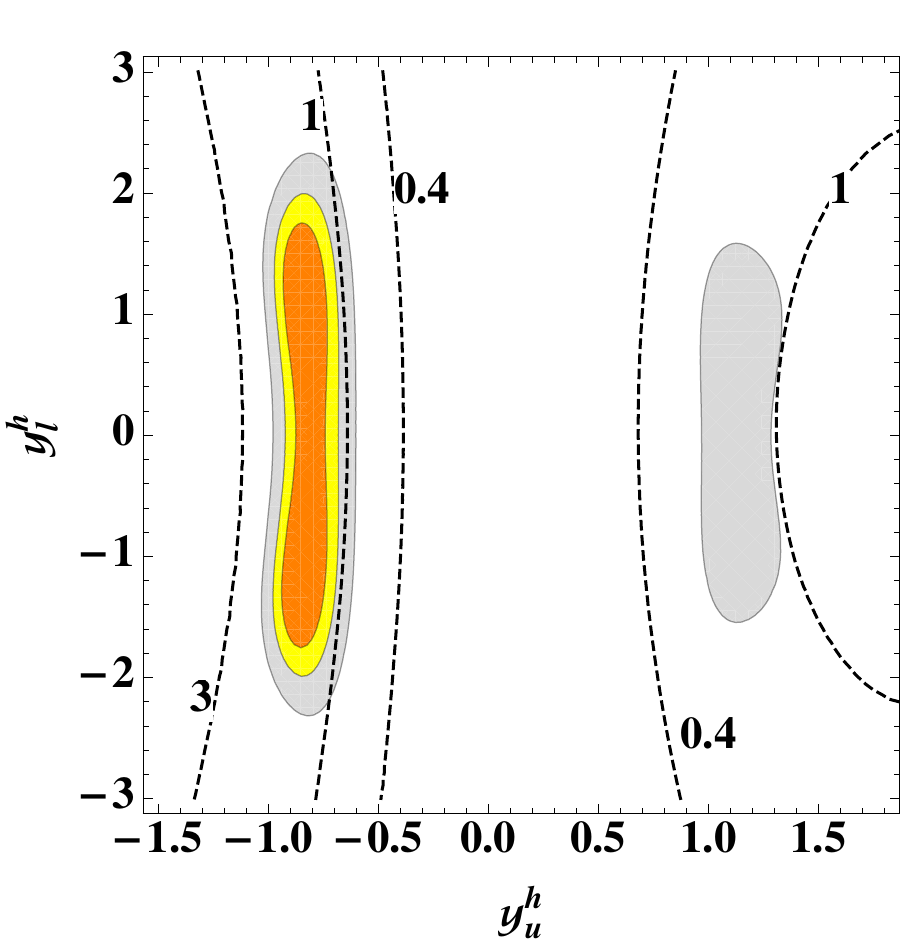}\\
\caption{\label{fig:Real} \it \small
Global fit to the A2HDM, in the CP-conserving case, in the planes $y^h_u - y_d^h$ (left) and $y^h_u - y_l^h$ (right). The parameters not shown in each case are fixed to the best global-fit point.  The orange, yellow and gray areas denote 68\%, 90\% and 99\%~CL regions.
The dashed lines correspond to fixed values of $\mu^h_{\gamma \gamma}$.}
\end{figure}

In Fig.~\ref{fig:Real} we show graphically the results of this global fit, giving the allowed regions in the $y^h_u - y_d^h$ (left) and $y^h_u - y_l^h$ (right) planes at 68\%, 90\% and 99\%~CL. The parameters that are not shown are, in each case, set to the best global-fit point. The sign degeneracy in the $\tau$ and $b$ Yukawa couplings is clearly observed. Moreover, the right panel shows a somewhat reduced sensitivity to the leptonic coupling $y_l^h$.
The SM-like solution $(y_u^h, y_{d,l}^h) = (1,1)$ lies inside the 90\% CL allowed region; however, at 68\% CL the top Yukawa has the sign flipped with respect to the SM, {\it i.e.}, only the solution (\ref{fitreal2}) remains. Similar results have also been obtained in Ref.~\cite{Altmannshofer:2012ar,Cheung:2013kla}.

\begin{figure}[tb]
\centering
\includegraphics[width=8.8cm,height=6cm]{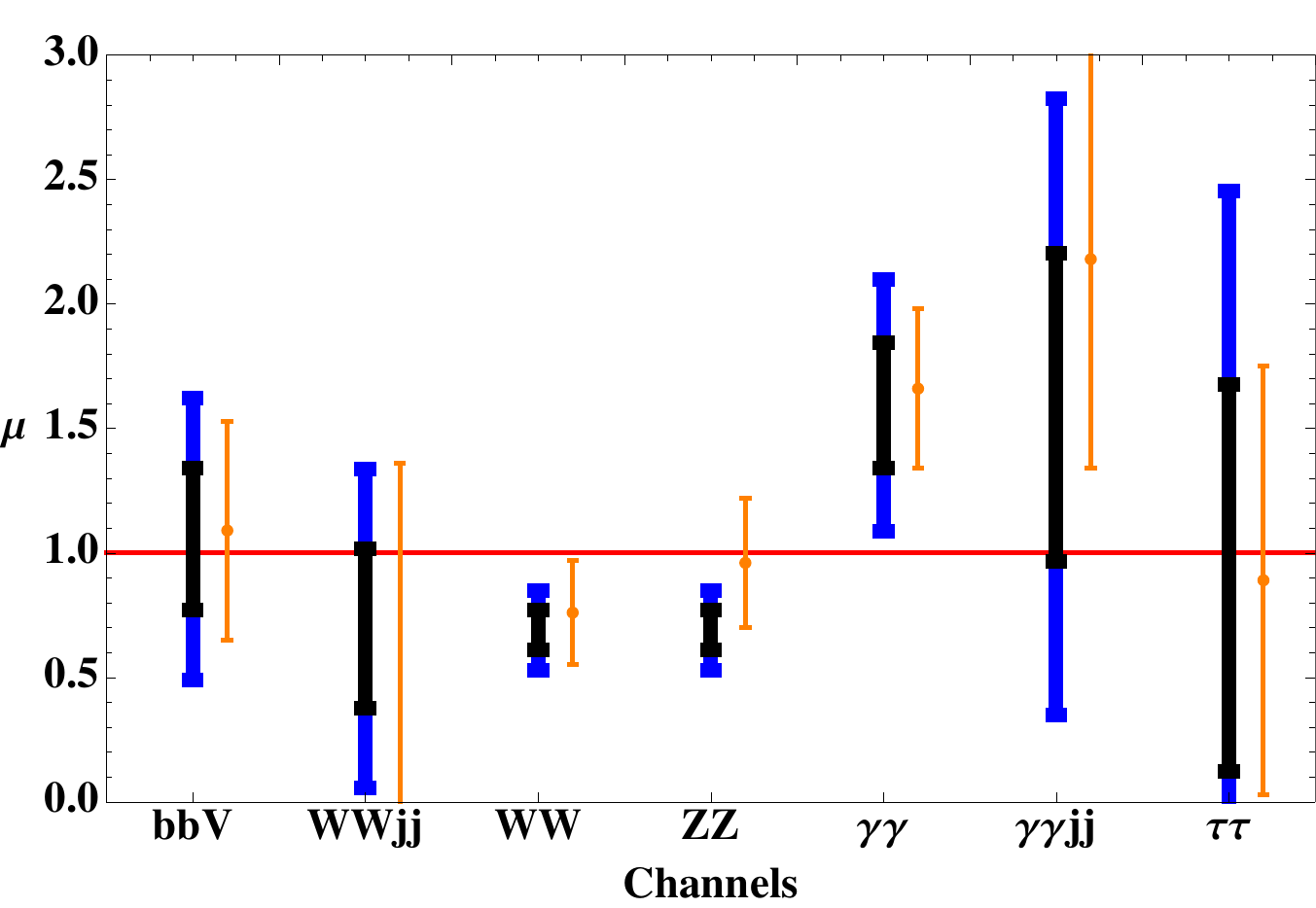}
\caption{\label{fig:RealH} \it \small  Allowed ranges for the Higgs signal strengths obtained from the fit \eqref{fitreal2} at $1 \sigma$ (black, dark) and $2 \sigma$ (blue, dark), together with the averaged experimental data from the ATLAS, CMS, CDF and D\O~collaborations with the corresponding $1\,\sigma$ errors (orange, light).}
\end{figure}

The allowed ranges, at the $1 \sigma$ and $2 \sigma$ level, for the different Higgs signal strengths in the fit (\ref{fitreal2}) are compared in Fig.~\ref{fig:RealH} with the experimental values. A good agreement with data is obtained in all cases.
Previous analyses within the CP-conserving A2HDM
have been performed  in Refs.~\cite{Altmannshofer:2012ar,Bai:2012ex},
using a different notation, also finding good agreement with the data.

Using the sum rules in Eqs.~\eqref{sumrules} and \eqref{sumrule}, we can extract constraints on the heavy CP-even Higgs couplings from our global fit with $M_h = 126$~GeV.   For the solution (\ref{fitreal2}) we find at $68\%$~CL that the coupling of $H$ to vector bosons is suppressed, $\sin{\ta} < 0.37$, while its coupling to top quarks is very large, $|y_u^H| > 4.6$.  This region of parameter space requires a very large value of $|\varsigma_u|$ in order to flip the sign of $y_u^{h}$, which is the top Yukawa of $h$.   Such large values of $|\varsigma_u|$ would then imply a significant enhancement of the production of $H$ via gluon fusion and can give rise to non-perturbative $H^+\bar t b$,  $H\bar t t$ and $A\bar t t$ couplings.   This was noted previously within the same context in Ref.~\cite{Altmannshofer:2012ar}.

\subsubsection{Global fit within $\mathcal{Z}_2$ models}

\begin{figure}[tb]
\centering
\includegraphics[width=6.3cm,height=6.3cm]{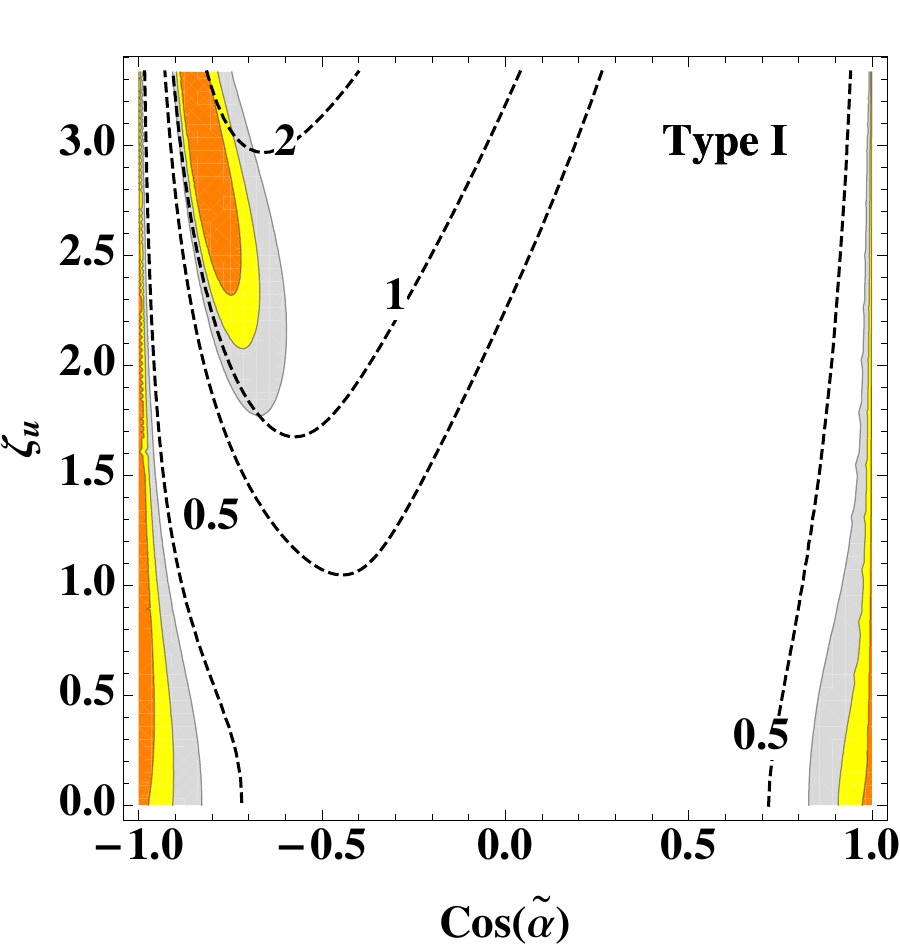}
~
\includegraphics[width=6.3cm,height=6.3cm]{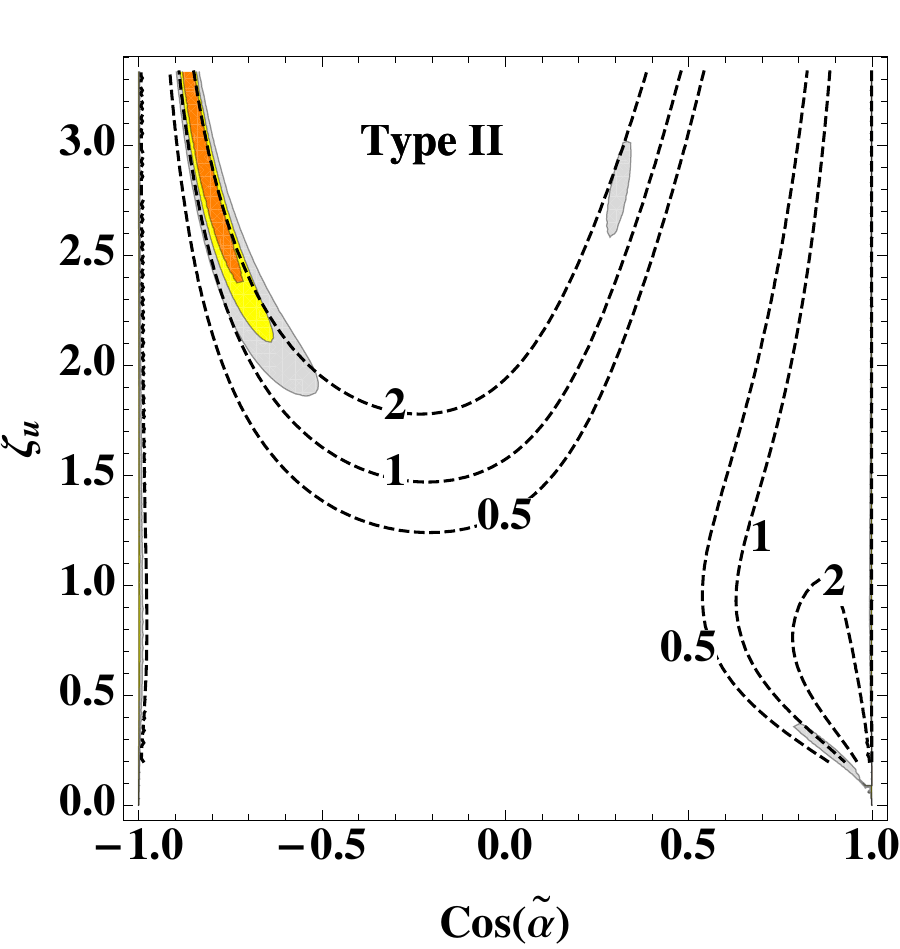} \\
\includegraphics[width=6.3cm,height=6.3cm]{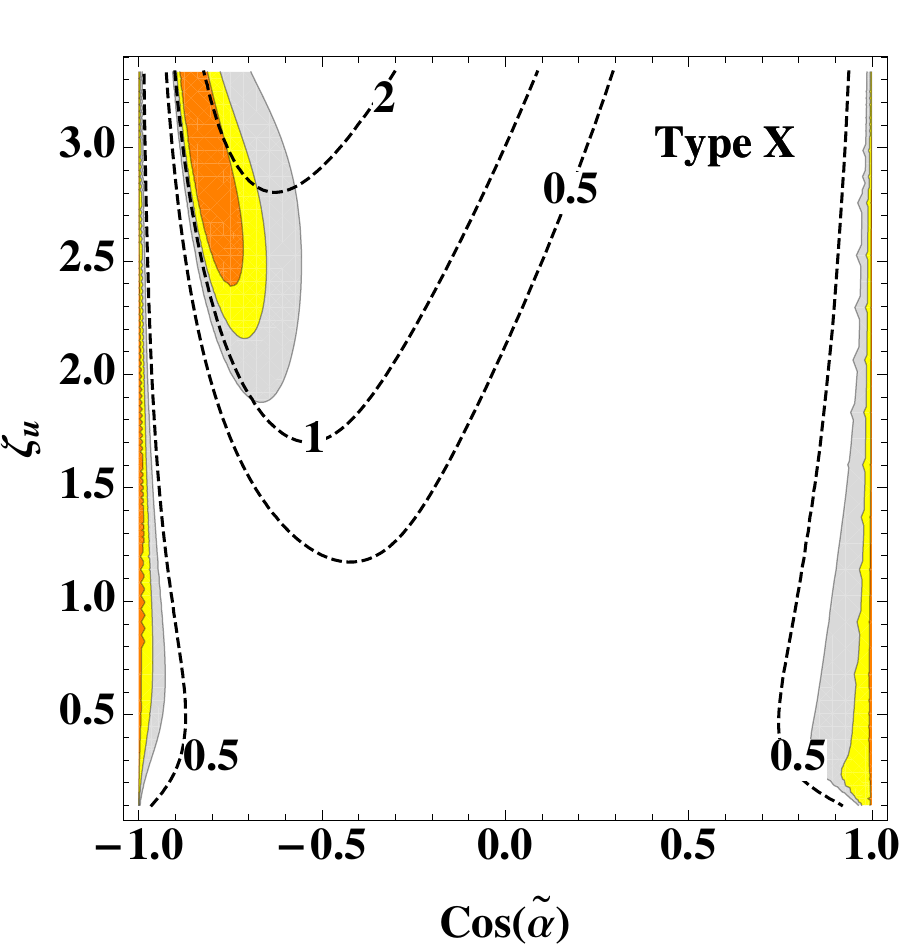}
~
\includegraphics[width=6.3cm,height=6.3cm]{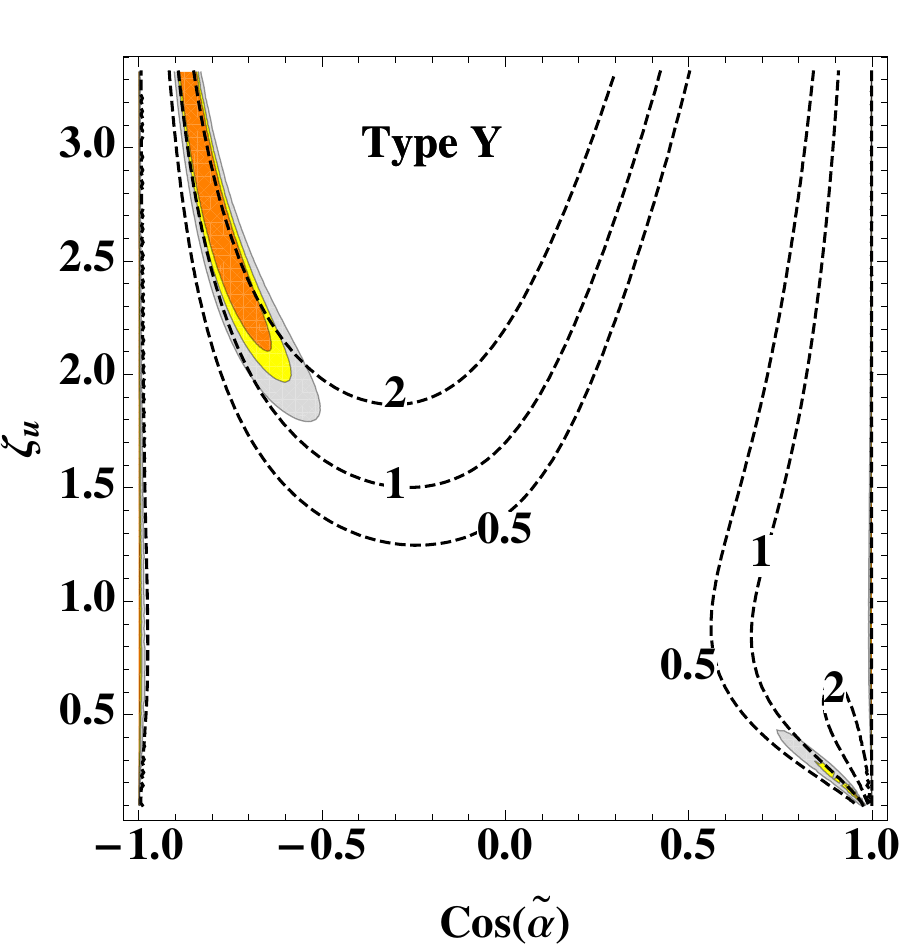}
\caption{\label{TypeZ2I} \it \small  Global fit within 2HDMs of types I (upper left), II (upper right), X (lower left) and Y (lower right), at 68\% (orange), 90\% (yellow) and 99\% (gray)~CL.  The dashed lines correspond to constant values of $ \mu^h_{\gamma \gamma}$.}
\end{figure}

The usual 2HDMs with natural flavour conservation, based on discrete $\mathcal{Z}_2$ symmetries, are particular cases of the CP-conserving A2HDM, with $\varsigma_f$ taking the values given in Table~\ref{tab:models}. Thus, the three alignment factors are determined by a single parameter through the constraints
$\varsigma_u = \varsigma_d = \varsigma_l = \cot{\beta}$ (type I),
$\varsigma_u = -\varsigma_d^{-1} = -\varsigma_l^{-1} = \cot{\beta}$ (type II),
$\varsigma_u = \varsigma_d = -\varsigma_l^{-1} = \cot{\beta}$ (type X) and
$\varsigma_u = -\varsigma_d^{-1} = \varsigma_l = \cot{\beta}$ (type Y), with
$\cot{\beta} = v_1/v_2 \ge 0$.
This leads to specific relations among the production cross sections and decay rates for the Higgs bosons that can be tested with the LHC data.    The separate measurement of the various Higgs signal strengths should allow to disentangle the different scalings of the three Yukawa couplings. In particular, exclusive Higgs production measurements in the final states $\tau^+ \tau^-$ and $b \bar b$ will be crucial to test the different $\mathcal{Z}_2$ versions of the 2HDM~\cite{Craig:2012vn,Chen:2013kt,Altmannshofer:2012ar}.

Figure~\ref{TypeZ2I} shows the results of the global fit for the 2HDMs of types I, II, X and Y,
assuming that the lightest neutral Higgs $h$ is the boson observed around 126 GeV.
Allowed regions at 68\%, 90\% and 99\%~CL are shown, together with lines of constant $\mu^h_{\gamma \gamma}$.  The relevance of the diphoton channel is evident from the figure. In models I and X, an allowed region around $\cos{\tilde \alpha} \approx 1$ appears, where there is no sensitivity to $\varsigma_u$ since its contribution to the neutral Yukawa couplings is suppressed by $\sin{\tilde \alpha}$; in this region the couplings of $h$ to vector bosons and fermions are close to the SM ones.
Another allowed region appears for negative values of $\cos{\ta}$, in which the $W^\pm$ and top-quark loops contribute with the same sign to the $h\to\gamma \gamma$ decay amplitude, thus allowing for a constructive interference. Both solutions with $\cos{\ta}\approx \pm 1$ are present for the inert model (type I with $\varsigma_u=0$). There is a third allowed region at large values of the top Yukawa and negative $\cos{\ta}$, which approaches $\cos{\ta}=-1$ as $\varsigma_u$ increases.

\begin{figure}[tb]
\centering
 \includegraphics[width=8.2cm,height=5cm]{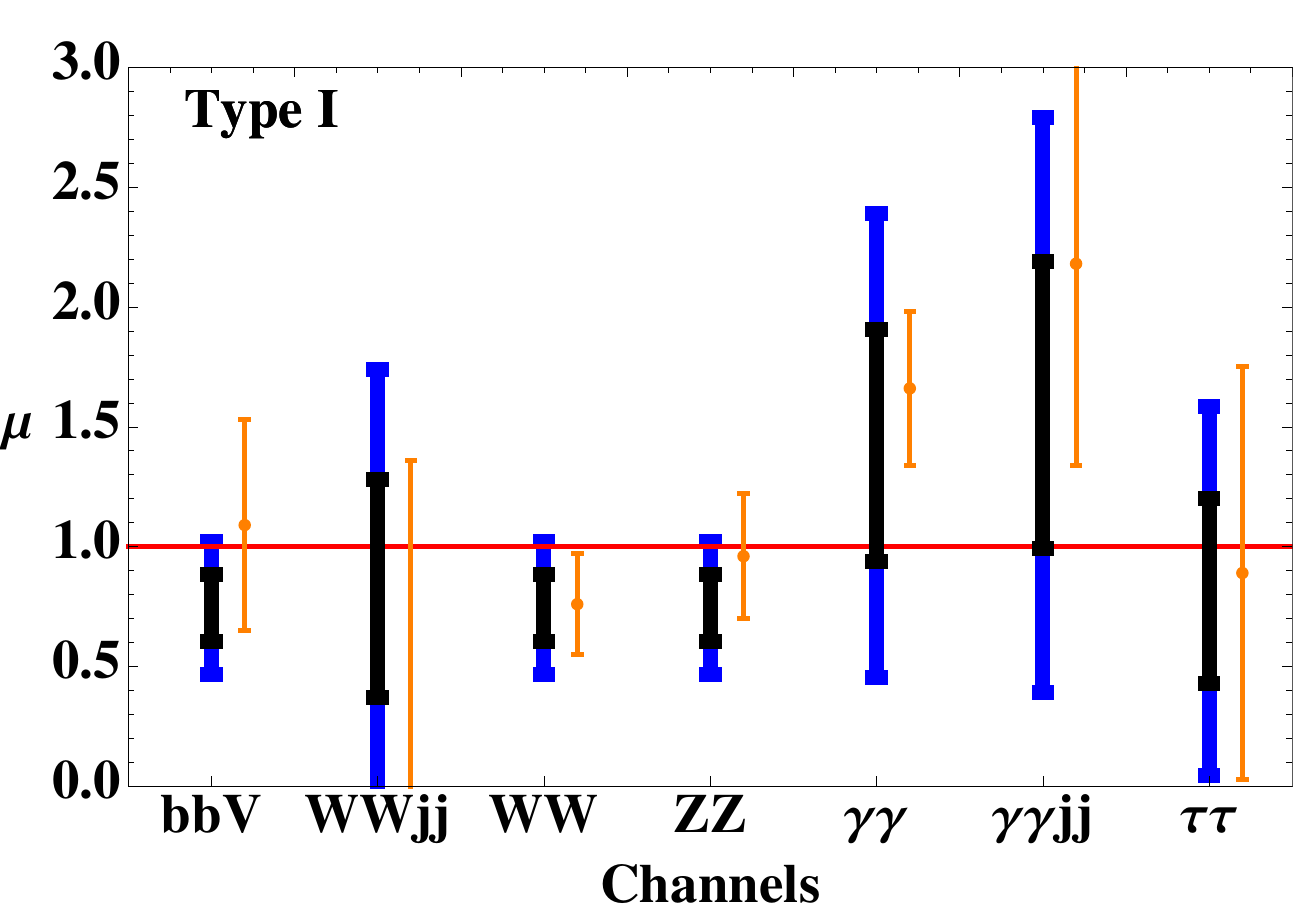}
~
\includegraphics[width=8.2cm,height=5cm]{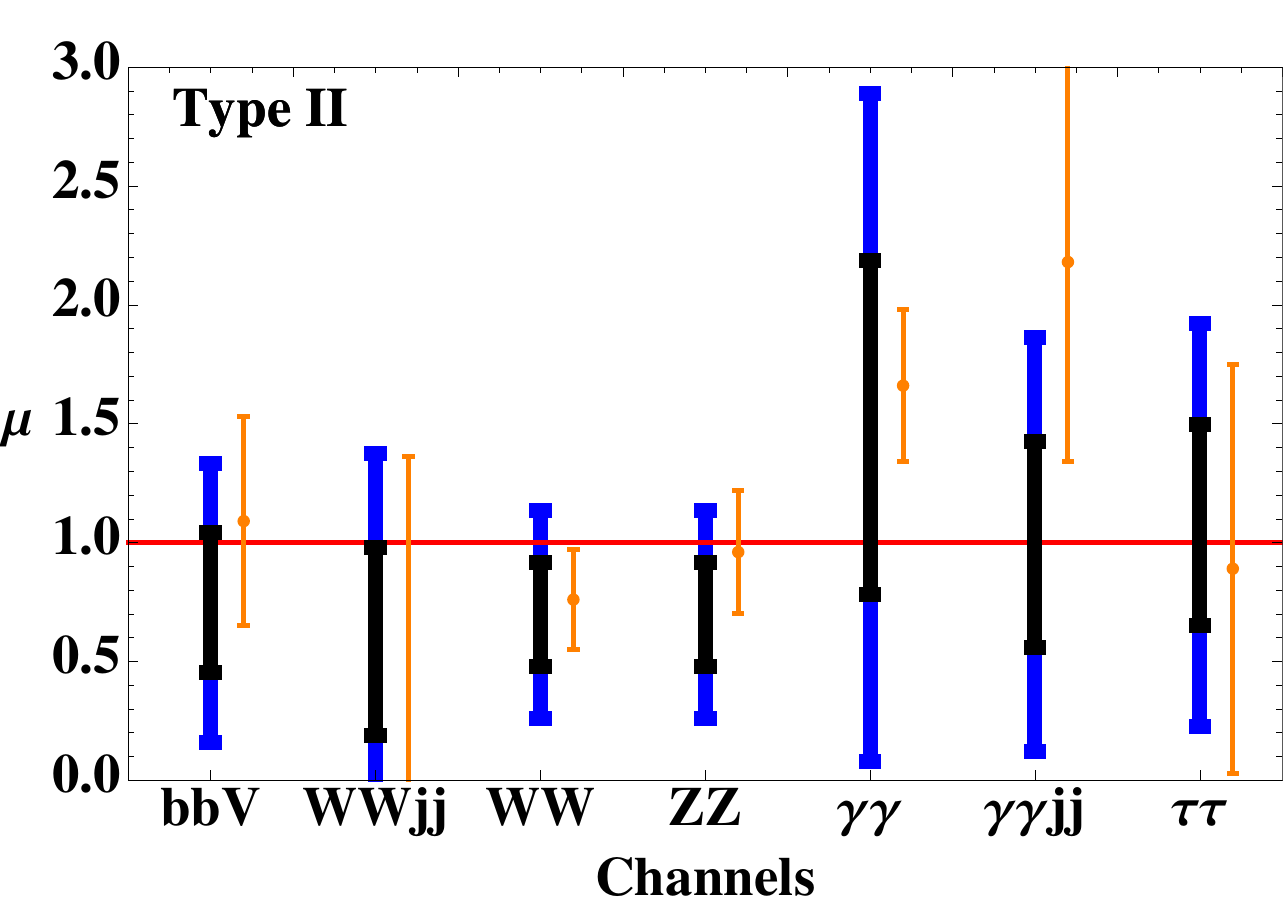}\\
\includegraphics[width=8.2cm,height=5cm]{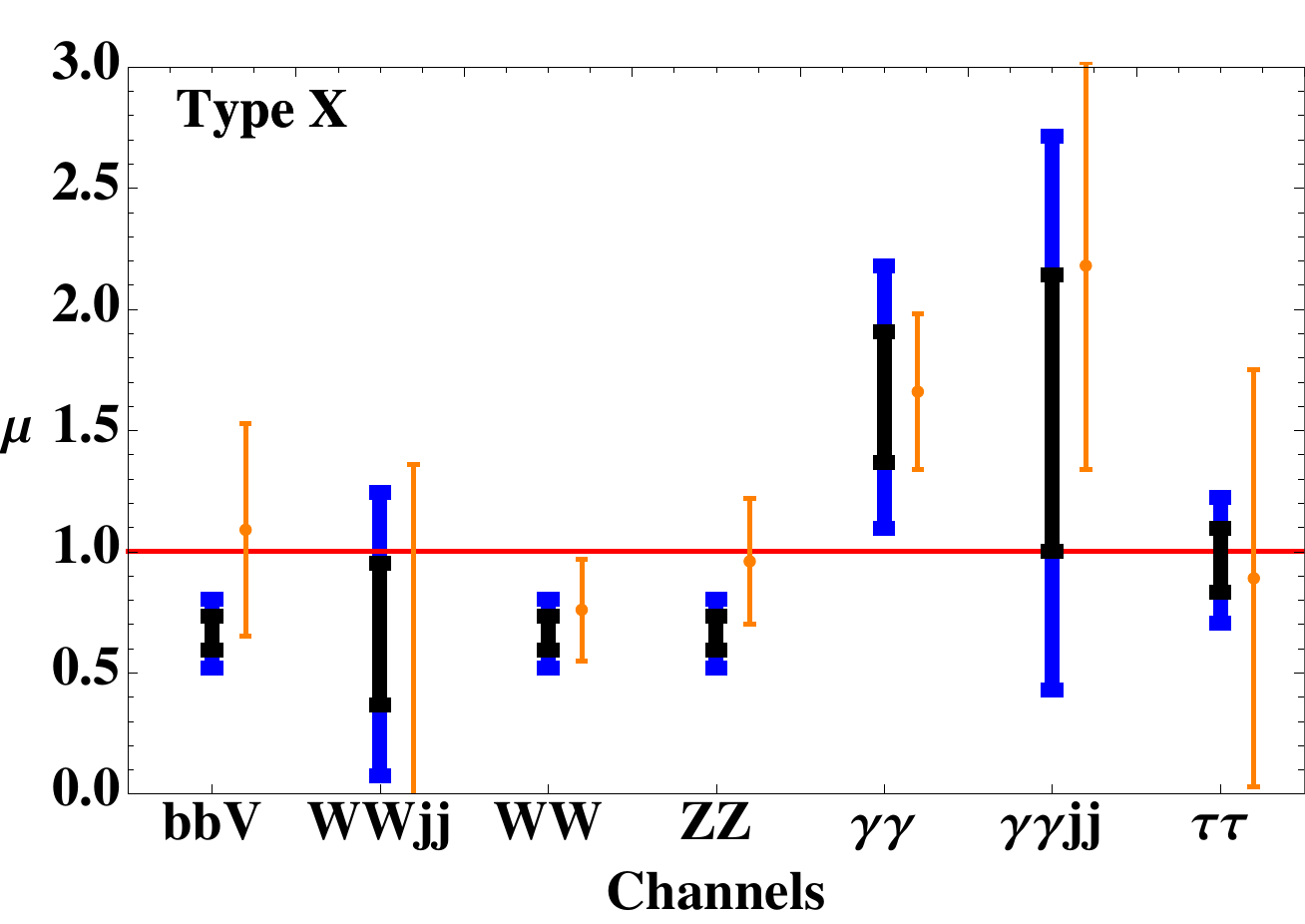}
~
\includegraphics[width=8.2cm,height=5cm]{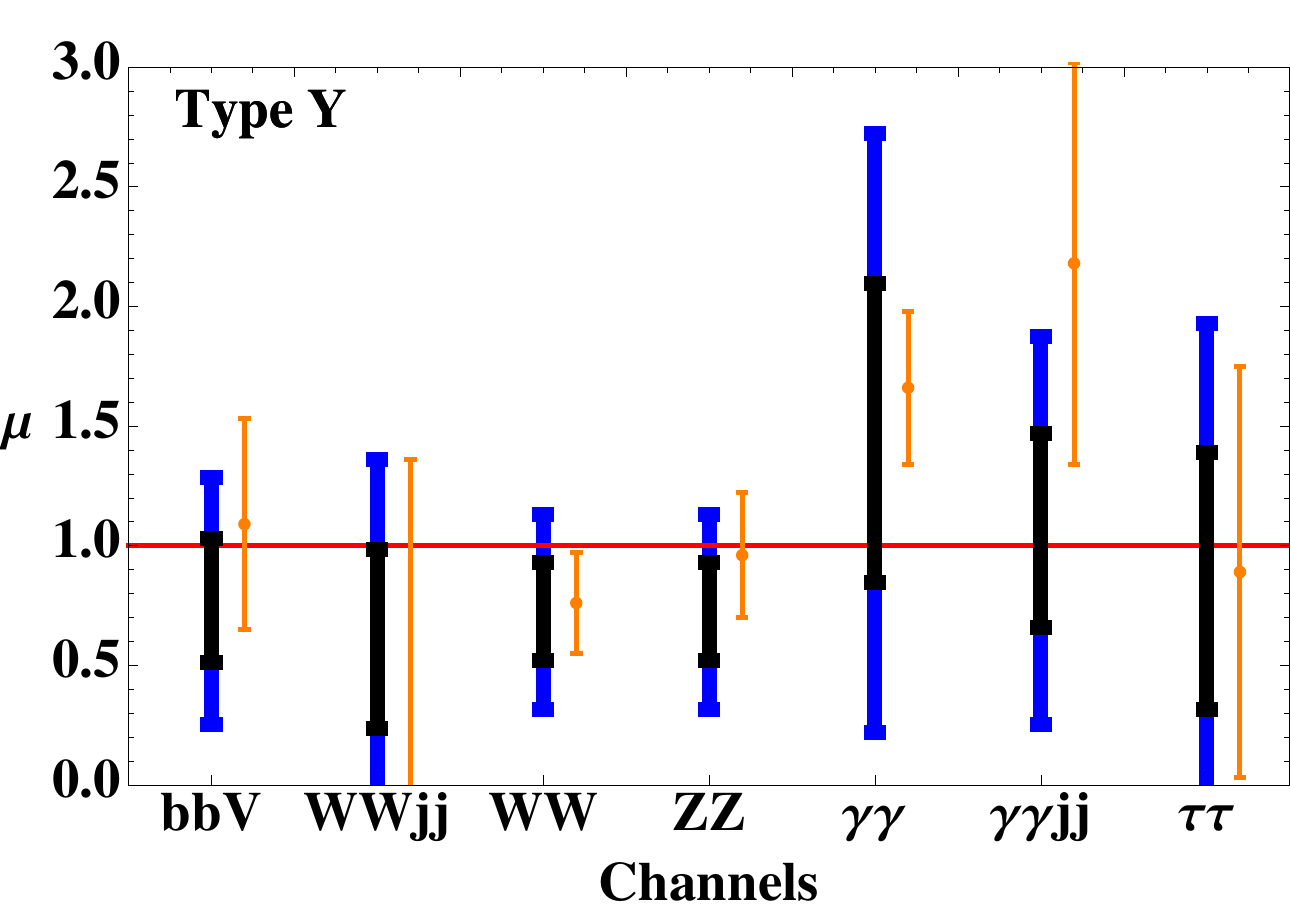}
\caption{\label{TypeZ2II} \it \small  Allowed ranges for the Higgs signal strengths in 2HDMs of type I, II, X and Y, at $1 \sigma$ (black, dark) and $2 \sigma$ (blue, dark). Other captions as in Fig.~\ref{fig:RealH}.}
\end{figure}

In models II and Y the solutions around $\cos{\ta}\approx \pm 1$ reduce to two extremely narrow vertical lines and one small region at low $\varsigma_u$ and positive $\cos{\ta}$, which remain allowed at 99\% CL but are not present at 90\%.   The solution at large values of the top Yukawa and negative $\cos{\ta}$ is also present, but in a region much smaller than in models I and X.

Figure~\ref{TypeZ2II} shows the allowed ranges for the Higgs signal strengths obtained in these four types of 2HDMs (I, II, X and Y). The agreement with the data is good; however, as already noted in Ref.~\cite{Chen:2013kt}, the preferred region has large values of $|\varsigma_u|$, which are ruled out from flavour physics constraints for a charged Higgs boson below the TeV scale.   Large values of $|\varsigma_u|$ can also make some top-quark Yukawa couplings non-perturbative, as commented in the previous section.

\subsubsection{A charged Higgs and the diphoton excess}
\label{sec:charged}

One of the most distinctive features of 2HDMs with respect to other alternative scenarios of electroweak symmetry breaking is the presence of a charged scalar boson in the spectrum. 
The present experimental lower bound on the $H^\pm$ mass is $M_{H^\pm} \gtrsim 80$ GeV (95\% CL)~\cite{Searches:2001ac}, assuming that the charged scalar $H^+$ only decays into the fermionic channels $H^+\to c\bar s$ and $H^+\to \tau^+\nu_\tau$. A slightly softer limit $M_{H^\pm} \gtrsim 72.5$ GeV is obtained, allowing for the decay $H^+\to W^+ A\to W^+ b\bar b$, with $M_A>12$~GeV, and assuming a type-I fermionic structure~\cite{Searches:2001ac}.
A model-independent bound can be extracted from the measured $Z$ width which constraints the $Z$ decays into non-SM modes, and in particular $Z\to H^+H^-$, to be below $\Gamma_Z^{\mathrm{non-SM}} < 2.9$~MeV (95\% CL); this implies $M_{H^\pm} \gtrsim 39.6$ GeV (95\% CL)~\cite{Searches:2001ac}.

Direct searches for charged Higgs bosons at the Tevatron~\cite{Abulencia:2005jd} and the LHC~\cite{Aad:2012tj} have also been performed with null results so far.

Current LHC data are sensitive to such charged scalar through the $h\to\gamma \gamma$ decay channel. The one-loop $H^\pm$ contribution can interfere with the $W^{\pm}$ and fermionic amplitudes, thus being able to enhance or suppress the decay rate. The  exact value of the charged Higgs contribution $C_{H^{\pm}}^{h}$ depends on the cubic Higgs coupling $\lambda_{hH^+H^-}$ and the charged Higgs mass $M_{H^\pm}$. One expects however that  $|C_{H^{\pm}}^{h}| \lesssim O(1)$ based on perturbativity arguments (see appendix \ref{perturbatvity}).

When considering a relatively light charged Higgs boson, one must take into account constraints from electroweak precision tests and the flavour sector; a light $H^\pm$ would contribute sizably to loop-induced processes, such as $Z \rightarrow \bar b b$, $b\to s\gamma$ or $B^0$--$\bar B^0$ mixing. These phenomenological constraints have been analyzed in detail within the framework of the A2HDM in~Refs.~\cite{Jung:2010ik,Jung:2010ab,Jung:2012vu,Celis:2012dk}, where it has been found that
a charged Higgs below the TeV scale would require $|\varsigma_u | \lesssim 2$ to be compatible with present data. This rules out the hypothetical scenario of a top Yukawa coupling with flipped sign, as found in \eqref{fitreal2} and also favoured by the fits shown in Fig.~\ref{TypeZ2I} within the four types of ${\mathcal Z}_2$ models. The reason is that current $h \rightarrow WW, ZZ, \gamma \gamma (jj)$ data require $|\cos{\ta} | \sim 1 $ ({\it i.e.}, the gauge coupling of the new neutral scalar should be close to the SM one). Since the top Yukawa coupling is given by $y_u^h = \cos{\tilde \alpha} + \varsigma_u\,\sin{\tilde \alpha}$, in order to flip the sign of $y_u^h$ one needs then a large value for $|\varsigma_u|$, which is excluded by the previous bound.

\begin{figure}[tb]
\centering
\includegraphics[width=8.2cm,height=6cm]{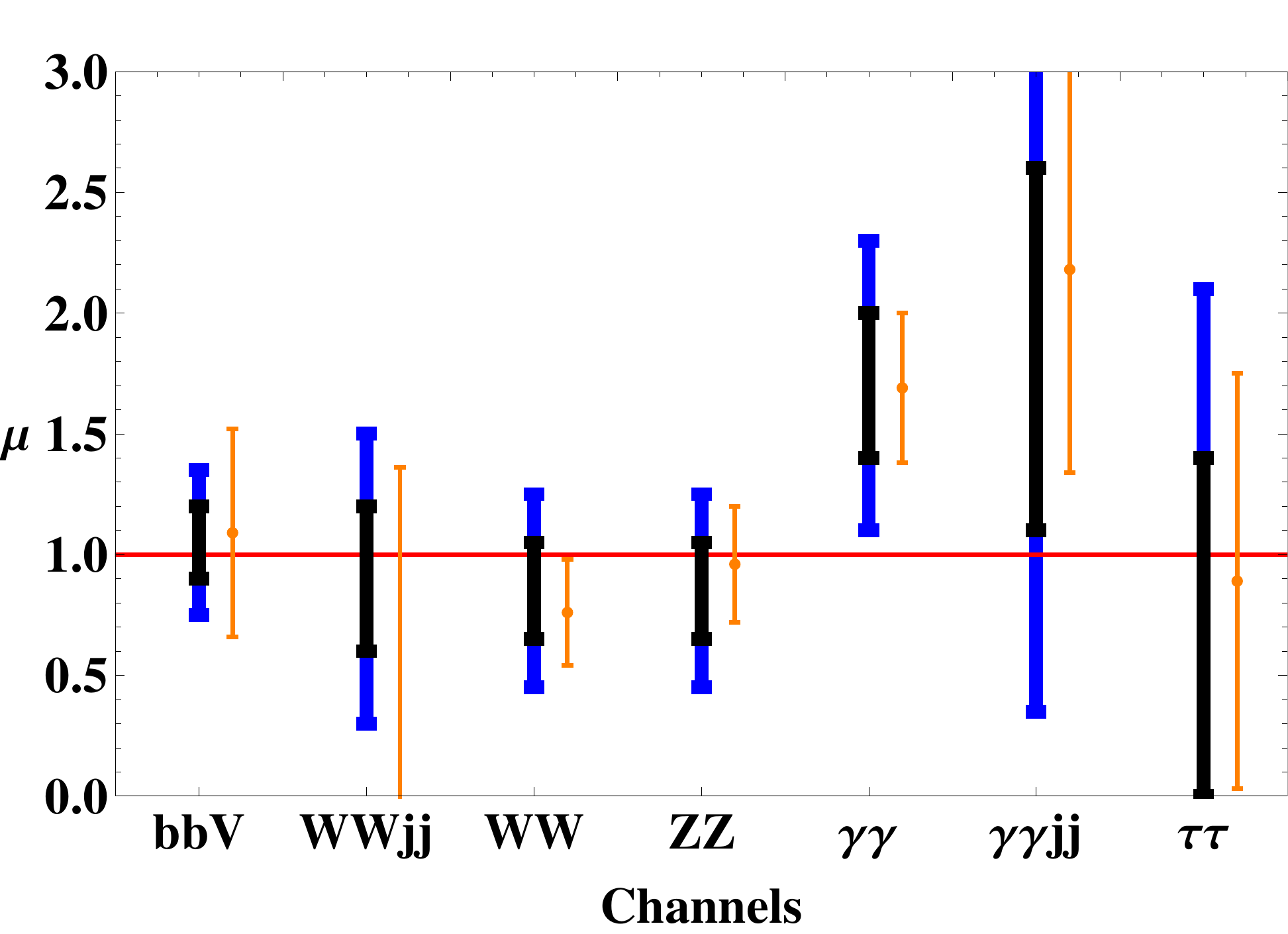}
\caption{ \it \small  Allowed ranges for the Higgs signal strengths from the global fit within the CP-conserving A2HDM, including the charged Higgs contribution to $h \rightarrow \gamma \gamma$, at $1 \sigma$ (black, dark) and $2 \sigma$ (blue, dark).  Other captions as in Figure~\ref{fig:RealH}.}
\label{firspl}
\end{figure}

Including the charged-Higgs contribution, it is no longer necessary to flip the sign of the top Yukawa in order to enhance the $h \rightarrow \gamma \gamma$ decay width.  The best fit region is now obtained for Yukawa and gauge couplings close to the SM limit:
\begin{eqnarray}
\cos{\ta} &=& 0.98^{+0.02}_{-0.06}\, ,
\qquad\qquad \qquad\quad
C_{H^\pm}^{h} \; =\; (-2.8 \pm 1.3)\,  \cup\, (16.0 \pm 1.3)\, ,
\nonumber\\
y_u^h &=& 1.0 \pm 0.2\, ,
\qquad\quad
\left| y_d^h\right|\; =\; 1.1 \pm 0.3 \, ,
\qquad\quad
\left| y_l^h \right|\; =\; 0.8 \pm 0.5\, .
\end{eqnarray}
The two disjoint $C_{H^\pm}^{h}$ solutions correspond to either a constructive interference of the $H^\pm$ and $W^\pm$ amplitudes or a destructive one but with a charged-Higgs contribution so large that it reverses the sign of the total $h\to 2\gamma$ amplitude.
In both cases, one obtains a better fit than in the SM and also better than the previous A2HDM fits (except for (\ref{fitreal2}) which is comparable to this one).   The presence of the charged Higgs allows one to easily explain the $h \to \gamma\gamma(jj) $ excess without large modifications of the total decay rate ({\it i.e.}, $\rho(h)\approx 1.1$).
The fit predictions for the $\mu_k$ ratios and their one and two-sigma statistical errors are shown in Fig.~\ref{firspl}. In all cases, good agreement with the data is obtained.

\begin{figure}[tb]
\centering
\includegraphics[scale=0.65]{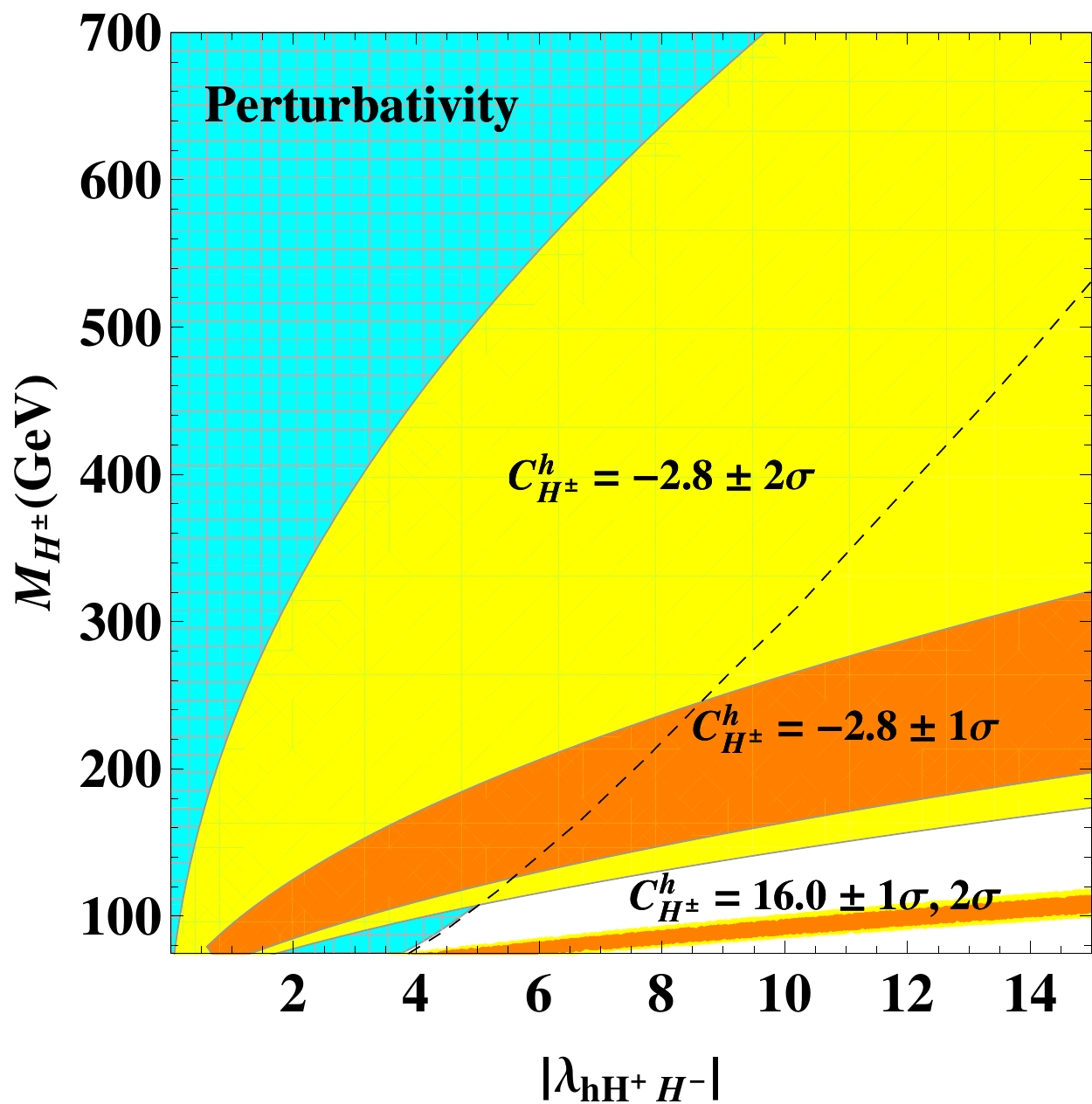}
\caption{  \it \small
Allowed regions of the $(\left|\lambda_{hH^+H^-}\right|,M_{H^\pm})$ plane, corresponding to the two possible fitted values of $C^h_{H^{\pm}}$, at 68\% (orange, dark) and 90\% CL (yellow, light).
The blue (hashed) area, between the left vertical axis and the dashed line, is the domain where the theory remains perturbative.}
\label{gbounds}
\end{figure}

In Fig.~\ref{gbounds} we show the allowed regions of the $(\left|\lambda_{hH^+H^-}\right|,M_{H^\pm})$ plane, corresponding to the two possible fitted values of $C^h_{H^{\pm}}$, at 68\% and 90\% CL, together with the perturbativity bounds discussed in appendix~\ref{perturbatvity}.
Clearly, the solution with a very large contribution to $h\to \gamma\gamma$ from the charged Higgs ($C_{H^{\pm}}^h \approx 16$) is excluded if one requires the theory to be perturbative.
We obtain an upper bound for the mass of the charged Higgs around 300~GeV, at the one-sigma level. However, the bound disappears at the two-sigma level because the charged-Higgs contribution becomes compatible with zero.

\subsubsection{Inert 2HDM}
In the inert 2HDM a ${\cal{Z}}_2$ symmetry is imposed, in the Higgs basis~\eqref{Higgsbasis}, under which all SM fields and $\Phi_1$ are even while $\Phi_2 \rightarrow - \Phi_2$.   Terms with an odd number of $\Phi_2$ fields in the scalar potential~\eqref{eq:potential} are then forbidden by the ${\cal{Z}}_2$ symmetry, therefore $\mu_3= \lambda_6 = \lambda_7 =0$. In this case there is no mixing between the CP-even neutral states $h$ and $H$, and the scalars $H$, $A$ and $H^{\pm}$ decouple from the fermions. The couplings of the remaining Higgs field $h$ to fermions and to vector bosons are the same than in the SM ({\it i.e.}, $\cos \tilde \alpha =1$ and $y_{f}^h=1$).
Thus, only the diphoton channels can show a deviation from the SM prediction (assuming that there are no open decay channels other than the SM ones). From the global fit of this scenario, we find a charged-Higgs contribution to the $h\to\gamma \gamma$ amplitude in the range $C^h_{H^{\pm}} \in [-1.7,-0.89]$ at 68\%~CL  and $C_{H^{\pm}}^h \in [-2.4,-0.1]$ at  90\%~CL. We have assumed that $M_{H^\pm}$ is greater than $M_h/2\approx 63$ GeV so that $C^h_{H^{\pm}}$ is real; for lower charged-Higgs masses, it would develop and imaginary absorptive part. The fitted negative sign of $C^h_{H^{\pm}}$ causes a constructive interference with the $W^{\pm}$ amplitude in the $h\to\gamma \gamma$ decay width.

Note that in the limit $\varsigma_f = 0$, the charged Higgs does not couple to fermions independently of any assumption on the scalar potential, see Eq.~\eqref{lagrangian}.  The implications of this more general case for the neutral Higgs boson phenomenology as well as the possibility of a very light charged Higgs boson are discussed in section~\ref{fermiophobic}.   Detailed analyses of the inert 2HDM and the possibility of a Dark Matter candidate within this model, in light of the LHC data, can be found in Refs.~\cite{Arhrib:2012ia,Gustafsson:2012aj}.  An enhancement of the $h\to\gamma \gamma$ decay rate has also been discussed in Ref.~\cite{Wang:2012zv}  within the Quasi-Inert 2HDM in connection with the top forward-backward asymmetry observed at the Tevatron; the limit on $C_{H^{\pm}}^h$ obtained in this section also applies to this scenario.

\subsubsection{A heavy CP-even Higgs at 126 GeV \label{heavy}}
\label{sec:heavy}

We have discussed so far the phenomenology of the lightest Higgs boson, but there is nothing a priori preventing the boson discovered by ATLAS and CMS to be identified with the heaviest CP-even state $H$ or with the CP-odd Higgs $A$.   These possibilities have been already discussed in Refs.~\cite{Ferreira:2012my,Altmannshofer:2012ar,Burdman:2011ki}.   An analysis in terms of the more general CP-violating scalar potential, setting limits on the scalar-pseudoscalar mixing, has been done in Ref.~\cite{Barroso:2012wz}.

Using the previous fits for $h$, it is straightforward to analyze the possibility of having a heavy Higgs with $M_H = 126$ GeV. Assuming that non-SM decays like $H\to hh$ are kinematically forbidden or very suppressed, the constraints on the heavy Higgs boson couplings can be easily obtained from those of $h$ through an appropriate change of the mixing angle: $\tilde \alpha \rightarrow \tilde \alpha - \pi/2 $.      In this case the coupling of the heavy Higgs to vector bosons is close to the SM limit ($\sin{\ta} \approx 1$), while the light-scalar $g_{hVV}$ couplings are
suppressed by $\cos{\ta} \approx 0$.  The absolute values of the Yukawa couplings and all the other parameters remain unchanged.    A solution analogous to the one in Eq.~\eqref{fitreal2}, where a large value of $|\varsigma_u|$ is required to flip the sign of the top Yukawa coupling, is excluded by low energy  flavour constraints for a charged Higgs below the TeV scale ($Z \rightarrow \bar b b$, $B^0 - \bar B^0$ mixing and neutral Kaon mixing~\cite{Jung:2010ik}).

\begin{figure}[tb]
\centering
\includegraphics[scale=0.29]{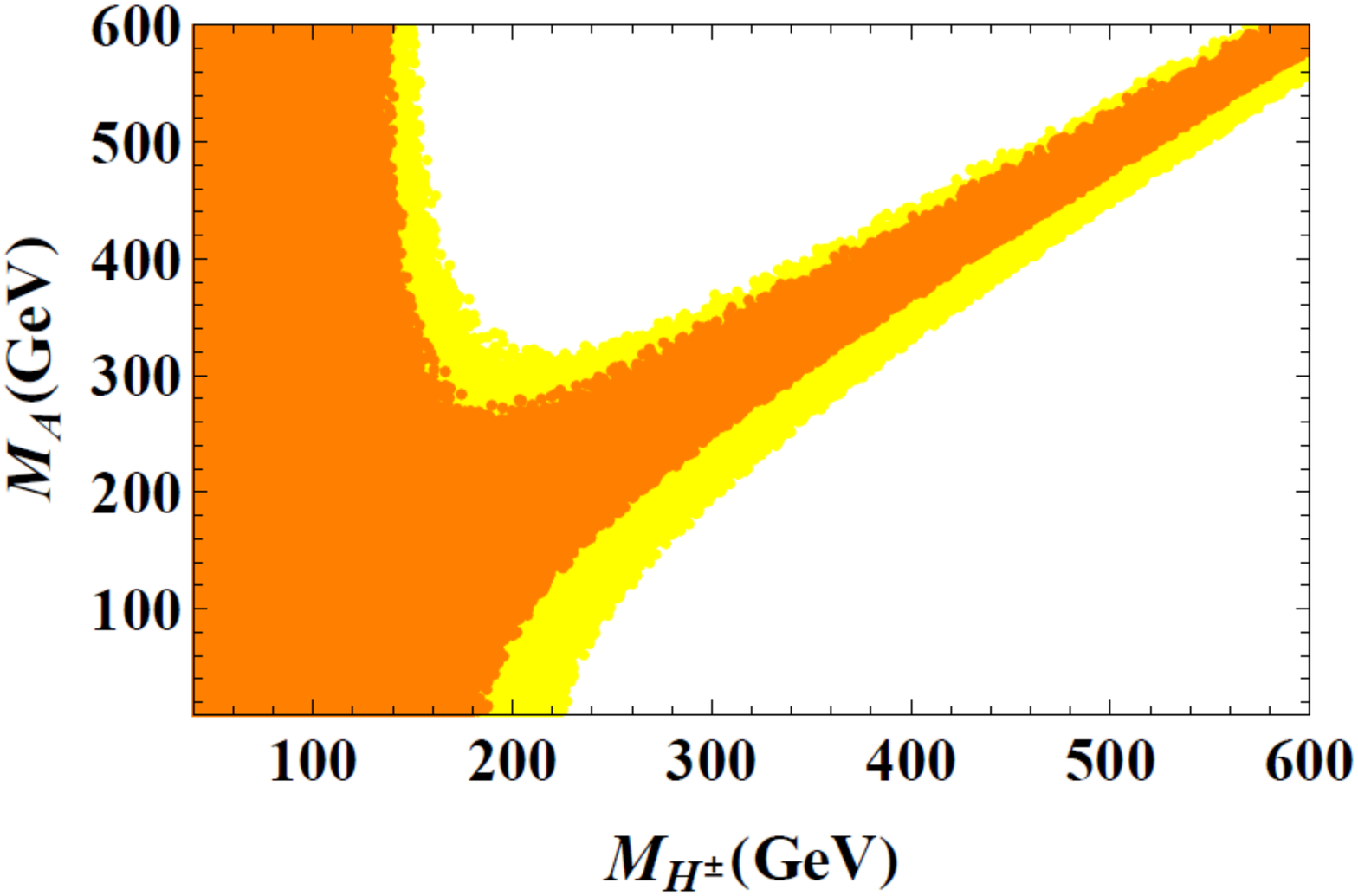}
\hskip 1.6cm
\includegraphics[scale=0.63]{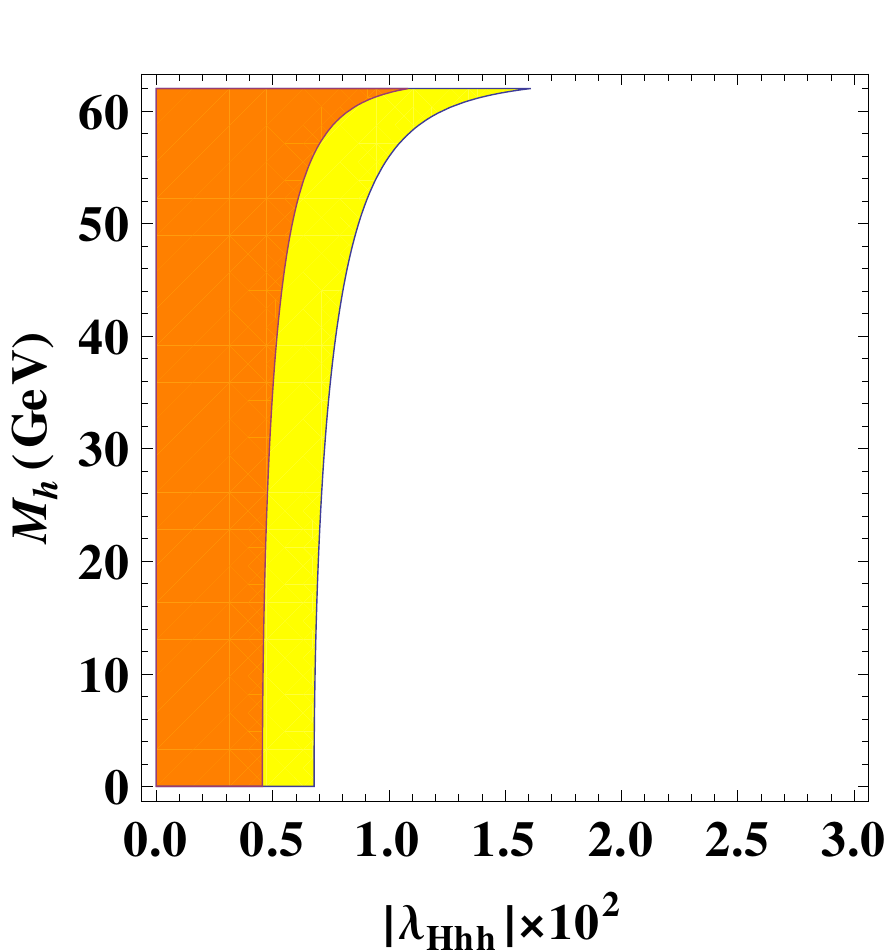}
\caption{\it \small Left-panel:
Constraint in the $(M_{H^\pm}, M_A)$ plane from the oblique parameters $S$, $T$ and $U$. Right-panel: Constraints from the invisible Higgs decay width in the $(\left|\lambda_{Hhh}\right|, M_h)$ plane, assuming SM couplings of $H$ to fermions and vector bosons. The orange (dark) and yellow (light) regions are allowed at 68\% and 90\%~CL.}
\label{ewfits}
\end{figure}

The LEP searches for neutral Higgs particles could have missed the light scalar $h$, since the associated production with a vector boson would be strongly suppressed. Moreover, $\left| y_d^h\right|\sim |\varsigma_d|$ could be small enough to avoid the constraints from the usual $h\to b \bar b$ search mode.
The OPAL collaboration performed a decay-mode-independent search for a light neutral scalar and found upper limits for the Higgs-strahlung cross section in units of the SM: $ (\mathcal{R}_{11})^2 \equiv (g_{hVV}/g_{hVV}^{\mathrm{SM}})^2 < 0.1$ for $M_{h} < 19$~GeV, and  $(\mathcal{R}_{11})^2 < 1 $ for $M_{h} < 81~\text{GeV}$~\cite{Abbiendi:2002qp}.
Together with the constraints from electroweak precision tests at the $Z$ peak,
this provides useful information on the allowed mass spectrum for the remaining scalars. Using the current bounds from the oblique parameters $S$, $T$ and $U$ \cite{Baak:2012kk,LEPEWWG} (the corresponding A2HDM formulae are given in appendix~\ref{oblique}), we show in the left panel of Fig.~\ref{ewfits} the allowed regions in the $(M_{H^\pm}, M_A)$ plane. We have set $M_H = 126$~GeV and
$\sin{\ta} \in [0.7,  \,  1]$. The constraints shown in the figure turn out to be determined by the $T$ parameter, since $S$ and $U$ give weaker restrictions.
The charged scalar mass is of course constrained by the direct experimental lower bound discussed before, but its exact value depends on the assumed decay channels. The region where both $M_{H^\pm}$ and $M_A$ become very heavy corresponds to uncomfortably large values of the quartic couplings $\lambda_i$ of the scalar potential and the theory is no longer perturbative.

A light neutral boson $h$ or $A$ below the kinematical threshold of $M_H/2 \approx 63$~GeV would have important phenomenological consequences, because the $126$~GeV Higgs could decay into lighter scalars. These decay channels can be included in our fit in terms of an invisible decay width as long as we neglect possible contributions from cascade decays into the observed final states.\footnote{These effects are beyond the scope of the present work, but they could be relevant. For example, $H \rightarrow AA \rightarrow \gamma \gamma + \gamma \gamma$ could be mistaken by a two-photon signal when the photon pairs are very collimated~\cite{ATL079}}
In general one would expect in this case a suppression of the measured Higgs decay rates compared with the SM, due to the larger total width of the scalar $H$.   Current data for the $\gamma \gamma$ channel, however, shows a slight enhancement over the SM prediction, thus placing strong bounds on possible invisible decays of the $126$~GeV Higgs boson. Assuming that the heavy-Higgs couplings to fermions and vector bosons are SM-like ({\it i.e.}, $y_f^H =1$ and $\sin{\ta} = 1$), the best fit point is obtained for a null invisible $H$ decay width;
at 68\%~CL (90\%~CL) we obtain an upper bound of 9\% (20\%) on the invisible
$H$ decay width (in units of the SM total decay width).

Considering the scenario of a very light CP-even Higgs $h$, the decay width of the heavier CP-even scalar into $hh$ is given by
\be  \label{eq:decayhl}
\Gamma(H \rightarrow h h )\;  =\; \dfrac{v^2 \lambda^2_{Hhh} }{8 \pi M_{H}} \; \left( 1 - \dfrac{4 M^2_{h }}{M_{H}^2} \right)^{1/2}\, ,
\ee
where the cubic scalar coupling $\lambda_{Hhh}$ is expressed in units of $v$ and can be obtained from Eq.~\eqref{cubic}.    In the right panel of Fig.~\ref{ewfits} we show the constraints from our $\Gamma(H \rightarrow h h )$ fit in the $(|\lambda_{Hhh}|, M_h)$ plane. Strong bounds are obtained for the cubic Higgs coupling, $|\lambda_{Hhh}| \lesssim 10^{-2}$, as expected.

Recent updates from the ATLAS collaboration in the high-resolution channels report a significant difference in the mass of the neutral boson as determined from $H \rightarrow ZZ^{(*)} \rightarrow 4 \ell$ ($123.5\pm 0.8\pm0.3$~GeV) and $H \rightarrow \gamma \gamma$ ($126.6\pm0.3\pm0.7$~GeV) events  \cite{:2012gk}. Here we do not consider as a possible explanation for this discrepancy, the possibility of having two quasi-degenerate Higgs bosons, since the current mass value in the $H \rightarrow ZZ^{(*)} \rightarrow 4 \ell$ channel obtained by CMS, $126.2\pm0.6\pm0.2$~GeV~\cite{Chatrchyan:2012dg}, does not support this hypothesis.

\subsubsection{Degenerate CP-even and CP-odd Higgs bosons at 126 GeV \label{CPodd}}

A CP-odd scalar does not couple at tree level to two vector bosons; its decay to gauge bosons starts at the one-loop level and it is therefore very suppressed.  For this reason, a pure CP-odd Higgs boson is already strongly disfavoured by present data as a candidate for the 126 GeV boson. However, the observed signal could result from two Higgs bosons with quasi-degenerate masses; this could explain the excess of $\gamma \gamma$ events observed by ATLAS and CMS.   This possibility was proposed in Ref.~\cite{Gunion:2012gc} within the non-minimal supersymmetric extension of the SM, and has also been considered within the context
of 2HDMs, both for $\mathcal{Z}_2$ versions \cite{Drozd:2012vf,Chang:2012ve,Ferreira:2012nv} and
with a more general Yukawa structure~\cite{Cervero:2012cx,Altmannshofer:2012ar}. Model-independent methods to test experimentally for such possibility have also been proposed recently in Refs.~\cite{Gunion:2012he,Grossman:2013pt}.

\begin{figure}[tb]
\centering
\includegraphics[scale=0.27]{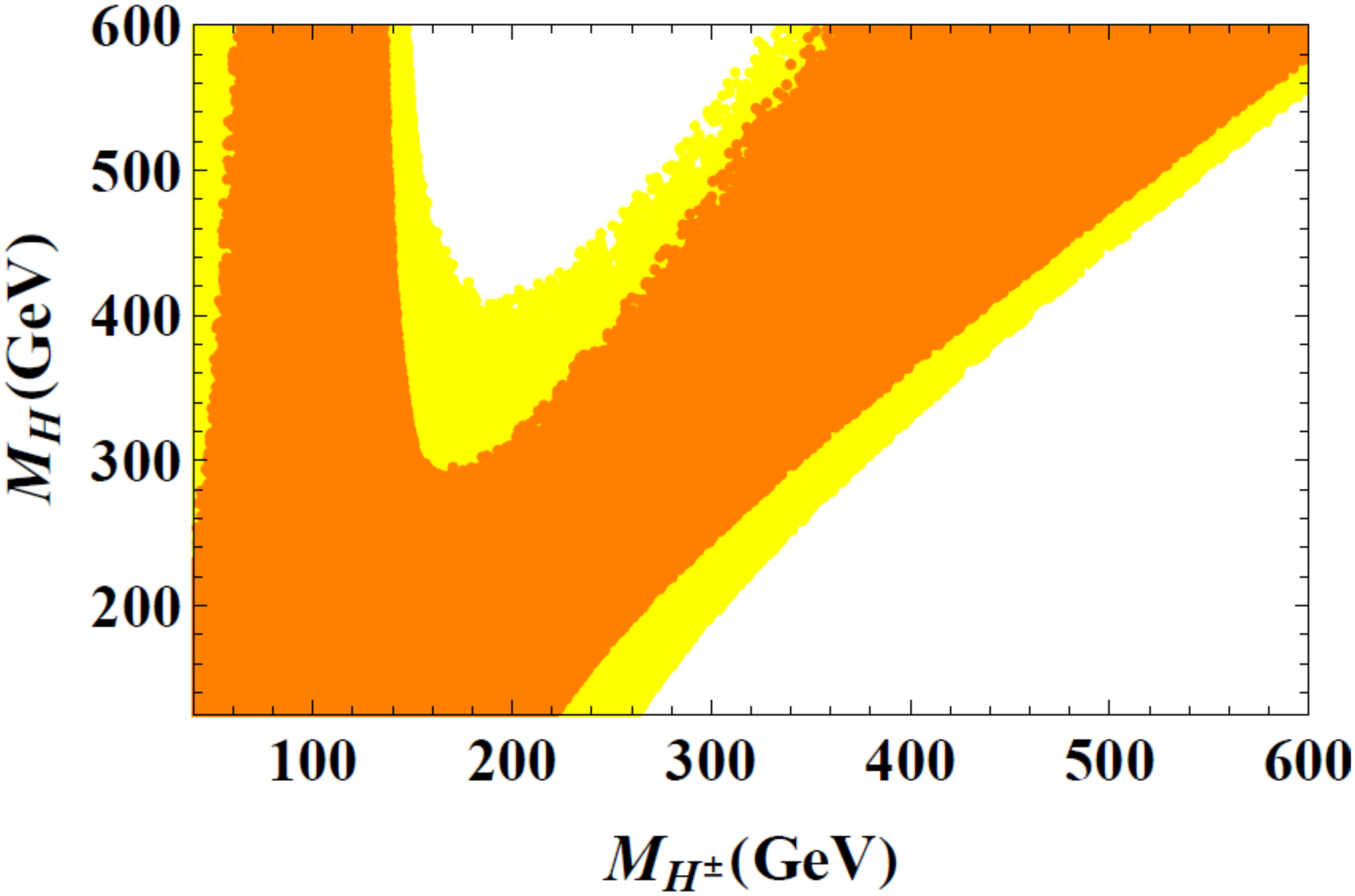}
\hskip 0.75cm
\includegraphics[scale=0.27]{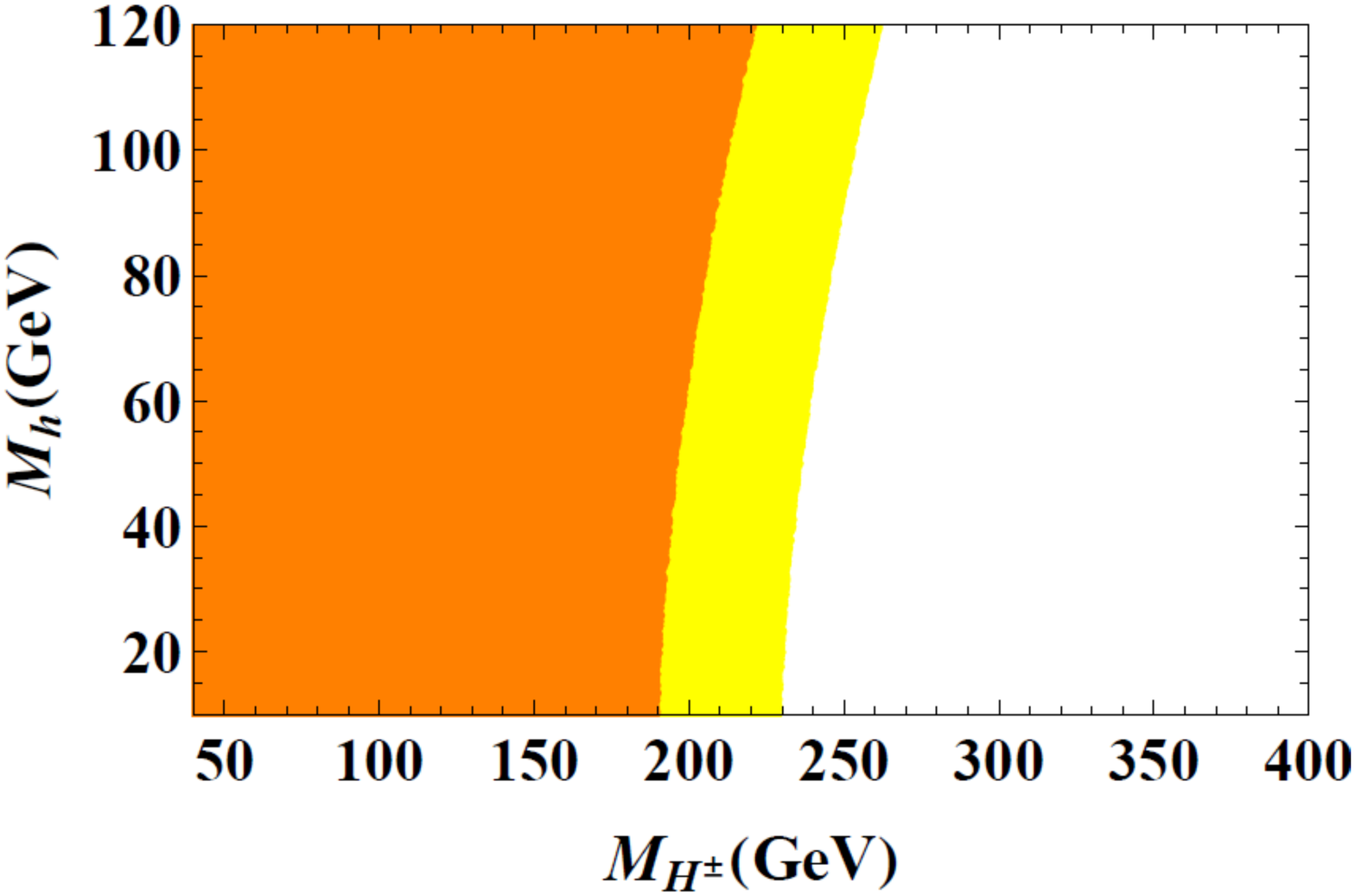}
\caption{\it \small Constraints in the $(M_{H^\pm}, M_H)$ plane for the case $M_h = M_ A = 126$~GeV (left) and in the $(M_{H^\pm}, M_h)$ plane for the case $M_H = M_ A = 126$~GeV (right), from the oblique parameters $S$, $T$ and $U$. The orange (dark) and yellow (light) regions are allowed at 68\% and 90\%~CL.}
\label{ewfits2}
\end{figure}

We consider in this section the possibility of two Higgs bosons with quasi-degenerate masses around $126$~GeV, one of them being CP-even and the other one CP-odd.   We perform a global fit of the data with $M_h = M_A \approx 126$~GeV,
and comment on the alternative possibility of quasi-degenerate $H$ and $A$.    The observed Higgs signals strengths will then receive contributions from both particles:
\begin{equation}
\mu_k^{(h+A)}\; = \;\mu_k^h + \mu_k^A \, .
\end{equation}
Given the presently large experimental uncertainties, we neglect the small $AVV$ coupling generated at one loop. Therefore, among all the channels considered in this work, the CP-odd Higgs will only contribute to $A \rightarrow \tau \tau$ and $A \rightarrow \gamma \gamma$. In both cases the dominant production channel is the gluon-fusion one. The loop-induced decay $A \to \gamma\gamma$ is only mediated by fermions. In Fig.~\ref{ewfits2} (left) we show the constraints on $M_{H^{\pm}}$ and $M_H$ obtained from the oblique parameters.   These masses are varied in the ranges $M_{H^{\pm}} \in [50, 600]$~GeV and $M_H \in [126, 600]$~GeV, while the coupling of $h$ to vector bosons is kept close to the SM limit ({\it i.e.}, $|\cos{\ta}| \in [0.8,1]$), as suggested by the current experimental data.   In the right panel of Fig.~\ref{ewfits2} we show similar bounds on the plane  $(M_{H^{\pm}}, M_h)$, keeping the light scalar mass below $M_H = M_A = 126$~GeV and taking $\sin{\ta} \in [0.8,1]$; in this case the oblique parameters require the existence of a charged Higgs below the electroweak symmetry breaking scale $v = 246$~GeV.

\begin{figure}[tb]
\centering
\includegraphics[width=8.2cm,height=6cm]{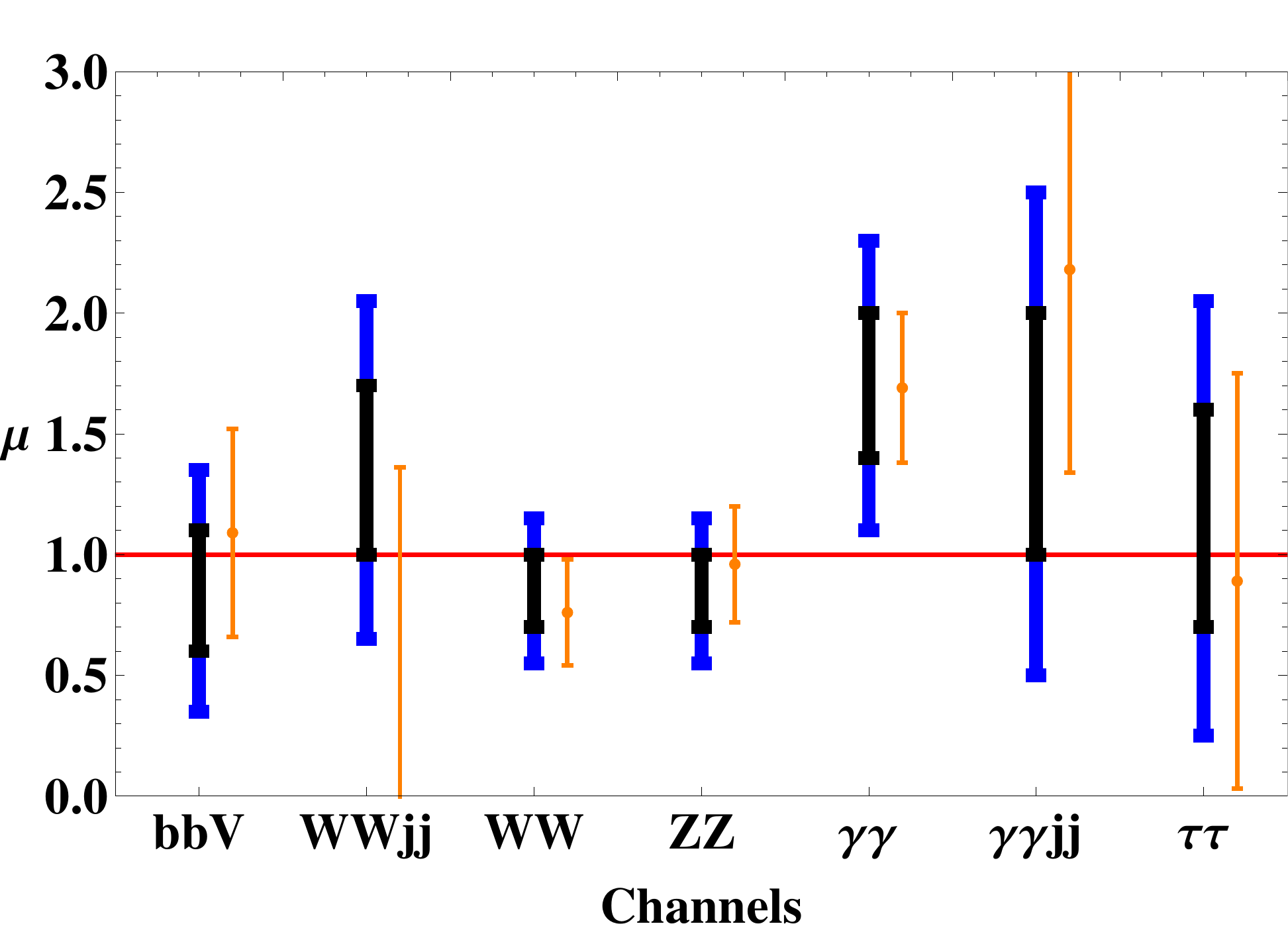}
\caption{ \it \small  Allowed ranges for the Higgs signal strengths from the global fit within the CP-conserving A2HDM for the case of degenerate Higgs bosons with $M_h = M_A = 126$~GeV, at $1 \sigma$ (black, dark) and $2 \sigma$ (blue, dark).  Other captions as in Figure~\ref{fig:RealH}.}
\label{fHShAfit}
\end{figure}

In the scenario $M_h = M_A = 126$~GeV, the
best fit region in the A2HDM parameter space, assuming the charged Higgs contribution to the $2 \gamma$ channel to be negligible, is given by:
\begin{eqnarray}  \label{eq:fitwc}
\cos{\ta}\, =\, 0.98\pm 0.2 \, , \qquad\quad
\varsigma_u \, =\, -1.1\,{}^{+\, 0.5}_{-\, 0.4} \, , \qquad\quad
|\varsigma_d|\, =\, 1.2 \pm 1.2 \, , \qquad\quad
\varsigma_l\, =\, -0.2\,{}^{+\, 0.6}_{-\, 0.4} \, .\quad
\end{eqnarray}
The corresponding allowed ranges for the Higgs signal strengths, at $1 \sigma$ and $2 \sigma$, are shown in  Fig.~\ref{fHShAfit}. We obtain a smaller total decay width of the CP-even boson, $\rho(h)\approx 0.7$, which produces a sizeable enhancement of the $\mu^h_{\gamma\gamma jj}$ signal strength (the CP-odd boson $A$ does not contribute to this channel). On the other hand, the excess in the two photon channel comes from the decays of both $A$ and $h$, which give contributions of similar size ($\mu^{h}_{\gamma\gamma}\approx\mu^A_{\gamma\gamma}\approx 0.7$). The remaining contribution of $A$ is to the $\tau^+\tau^-$ decay channel, which is small ($\varsigma_l$ is small). We must also notice that solutions with a flipped relative sign between the $W$ and top contributions to $h\to \gamma\gamma$
are not allowed because they would require large values of $\varsigma_u$; this would increase $C_{gg}^A$ and $C_{\gamma\gamma}^A$ generating a large excess in the $\tau^+ \tau^-$ and $\gamma\gamma$ channels, exceeding the current experimental bounds.

It is important to note that for a light charged Higgs boson, very strong flavour constraints in the $\varsigma_u - \varsigma_d$ plane can be obtained from $\bar B \rightarrow X_s \gamma$~\cite{Jung:2010ik}. The allowed ranges at $68\%$~CL shown in Eq.~\eqref{eq:fitwc} were obtained assuming that the charged Higgs contribution to the diphoton channel is negligible (this is true even for a light charged Higgs if $\lambda_{h H^+ H^-} \simeq 0$).   Including the charged Higgs contribution to the $2 \gamma$ channel in the fit one obtains at $68\%$~CL that $C_{H^{\pm}}^{h} = -3.0 \pm 1.4$, while the alignment parameters $\varsigma_f$ remain weakly constrained and compatible with zero. In the limit $\varsigma_{f} = 0$, the stringent flavour constraints for a light charged Higgs, in particular $\bar B \rightarrow X_s \gamma$, are avoided since the charged Higgs decouples from the fermions.  These constraints would be particularly relevant in the scenario $M_H = M_A = 126$~GeV for which the charged Higgs mass is bounded to lie below the electroweak scale, see Figure~\ref{ewfits2} (right). 

\subsection{The CP-violating A2HDM}
\label{sec:complex}
In the A2HDM the up and down-quark as well as the leptonic Yukawa couplings are all independent complex parameters. Thus, one can expect a very rich phenomenology associated to the Higgs sector responsible for the breaking of the electroweak symmetry. Moreover, if one considers the most general scalar potential, the neutral scalars $h$, $H$ and $A$ are not CP eigenstates but rather a mixture of CP-even and CP-odd fields, parametrized by the general orthogonal matrix ${\cal{R}}$ introduced in section~\ref{sec:A2HDM}. Thus, there are new sources of CP violation, both from the Yukawa sector and the scalar potential, which could lead to interesting phenomenological predictions.

The study of CP-violating observables is beyond the scope of the present work and we will defer it to future publications.\footnote{
For theoretical studies about the CP-properties of extended Higgs sectors at the LHC and in possible future colliders see Ref.~\cite{Accomando:2006ga} and references therein.}
Nevertheless, we shall investigate next, the sensitivity of the different (CP-conserving) Higgs signal strengths to the CP-violating phases.
Since the present data are consistent with the SM within rather large uncertainties, we will consider separately the different CP-odd possibilities, by fitting some complex coupling constants to the Higgs-signal-strength data while
setting the remaining parameters to their SM-like values.    A similar analysis has also been performed within a model independent framework in Ref.~\cite{Cheung:2013kla}.

\subsubsection{Complex Yukawa couplings}

\begin{figure}[tb]
\centering
\includegraphics[width=6cm,height=6cm]{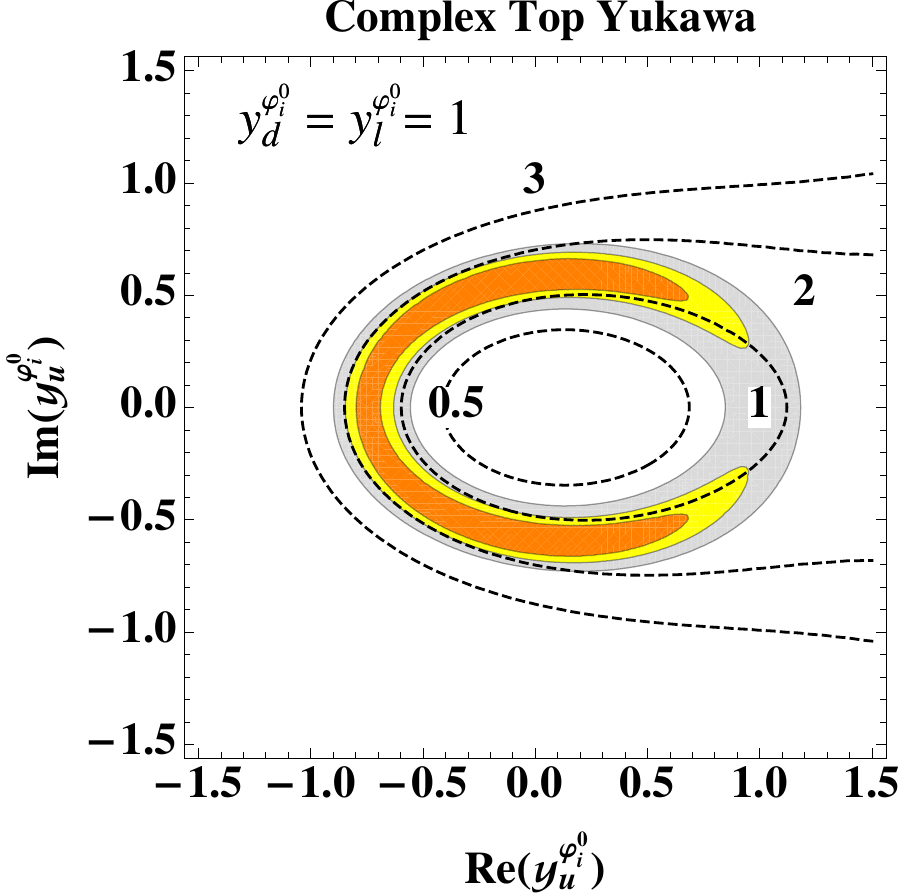}
\hskip 1.5cm
\includegraphics[width=6cm,height=6cm]{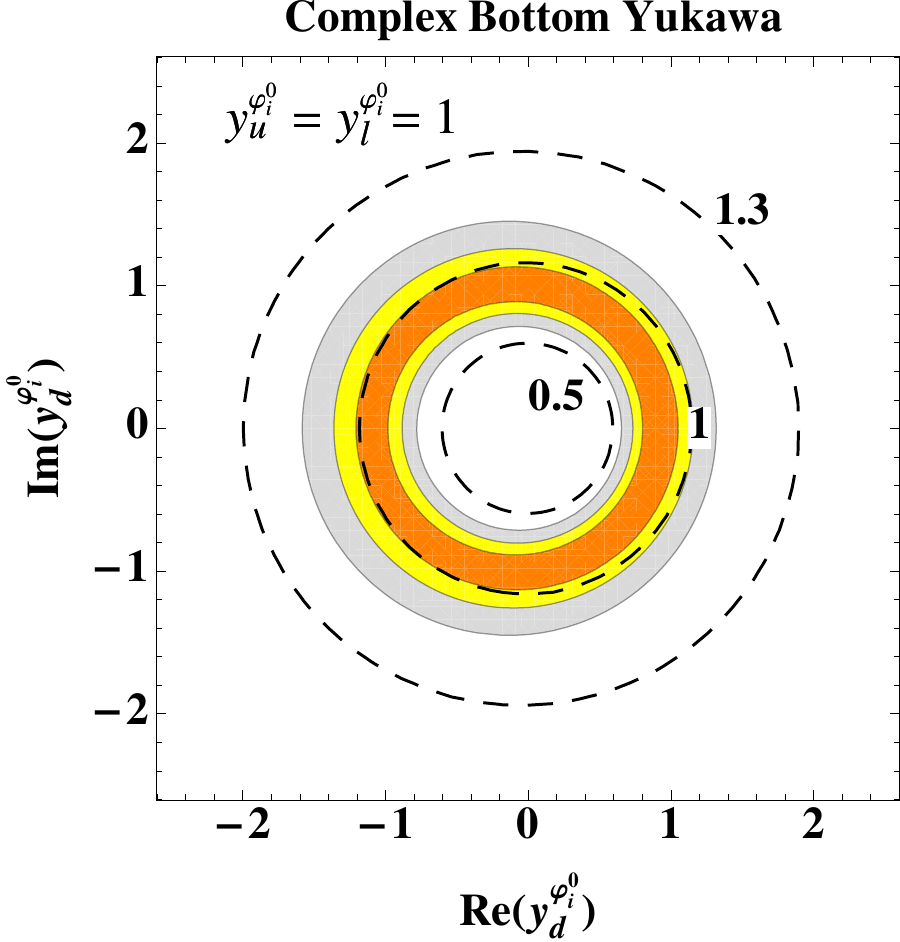}
\vskip .5cm
\includegraphics[width=6cm,height=6cm]{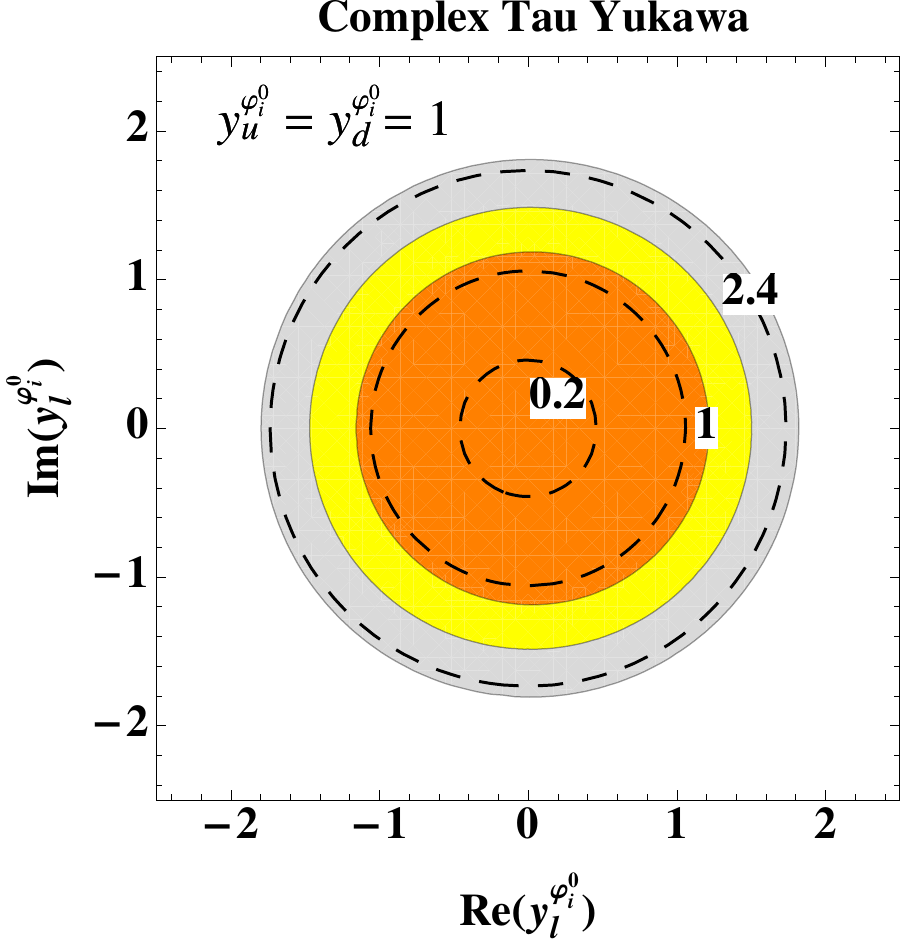}
\caption{\label{CYuka} \it \small
Allowed regions at 68\% (orange), 90\% (yellow) and 99\% (grey)~CL
for the complex top (upper-left), bottom (upper-right) and tau (lower) Yukawa couplings. In each plot the two Yukawa couplings not shown are set to their SM value and the coupling to vector bosons is taken to be $\cR_{i1} = 0.95$.
The dashed lines show contours of constant values for $\mu_{\gamma\gamma}^{\varphi_i^0}$ (top plot), $\mu_{bbV}^{\varphi_i^0}$ (bottom plot) and $\mu_{\tau \tau V}^{\varphi_i^0}$ (tau plot).}
\end{figure}

Let us consider $\varphi_i^0$ to be the observed boson with a mass of 126 GeV.
We will analyze three simple scenarios that will serve to determine the sensitivity to its complex Yukawa couplings and to what extent the SM limit is preferred by present data. We will set two Yukawa couplings to their SM values ($y_f^{\varphi_i^0} =1$), and find the preferred values for the remaining Yukawa coupling by minimizing the $\chi^2$ function.  Figure~\ref{CYuka} shows the resulting allowed regions for the top, bottom and tau Yukawa couplings when the coupling of $\varphi_i^0$ to vector bosons is fixed to $\cR_{i1}  = 0.95$; this value lies well within the $90\%$ CL allowed band obtained from our previous fits.

Since all the observables considered are CP-even, the bounds obtained are symmetric under $\im(y_f^{\varphi_i^0}) \rightarrow - \im(y_f^{\varphi_i^0})$.  Moreover, the real and imaginary parts of the Yukawa couplings do not interfere.  The sensitivity to $\im(y_f^{\varphi_i^0})$ is similar to that obtained previously, when considering only real couplings. For tree-level decays this is obvious from Eq.~\eqref{ratios2}, given that the parameter $\beta_f$ is very close to one for $f= b, \tau$. For loop-induced decays this can be understood by observing that the loop functions \eqref{functions} are closely related, $\cF(\tau) =  2 \tau + \frac{\tau^2}{2} f(\tau) + \mathcal{K}(\tau)$. For $b$ quarks and $\tau$ leptons, $\cF(\tau_f) \approx \mathcal{K}(\tau_f)$; for the top quark there is a small but sizable difference between the contributions of its real and imaginary Yukawa parts.  Note that in the limit $\cR_{i1} = 1$ the Yukawa couplings of $\varphi_i^0$ become SM-like ($y_{f}^{\varphi_i^0} =1$) due to the orthogonality of $\cR$; thus, there is no sensitivity to the $\varsigma_f$ parameters when considering the neutral Higgs couplings. The charged Higgs couplings on the other hand are proportional to $\varsigma_f$ and do not depend on the mixing matrix $\cR$.

In the left upper panel of Fig.~\ref{CYuka} we show the results of the fit for a complex top Yukawa coupling, while setting $y^{\varphi_i^0}_{d} = y^{\varphi_i^0}_l = 1$.  The dashed lines show contours of constant value for $\mu^{\varphi_i^0}_{\gamma \gamma}$. The SM-like point  $(\re(y^{\varphi_i^0}_u), \im(y^{\varphi_i^0}_u))= (1,0)$ lies outside the 90\%~CL region, but becomes allowed at 99\%~CL.
%
%
It can be seen that the allowed region at 90\%~CL accommodates an enhanced $\gamma \gamma$ rate between one and two times that of the SM. Within this 90\%~CL region, $\rho(\varphi_i^0) = 1.00 \pm 0.03$ as expected, since the dominant decay channel is $\bar b b$; the gluon fusion cross section is slightly reduced compared with the SM ($C_{gg}^{\varphi_i^0} = 0.87 \pm 0.28$), while the $\gamma \gamma$ partial decay width is enhanced ($C_{\gamma \gamma}^{\varphi_i^0} = 1.67 \pm  0.56$). The preferred allowed region is that for which the top Yukawa coupling has opposite sign to $\cR_{i1}$, thus, creating a constructive interference with the vector boson contribution for the $\varphi_i^0 \rightarrow \gamma \gamma$ amplitude. The other option would be to have a significant imaginary component $\mathrm{Im}(y_{u}^{\varphi_i^0})$, which would also enhance the $\gamma \gamma$ rate.   Similar results were obtained in Ref.~\cite{Cheung:2013kla}.

The right upper panel of Fig.~\ref{CYuka} shows the fitted values for the complex bottom coupling, with the top and tau Yukawa couplings set to their SM values. The dashed lines indicate contours of constant value for $\mu_{bbV}^{\varphi_i^0}$. In this case the SM limit  $(\re(y^{\varphi_i^0}_d), \im(y^{\varphi_i^0}_d) )= (1,0)$ lies inside the 90\%~CL allowed region, which accommodates $ 0.7 < \mu_{bbV}^{\varphi_i^0} < 1.2$.   In this 90\% CL region, the total decay width is rescaled by
$\rho(\varphi_i^0) = 1.11 \pm 0.67$; the gluon-fusion cross section ratio is $C_{gg}^{\varphi_i^0} = 1.15 \pm 0.10$, while the $\gamma \gamma$ partial decay width turns out to be slightly suppressed with respect to the SM, $C^{\varphi_i^0}_{\gamma \gamma} = 0.89 \pm  0.10$. Since the total decay width depends strongly on the value of $|y_d^{\varphi_i^0}|^2$, a large variation range is obtained for $\rho(\varphi_i^0)$.

In the lower panel of Fig.~\ref{CYuka}, we show the fitted values of the complex $\tau$ Yukawa coupling assuming $y^{\varphi_i^0}_{u} = y^{\varphi_i^0}_d = 1$.
Contours of constant value for $\mu_{\tau \tau V}^{\varphi_i^0}$ are also shown as dashed lines. We obtain that the signal strength $\mu_{\tau \tau V}^{\varphi_i^0} <1.5$ lies within the 68\%~CL allowed region.  The total Higgs decay width and the gluon-fusion cross section are equal in this case to the SM ones, while some suppression is observed in the $\gamma \gamma$ partial decay width: at 90\%~CL, $C^{\varphi_i^0}_{\gamma \gamma } = 0.90 \pm 0.11$ is obtained. This scenario is therefore disfavoured by the observed excess in the two-photon channel.

\subsubsection{A fermiophobic charged Higgs }
\label{fermiophobic}

In the limit $\varsigma_f \rightarrow 0$ the charged Higgs does not couple to fermions, independently of any assumption about the scalar potential.   Such {\it fermiophobic} charged Higgs could have avoided detection at LEP while being very light.   Current LHC searches, as well as searches at the Tevatron, would have also missed such particle since it can neither be produced via top decay nor decay into fermions.  Flavour constraints on this charged Higgs are also avoided trivially.    Detecting such particle in an experiment is therefore quite challenging, since it can only be produced in processes involving vector bosons and/or neutral Higgs particles; the same occurs for its decay channels.

The case of a fermiophobic charged Higgs is however highly predictive in the neutral Higgs sector, since all the channels which do not involve the $\gamma \gamma$ ($\gamma Z$) final state only depend on one free parameter, $\mathcal{R}_{i1}$. The rescaling of the Higgs coupling to vector bosons in this case is the same as that of the neutral Yukawa ones,
$y^{\varphi_i^0}_f = g_{\varphi_i^0 VV}/g_{\varphi_i^0 VV}^{\mathrm{SM}} = \mathcal{R}_{i1}$,
which implies that all Higgs signal strengths are rescaled by a factor $\mathcal{R}_{i1}^2$ with respect to the SM, meaning that $\mu^{\varphi_i^0}_{bb} = \mu^{\varphi_i^0}_{\tau \tau} = \mu^{\varphi_i^0}_{WW, ZZ}  = \rho(\varphi_i^0)^{-1}\mathcal{R}_{i1}^4 = \mathcal{R}_{i1}^2$, in any of the relevant production mechanisms. Therefore, in this scenario the signal strengths of the three neutral scalars are correlated:
\begin{equation}\label{eq:fermiophobic_rel}
\sum_{\varphi_i^0= h, H, A} \mu_{ff}^{\varphi_i^0}\;\, =\; \sum_{\varphi_i^0= h, H, A} \mu_{WW,ZZ}^{\varphi_i^0}\; =\; 1\, .
\end{equation}

\begin{figure}[tb]
\centering
\includegraphics[width=7cm,height=6.8cm]{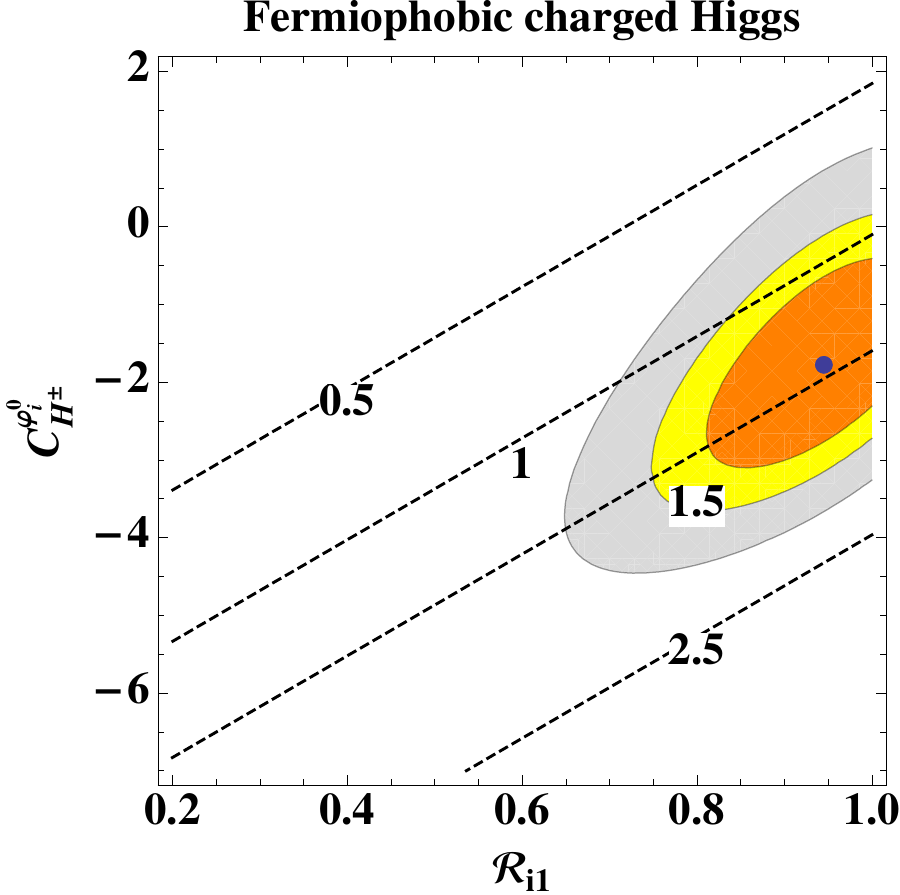}
\hskip 1.25cm
\includegraphics[width=7cm,height=6.8cm]{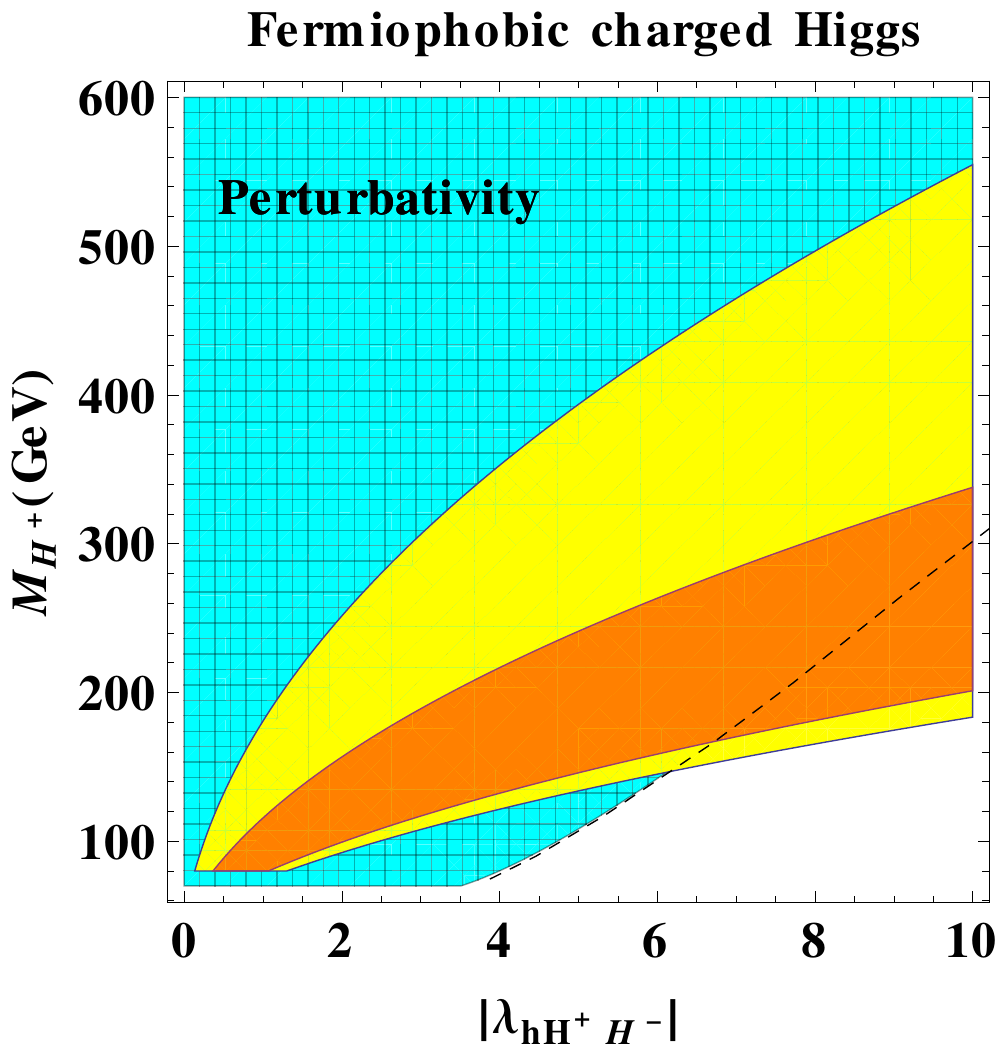}
\caption{\label{Higgs125Charged} \it \small
Allowed regions at 68\% (orange), 90\% (yellow) and 99\%~CL (gray) for a fermiophobic charged Higgs on the parameter space $(\mathcal{R}_{i1},\mathcal{C}_{H^\pm}^{\varphi^0_i})$; dashed lines denote contours of constant $\mu_{\gamma\gamma}^{\varphi_i^0}$ (left). The right plot shows the corresponding 68\% and 90\%~CL regions in the parameters $\lambda_{\varphi^0_i H^+H^-}$ and $M_{H^\pm}$, setting the value of $\mathcal{R}_{i1}$ at its best fit point.  The region where perturbation theory remains valid is indicated in blue (hashed).}
\end{figure}

Present data on the neutral Higgs boson are sensitive to a fermiophobic charged Higgs through the loop-induced decay $\varphi_i^0\to\gamma \gamma$. The charged-scalar contribution to this decay can be sizeable for a light $H^\pm$, and this is a quite interesting situation in view of the possibility to detect such particle in the future.  Assuming that the scalar with a mass of 126 GeV  does not decay into lighter scalars, we show in Fig.~\ref{Higgs125Charged} the allowed region in the parameter space $(\mathcal{R}_{i1},\mathcal{C}_{H^\pm}^{\varphi^0_i})$. For the $\chi^2$ fit we have only considered real values of $\mathcal{C}_{H^\pm}^{\varphi^0_i}$, which is true above the kinematical threshold $M_{H^\pm} > M_{\varphi_i^0}/2 \approx 63$~GeV, as we have mentioned before.  In the figure we also show dashed contour lines of constant $\mu_{\gamma\gamma}^{\varphi_i^0}$. It can be observed that the preferred relative sign between the charged Higgs and the $W^{\pm}$ contributions to the $\gamma \gamma$ decay rate is such that it causes a constructive interference, thus enhancing slightly the $\gamma \gamma$ decay rate. The fit prefers a gauge coupling close to the SM one ($\chi^2_{\mathrm{min}}$ is obtained for $\mathcal{R}_{i1}\approx 0.95$) and puts the 90\% CL lower bound $|\mathcal{R}_{i1}| > 0.79$. The SM-like point $(\mathcal{R}_{i1},\mathcal{C}_{H^\pm}^{\varphi^0_i}) = (1,0)$ lies outside the
68\% CL region, but is allowed at 90\% CL (although close to the boundary). The presence of a non-zero (and negative) $\mathcal{C}_{H^\pm}^{\varphi^0_i}$ contribution is clearly favoured, while the preference for a slightly reduced gauge coupling implies a small suppression of the total decay width compared with the SM ({\it i.e.}, $\rho(\varphi_i^0) = 0.85 \pm 0.19$, at $90\%$~CL). From the global fit, $\mu^{\varphi_i^0}_{\gamma \gamma} = \mu^{\varphi_i^0}_{\gamma \gamma j j} = 1.45 \pm 0.49$ is obtained at $90\%$~CL; all the other Higgs signal strengths that are not affected by the charged Higgs contribution are equal to $\mu = 0.8 \pm 0.2$.

The right panel in Fig.~\ref{Higgs125Charged} shows the corresponding allowed regions
in terms of the variables $\lambda_{\varphi^0_i H^+H^-}$ and $M_{H^\pm}$.
The value of $\mathcal{R}_{i1}$ has been set to its best fit point. Also shown in the figure, is the region satisfying the perturbativity constraints discussed in appendix~\ref{perturbatvity}.

For the previous discussion we have not made any assumptions on the quantum numbers of the scalar field $\varphi_i^0$; we have only assumed that $M_{\varphi_i^0} = 126$~GeV and that its decay into lighter scalars is not allowed. Thus, the obtained results are general and apply both to a CP-conserving and to a CP-violating scalar potential.
It must be noted that in the limit $|\mathcal{R}_{i1}| =1$ the phenomenology of $\varphi_i^0$ becomes identical to that of the SM in every channel, except for $\gamma \gamma$ and $\gamma Z$ which are affected by the $H^\pm$ contribution. For a fermiophobic charged Higgs lighter than $M_{\varphi_i^0}/2 \approx 63$~GeV, $\mathcal{C}_{H^\pm}^{\varphi^0_i}$ develops an imaginary absorptive part. If kinematically open, the channel $\varphi_i^0 \rightarrow H^+ H^-$ would increase the total width of the Higgs boson; furthermore, in this scenario the production cross section is always less or equal to the SM. Therefore, the signal strengths would be reduced in every channel, with respect to the SM. This is in clear contradiction with the data, specially with the measurements for the two-photon channel.

\subsubsection{CP-even and CP-odd neutral scalar mixing}

A CP-violating scalar potential generates mixings among the three neutral scalars, which are no longer CP eigenstates. Here, we are interested in exploring the possibility that the observed 126 GeV state could be the CP-odd scalar with a small CP admixture of the CP-even ones.   A similar analysis within 2HDMs of types I and II, with explicit CP violation and soft breaking of the $\mathcal{Z}_2$ symmetry has been done in Ref.~\cite{Barroso:2012wz}, placing numerical bounds on the size of a possible CP-odd component for the scalar particle with $126$~GeV of mass.

In the presence of CP violation, the admixture between the three neutral scalar fields is described by the 3-dimensional orthogonal matrix $\cR$ which diagonalizes their mass matrix. This diagonalization can be done numerically, once the parameters of the scalar potential are known, but a simple analytical solution is not available for the most general case.  It is well known, on the  other hand, that in the CP-conserving limit the mass-matrix simplifies and it is possible to give explicit expressions for the masses and physical states in terms of the scalar potential parameters.
A reasonable assumption when dealing with the general 2HDM scalar potential, is that the CP-violating terms are small; this makes a perturbative expansion in these parameters a valid approximation in principle. In appendix~\ref{app:potential} we provide explicit analytical expressions for the neutral scalar masses and the corresponding eigenstates to leading order in the CP-violating parameters of the scalar potential $\lambda_{5,6}^{\mathrm{I}}$. The corrections to the masses are quadratic in $\lambda_{5,6}^{\mathrm{I}}$, while the mixing between the CP-even and CP-odd states is only suppressed by one power of $\lambda_{5,6}^{\mathrm{I}}$, making this effect the dominant one.

Let us assume that the discovered boson is the state $A = S_3 + \cR_{31} S_1 + \cR_{32} S_2$, with $\cR_{31}$ and $\cR_{32}$ the small CP-even admixture coefficients. To simplify the discussion, we consider a simple scenario in which we set the parameters $\varsigma_{u,d,l}=0$.  The Yukawa couplings, as well as the coupling to vector bosons, are equal in this case, $y_{f}^A = \cR_{31}$.  From a global fit to the data, we find a lower bound on the admixture coefficient: $\cR_{31}  > 0.83$, at $99\%$~CL.  This result is  mainly driven by the measurements in the $W^+W^-, ZZ$ and $\gamma \gamma$ channels, which are SM-like to a good degree.

We can analyze whether such large values for the correction $\cR_{31}$ can be obtained for natural values of the scalar potential parameters.
From Eq.~\eqref{eq:mixA}, one has:
\bel{eq:R31}
\cR_{31}\; \approx\; \frac{v^4\,\left( 2 \lambda_5^{\mathrm{R}}\lambda_6^{\mathrm{I}}-\lambda_6^{\mathrm{R}}\lambda_5^{\mathrm{I}}\right)}{\left( \bar M_A^2 -  \bar M_h^2\right) \left(\bar M_A^2- \bar M_H^2 \right)}
\, .
\ee
Thus, large mass differences between the scalar states suppress the effect of mixing due to CP violation in the scalar potential; on the other hand if the scalar bosons have very similar masses these effects could be considerably enhanced.
Assuming that $|\lambda_{5,6}^{I,R}| \lesssim 10^{-1}$ we obtain
\mbox{$\cR_{31} \lesssim  \left[ \left(\bar M_A^2 - \bar M_H^2 \right) \left(\bar M_A^2 - \bar M_h^2 \right)  \right]^{-1}   \, 10^{8} \; \text{GeV}^4$}, which implies that $|\cR_{31}| \lesssim 10^{-2}$ for $\bar M_H > \bar M_h \gtrsim 300$~GeV.
Of course, when either $\bar M_h \sim \bar M_A$ or $\bar M_H \sim \bar M_A$ the coefficient $\cR_{31}$ diverges and the approximations used in appendix~\ref{app:potential} are no longer valid.
The general formalism to describe the dynamics of CP violation near degenerate neutral Higgs bosons has been developed in Refs.~\cite{Pilaftsis:1997dr,Ellis:2004fs}.  In Ref.~\cite{Choi:2004kq} the effect of resonant enhancement of $H$ and $A$ mixing was studied for the CP-violating 2HDM in the decoupling limit, $\bar M_A^2 \gg | \lambda_i |\,v^2 $.    In this case the heavy states $H$, $H^{\pm}$ and  $A$ are nearly mass degenerate and decouple from the light state $h$.

\begin{figure}[tb]
\centering
\includegraphics[width=7cm,height=6.4cm]{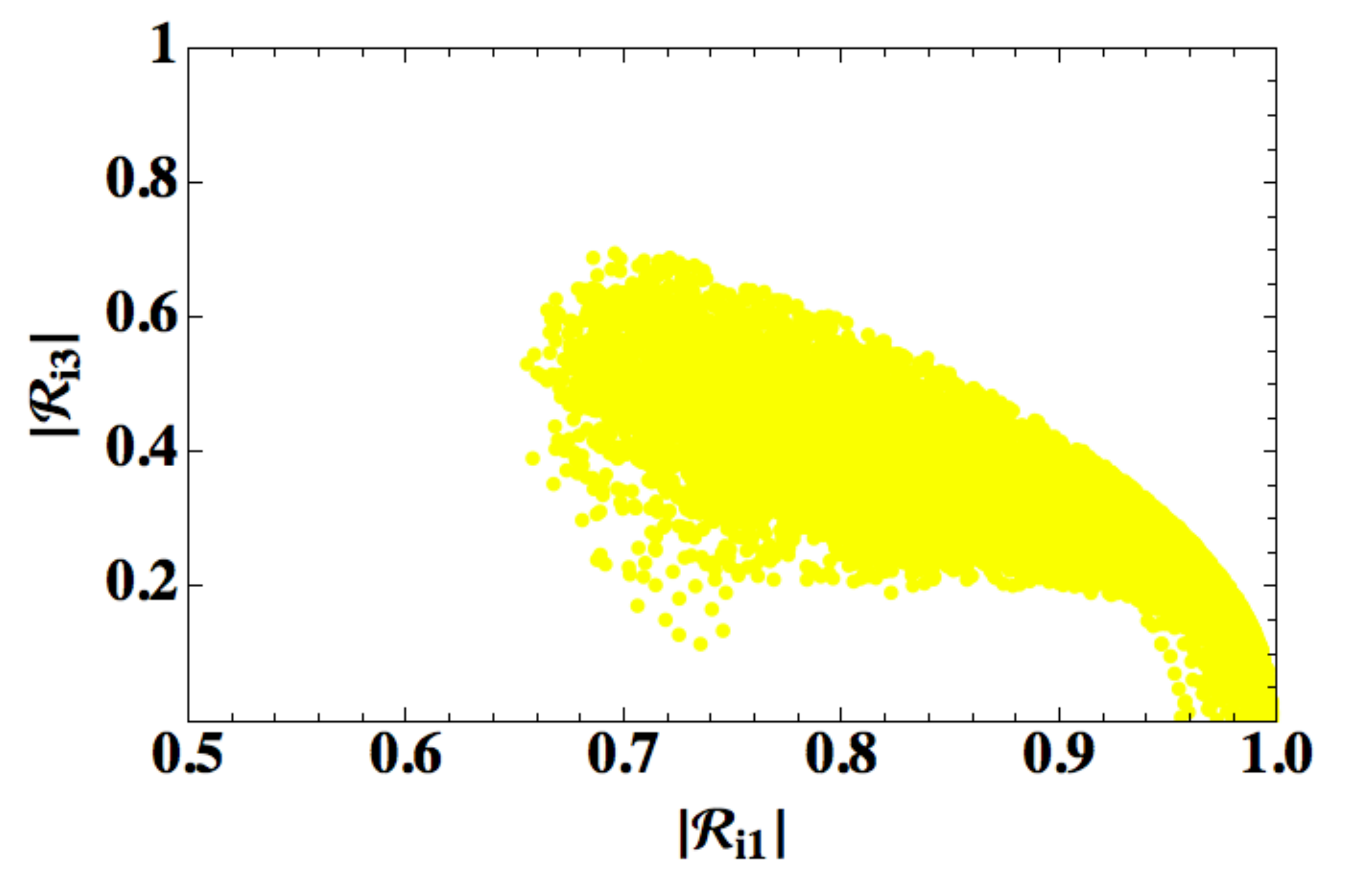}
\hskip 1.25cm
\includegraphics[width=7cm,height=6.4cm]{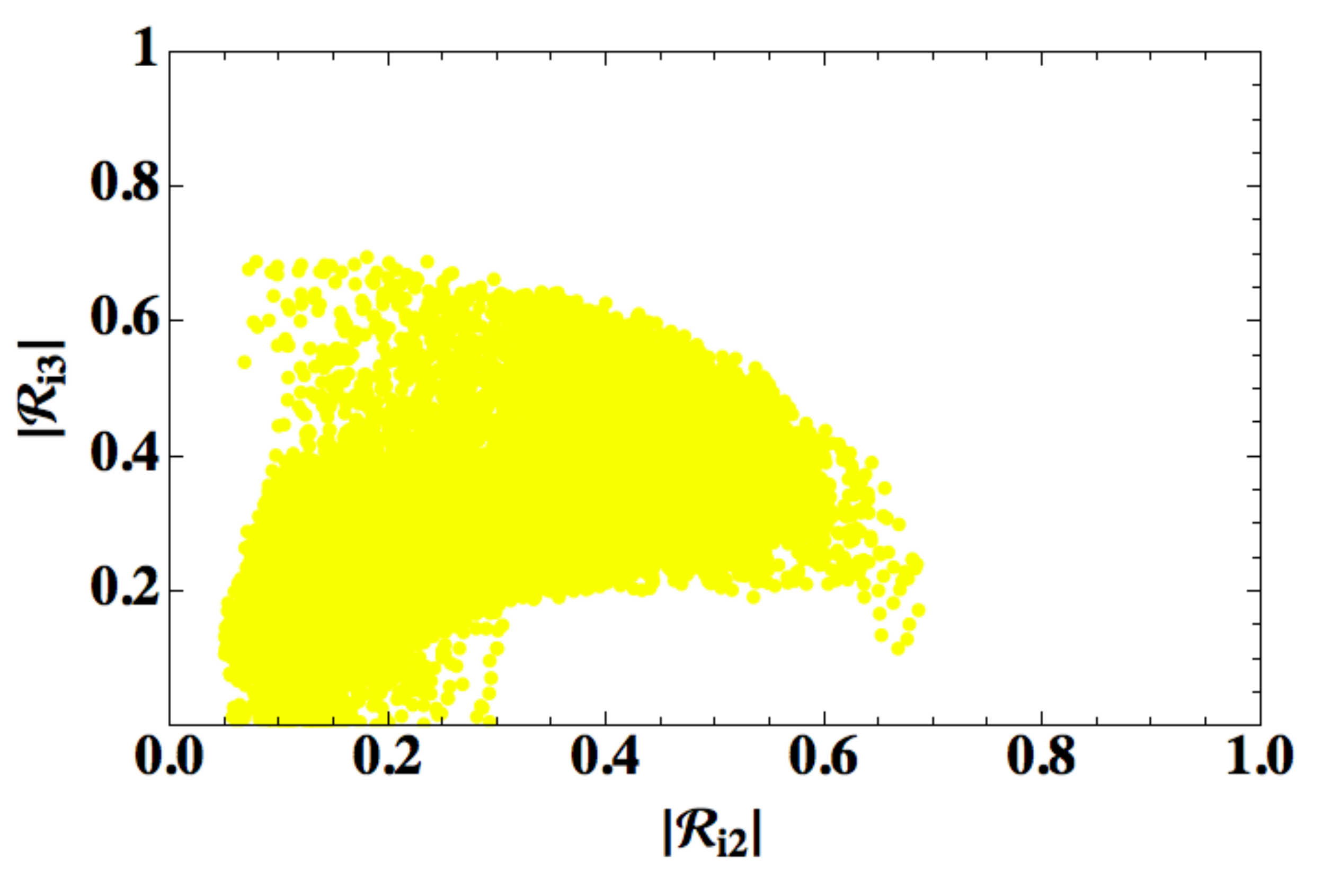}
\caption{\label{Higgsmixing} \it \small
Allowed regions at $90\%$~CL (yellow) on the parameter space $(\mathcal{R}_{i1}, \mathcal{R}_{i3})$, for real alignment parameters in the ranges $|\varsigma_u| < 2$ and  $|\varsigma_{d,l}| < 10$ (left). The right plot shows the corresponding $90\%$~CL region for the parameters $(\mathcal{R}_{i2}, \mathcal{R}_{i3})$.}
\end{figure}

In Fig.~\ref{Higgsmixing} we show the allowed values at $90\%$~CL for $(\cR_{i1}, \cR_{i2}, \cR_{i3})$ for a general scalar state $\varphi_i^0$ with $m_{\varphi_i^0} =126$~GeV, assuming that the alignment parameters $\varsigma_f$ $(f=u,d,l)$ are real. We have imposed $|\varsigma_u| <2$, in order to
satisfy the flavour constraints for a charged Higgs below the TeV scale, and moreover we have set $|\varsigma_{d,l}| < 10$.   It is seen that the CP-odd admixture in the $126$~GeV state has an upper bound $\cR_{i3} \lesssim 0.7$, similar to that obtained in Ref.~\cite{Barroso:2012wz} within 2HDMs of types I and II, with explicit CP violation and soft breaking of the $\mathcal{Z}_2$ symmetry.

\section{Summary}
\label{sec:conclusion}

The recent LHC discovery of a new neutral boson, with mass close to $126$~GeV, provides for the first time direct information on the electroweak symmetry breaking mechanism. The current data are so far compatible with the SM Higgs hypothesis, although a slight excess in the diphoton channel has been observed by the ATLAS and CMS collaborations. This channel is particularly interesting since the decay of the Higgs into two photons occurs at the one-loop level and is therefore sensitive to new charged particles that couple directly to the Higgs.

As new and more precise data become available, we shall test whether the properties of the 126~GeV particle correspond indeed to the SM Higgs boson or they manifest evidences for new phenomena, perhaps signalling the existence of a much richer scalar sector. Present experimental errors are still large but, nevertheless, they already allow us to extract useful constraints on alternative scenarios of electroweak symmetry breaking.

2HDMs constitute the simplest extension of the SM scalar sector, satisfying the electroweak precision tests, and give rise to interesting new phenomena through their enlarged scalar spectrum containing five physical scalars. In order to avoid dangerous FCNCs, the 2HDM phenomenology has been usually particularized to a few specific implementations, based on discrete $\mathcal{Z}_2$ symmetries, which severely restrict the fermionic couplings of the scalar bosons. The most widely used scenario is the so-called type II 2HDM, since it corresponds to the tree-level scalar sector of the minimal supersymmetric SM.  However, the phenomenological FCNC requirements can be easily satisfied imposing a much softer alignment condition on the Yukawa couplings. The resulting A2HDM provides a general framework to describe an extended scalar sector with two Higgs doublets and no FCNCs at tree level, which includes as particular cases all previously considered 2HDM variants. It has a much larger parameter space with plenty of new phenomenological possibilities, such as new sources of CP violation and tunable strengths of the (family universal) Yukawas. Thus, it is the appropriate framework to perform an unbiased phenomenological analysis of the Higgs data.

In this paper, we have analyzed the present data on the Higgs signal strengths from the ATLAS, CMS, CDF and D\O~collaborations, within the framework of the A2HDM. Even with the currently large experimental uncertainties, interesting conclusions can be obtained regarding the preferred regions in the parameter space of the model. We have considered a variety of possible departures from the SM predictions, within this framework, including the effects from new CP-violating phases. In particular, we have searched for possible ways to enhance the diphoton channel while being compatible with the rest of the data.

The measured $WW$, $ZZ$ and $\gamma \gamma$ decay channels of the new boson suggest that its coupling to the weak vector bosons ($W^+ W^-, ZZ$) is close to the SM one.  This rules out the possibility of a pure CP-odd assignment for the quantum numbers of the new Higgs-like boson. A CP-even scalar, either pure or with a CP-odd admixture arising from CP-violating terms in the scalar potential, however, can accommodate the data rather well.

By flipping the relative sign of the top Yukawa coupling, the top-quark contribution to the Higgs decay amplitude into $2\gamma$ interferes constructively with the dominant $W^{\pm}$ contribution.   This can only be realized in the A2HDM for large values of $|\varsigma_u|$, given that $ g_{\varphi^0_i VV}\; \approx \;\; g_{h VV}^{\mathrm{SM}}\,$. However, flavour constraints on a charged Higgs below the TeV scale (from $Z \rightarrow \bar b b$, $b\to s\gamma$ and $B^0$--$\bar B^0$ mixing) require that $| \varsigma_u| < 2$, even in the most general CP-violating A2HDM. Thus, a $2\gamma$ enhancement through a constructive interference of the top and $W^\pm$ contributions could only be possible in a decoupling scenario with an enormously large $H^\pm$ mass.

Including the charged scalar contribution to the Higgs decay amplitude into two photons, one can explain the observed excess without significant deviations of the neutral scalar couplings from the SM limit, and satisfying at the same time the flavour constraints. This appears to be the most natural and likely possibility to accommodate current data within the A2HDM framework. The confirmation by future data of a significatively enhanced $2\gamma$ decay width could be a strong indication that a light charged scalar is around the corner, within the LHC reach.

The possibility that a CP-even and a CP-odd Higgs bosons have quasi-degenerate masses near $126$~GeV was also analyzed.  An excess in the $\gamma \gamma$ channel can occur in this case due to the contributions from both scalars (when signal strengths are added incoherently).
We have also considered the most general A2HDM with complex Yukawa couplings. Since the Higgs signal strengths are CP-even observables, there is no interference between the contributions from the real and imaginary parts of the Yukawa couplings.  It is then possible to enhance the $\gamma \gamma$ decay rate with a complex Yukawa coupling which has its real part close to the SM-like limit.

Future improvements of the present bounds on neutral and charged Higgs bosons, or perhaps their direct discovery, as well as more precise measurements of the current Higgs signal strengths are expected from the LHC in the next years. The complementarity between flavour constraints and collider searches for new scalar resonances will be crucial for the understanding of the mechanism of electroweak symmetry breaking.  We have shown different alternative scenarios within the A2HDM that can accommodate present data very well, placing bounds on the relevant parameter space and discussing possible consequences that could be tested in the near future.

\vskip .5cm

\noindent {\bf\large Note added:}
After the submission of this work for publication, updated experimental analyses of the LHC data have been made public \cite{ATLAS_New,CMS_New}. While an enhanced diphoton rate is still present in the ATLAS results, the CMS collaboration finds now a $2\gamma$ rate compatible with the SM prediction. The new CMS results would favour a SM-like scenario, similar to that obtained in Eq.~\eqref{fitreal1}, without any need for a charged scalar contribution to the $2\gamma$ decay mode. More data are needed to clarify this issue.

\section*{Acknowledgements}
We are grateful to Luca Fiorini for discussions about the experimental data.
This work has been supported in part by the Spanish Government and ERDF funds from the EU Commission~[grants FPA2007-60323, FPA2011-23778 and CSD2007-00042~(Consolider Project CPAN)].  The work of A. C. is funded through an FPU grant (AP2010-0308, MINECO, Spain).

\begin{appendix}

\section{Scalar Potential}
\label{app:potential}

In the Higgs basis, the most general scalar potential takes the form
\beqn\label{eq:potential}
V & = & \mu_1\; \Phi_1^\dagger\Phi_1\, +\, \mu_2\; \Phi_2^\dagger\Phi_2 \, +\, \left[\mu_3\; \Phi_1^\dagger\Phi_2 \, +\, \mu_3^*\; \Phi_2^\dagger\Phi_1\right]
\no\\ & + & \lambda_1\, \left(\Phi_1^\dagger\Phi_1\right)^2 \, +\, \lambda_2\, \left(\Phi_2^\dagger\Phi_2\right)^2 \, +\,
\lambda_3\, \left(\Phi_1^\dagger\Phi_1\right) \left(\Phi_2^\dagger\Phi_2\right) \, +\, \lambda_4\, \left(\Phi_1^\dagger\Phi_2\right) \left(\Phi_2^\dagger\Phi_1\right)
\no\\ & + & \left[  \left(\lambda_5\; \Phi_1^\dagger\Phi_2 \, +\,\lambda_6\; \Phi_1^\dagger\Phi_1 \, +\,\lambda_7\; \Phi_2^\dagger\Phi_2\right) \left(\Phi_1^\dagger\Phi_2\right)
\, +\, \mathrm{h.c.}\right]\, .
\eeqn
The Hermiticity of the potential requires all parameters to be real except $\mu_3$, $\lambda_5$, $\lambda_6$ and $\lambda_7$; thus, there are 14 real parameters.

The minimization conditions
$\langle 0|\Phi_1^T(x)|0\rangle =\frac{1}{\sqrt{2}}\, (0, v)$ and $\langle 0|\Phi_2^T(x)|0\rangle =\frac{1}{\sqrt{2}}\, (0, 0)$
impose the relations
\bel{eq:minimum}
\mu_1\; =\; -\lambda_1\, v^2\, ,
\qquad\qquad\qquad
\mu_3\; =\; -\frac{1}{2}\,\lambda_6\, v^2\, .
\ee

The potential can then be decomposed into a quadratic term plus cubic and quartic interactions
\bel{eq:potential2}
V\; =\; -\frac{1}{4}\,\lambda_1\, v^4\, +\, V_2 \, +\,  V_3 \, +\,  V_4\, .
\ee
The mass terms take the form
\beqn\label{eq:mass_term}
V_2 & = &
M_{H^\pm}^2\, H^+ H^-\, +\, \frac{1}{2}\, \left(S_1, S_2, S_3\right)\; \mathcal{M}\; \left(\ba S_1\\ S_2\\ S_3\ea\right)
\no\\[10pt] & = &
M_{H^\pm}^2\, H^+ H^-\, +\, \frac{1}{2}\, M_h^2\, h^2\, +\, \frac{1}{2}\, M_H^2\, H^2\, +\, \frac{1}{2}\, M_A^2\, A^2\, ,
\eeqn
with
\bel{eq:mplus}
M_{H^\pm}^2\; =\; \mu_2 + \frac{1}{2}\,\lambda_3\, v^2
\ee
and
\bel{eq:mass_matrix}
\mathcal{M}\; =\; \left(\begin{array}{ccc}
2\lambda_1 v^2 & v^2\, \lambda_6^{\mathrm{R}} & -v^2\, \lambda_6^{\mathrm{I}}\\
v^2\, \lambda_6^{\mathrm{R}} & M_{H^\pm}^2  + v^2\left(\frac{\lambda_4}{2} + \lambda_5^{\mathrm{R}}\right)
& -v^2\, \lambda_5^{\mathrm{I}}\\
-v^2\, \lambda_6^{\mathrm{I}} & -v^2\, \lambda_5^{\mathrm{I}} & M_{H^\pm}^2  +  v^2\left(\frac{\lambda_4}{2} - \lambda_5^{\mathrm{R}}\right)
\ea\right)\, ,
\ee
where $\lambda_i^{\mathrm{R}}\equiv \mathrm{Re}(\lambda_i)$ and $\lambda_i^{\mathrm{I}}\equiv \mathrm{Im}(\lambda_i)$.
The symmetric mass matrix $\mathcal{M}$ is diagonalized by an orthogonal matrix $\mathcal{R}$, which defines the neutral mass eigenstates:
\bel{eq:mass_diagonalization}
\mathcal{M}\; =\; \mathcal{R}^T\; \left(\begin{array}{ccc} M_h^2 & 0 & 0 \\ 0 & M_H^2 &  0 \\ 0 & 0 & M_A^2
\ea\right)\; \mathcal{R}\, ,
\qquad\qquad
\left(\ba h\\ H\\ A\ea\right)\; =\; \mathcal{R}\; \left(\ba S_1\\ S_2\\ S_3\ea\right)\, .
\ee
Since the trace remains invariant, the masses satisfy the relation
\bel{eq:mass_sum}
M_h^2 \, +\, M_H^2 \, +\, M_A^2\; =\; 2\, M_{H^\pm}^2\, +\, v^2\,\left(2\,\lambda_1 +\lambda_4\right)\, .
\ee

The minimization conditions allow us to trade the parameters $\mu_1$ and $\mu_3$ by $v$ and $\lambda_6$. The freedom to rephase the field $\Phi_2$ implies,
moreover, that only the relative phases among $\lambda_5$, $\lambda_6$ and $\lambda_7$ are physical; but only two of them are independent. Therefore,
we can fully characterize the potential with 11 parameters: $v$, $\mu_2$, $|\lambda_{1,\ldots,7}|$,
$\mathrm{arg}(\lambda_5\lambda_6^*)$ and $\mathrm{arg}(\lambda_5\lambda_7^*)$. Four parameters can be determined through the physical scalar masses.

In the CP conserving limit $\lambda_5^{\mathrm{I}}=\lambda_6^{\mathrm{I}} =\lambda_7^{\mathrm{I}} =0$ and $S_3$ does not mix with the other neutral fields. The scalar spectrum contains then a CP-odd field $A=S_3$ and two CP-even scalars $h$ and $H$ which mix through the rotation matrix~\eqn{eq:CPC_mixing}. In this case,
the scalar masses are given by
\bel{eq:CPC_masses}
\bar M_h^2\; =\;\frac{1}{2}\,\left( \Sigma-\Delta\right)\, ,
\qquad
\bar M_H^2\; =\;\frac{1}{2}\,\left( \Sigma+\Delta\right)\, ,
\qquad
\bar M_A^2 \; =\; M_{H^\pm}^2\, +\, v^2\,\left(\frac{\lambda_4}{2} - \lambda_5^{\mathrm{R}}\right)\, ,
\ee
where
\beqn\label{eq:Sigma}
\Sigma & =& M_{H^\pm}^2\, +\, v^2\,\left(2\,\lambda_1 +\frac{\lambda_4}{2}+ \lambda_5^{\mathrm{R}}\right)\, ,
\\ \label{eq:Delta}
\Delta & =&\sqrt{\left[M_{H^\pm}^2\, +\, v^2\,\left(-2\,\lambda_1 +\frac{\lambda_4}{2}+ \lambda_5^{\mathrm{R}}\right) \right]^2 + 4 v^4 (\lambda_6^{\mathrm{R}})^2}\, ,
\eeqn
and the mixing angle is determined through
\bel{eq:mixingCPC}
\tan{\tilde\alpha}\; =\; \frac{\bar M_h^2 - 2\lambda_1 v^2}{v^2\lambda_6^{\mathrm{R}}}\, .
\ee
We use the notation $\bar M_{\varphi_i^0}$ to emphasize that these are the neutral scalar masses in the CP-conserving limit.
The cubic and quartic Higgs couplings involving the charged and the neutral physical scalars (without Goldstone boson couplings) take the form,
\begin{align}   \label{cubic}
V_3\; &=\; v  \,  H^+ H^-\, \left( \lambda_3  \, S_1  +    \lambda_7^\text{R}  \,  S_2 -    \lambda_7^\text{I}  \, S_3\right)  - \frac{1}{2}\, v\,\lambda_7^\text{I}  \,  S_3^3  - \frac{1}{2}\, v\, \lambda_7^\text{I}  \, S_2^2  S_3 - \frac{3}{2}\, v\, \lambda_6^\text{I}  \,  S_1^2  S_3
\notag \\
    & + \lambda_1\, v  \,  S_1^3 + \frac{1}{2}\, v\,\lambda_7^\text{R}  \,  S_2^3 + \frac{3}{2}\, v\, \lambda_6^\text{R}  \,  S_1^2  S_2
 + \frac{1}{2}\, v  \, \left(2\lambda_5^\text{R} + \lambda_3 + \lambda_4\right)  \, S_1  S_2^2
\notag \\
 &-\frac{1}{2}\, v\, \left( 2\lambda_5^\text{R} - \lambda_3 - \lambda_4\right) \,  S_1  S_3^2 +
\frac{1}{2}\, v\, \lambda_7^\text{R}  \, S_2 S_3^2 - 2\, v\, \lambda_5^I  \, S_1 S_2 S_3 \, ,
\end{align}
\begin{align}\label{quartic}
V_4\; &=\;  H^+ H^-  \left( \lambda_2 \, H^+ H^-  +  \frac{\lambda_3}{2} \, S_1^2  + \lambda_2 \,  S_3^2   + \lambda_2 \,  S_2^2   -  \lambda_7^{\mathrm{I}}  \,  S_1 S_3  +  \lambda_7^{\mathrm{R}}  \,  S_1 S_2 \right)
\nonumber \\
&   + \frac{1}{4}\, \left( \lambda_3  + \lambda_4  + 2 \lambda_5^{\mathrm{R}} \right) \, (S_1 S_2 )^2  +  \frac{1}{4}\, \left( \lambda_3 + \lambda_4 -2 \lambda_5^{\mathrm{R}}  \right) \, (S_1 S_3)^2 + \frac{\lambda_2}{2}  \, (S_2 S_3)^2     \nonumber \\
& - \frac{1}{2}\,   \lambda_6^{\mathrm{I}}  \, S_1^3 S_3  - \lambda_5^{I} \, S_1^2 S_2 S_3  - \frac{\lambda_7^{\mathrm{I}}}{2}  \, S_1 S_2^2 S_3  - \frac{\lambda_7^{\mathrm{I}}}{2}  \, S_1 S_3^3  + \frac{ \lambda_6^{\mathrm{R}} }{2}\, S_1^3 S_2 + \frac{ \lambda_7^{\mathrm{R}}}{2} \, S_1 S_2^3 + \frac{\lambda_7^{\mathrm{R}}}{2}  \, S_1 S_2 S_3^2
\nonumber \\
& +  \frac{\lambda_1}{4} \, S_1^4 +  \frac{\lambda_{2}}{4} \, S_2^4 + \frac{ \lambda_2 }{4}\, S_3^4 \, .
\end{align}
In the CP-conserving limit all vertices involving an odd number of $S_3$ fields vanish.   A basis-independent discussion of the 2HDM scalar sector can be found in Ref.~\cite{Haber:2006ue}.

\subsection{Neutral scalar mass matrix to lowest order in CP violation}

Assuming that $\lambda_5^{\mathrm{I}}$ and $\lambda_6^{\mathrm{I}}$ are small, we can diagonalize the mass matrix \eqn{eq:mass_matrix} perturbatively as an expansion in powers of these CP-violating parameters. The leading corrections to the neutral scalar masses are quadratic in $\lambda_{5,6}^{\mathrm{I}}$:
\begin{equation}\label{eq:mix}
 M_{\varphi_i^{0}}^2\; =\; \bar M_{\varphi_i^0}^2 + \alpha_1^{\varphi_i^{0}}\, (\lambda_5^I)^2 + \alpha_2^{ \varphi_i^{0}}\, (\lambda_6^I)^2 + \alpha_3^{\varphi_i^{0}}\, (\lambda_5^I \lambda_6^{I}) \, ,
\end{equation}
where $\bar M_{\varphi_i^0}$ denote the corresponding masses in the CP-conserving limit given in \eqref{eq:CPC_masses} and
\begin{align}  \label{alpha_param}
\alpha_1^{ \varphi_i^{0}}\; &=\; \dfrac{ v^4 \left(\bar M_{\varphi_i^0}^2 - 2 \lambda_1 v^2 \right)  }{ \prod_{ j \neq i}  \left( \bar M_{\varphi_j^0}^2 - \bar M_{\varphi_i^0}^2  \right)  }  \, ,
\nonumber \\[5pt]
\alpha_{2}^{ \varphi_i^{0}}\; &=\;  \dfrac{ v^4  \left(  2 \lambda_1 v^2 + \bar M_{\varphi_i^0}^{2} - \bar  M_{H}^2 - \bar M_{h}^2     \right) }{ \prod_{j \neq i}  \left( \bar M_{\varphi_j^0}^2 - \bar M_{\varphi_i^0}^2  \right)  }  \, ,
\nonumber \\[5pt]
\alpha_{3}^{ \varphi_i^{0}}\; &=\;  \dfrac{2 v^6 \lambda_6^R}{ \prod_{j \neq i}  \left( \bar M_{\varphi_j^0}^2 - \bar M_{\varphi_i^0}^2  \right) } \, .
\end{align}

The physical states $\varphi_i^{0}= \{  h,  H, A \}$ receive corrections at first order in $\lambda_{5,6}^{\mathrm{I}}$, which are given by
\bel{eq:mix_matrix}
\left(\ba  h\\   H \\   A \ea\right)\; =\;    \left(\begin{array}{ccc}
\cos{\ta}  &  \sin{\ta} &  \epsilon_{13}  \\
- \sin{\ta} &  \cos{\ta} &  \epsilon_{23} \\
 \epsilon_{31}  & \epsilon_{32}  &    1
\ea\right)\,  \left(\ba S_1\\ S_2\\ S_3\ea\right) \, ,
\ee
where
\beqn \label{eq:mixA}
 \epsilon_{13} &\! =&\! \frac{v^2}{ \left( \bar M_A^2 - \bar M_h^2 \right) }\, \left(  \sin{\ta} \, \lambda_5^{\mathrm{I}}   + \cos{\ta} \, \lambda_{6}^{\mathrm{I}}   \right) \, , \qquad\quad
 \epsilon_{23}\; =\; \frac{v^2}{ \left( \bar M_A^2 - \bar  M_H^2 \right) }\, \left(  \cos{\ta} \, \lambda_5^{\mathrm{I}}   - \sin{\ta} \, \lambda_{6}^{\mathrm{I}}   \right) \, , \nonumber \\[5pt]
 \epsilon_{31} &\! =&\!  - \dfrac{1}{2 v^2}\, \left( \alpha_3^{ A}  \, \lambda_{5}^{\mathrm{I}}   + 2\,  \alpha_2^{ A}   \, \lambda_{6}^{\mathrm{I}} \right) \,,\qquad\quad
 \qquad\quad\quad \,  \epsilon_{32}\; =\;  -  \dfrac{1}{2 v^2}\, \left(   2 \,\alpha_1^{ A}  \, \lambda_{5}^{\mathrm{I}}   +    \alpha_{3}^{ A} \, \lambda_{6}^{\mathrm{I}}  \right)   \,.
\eeqn
Note that for the case of a scalar potential with a softly-broken $\mathcal{Z}_2$ symmetry in the Higgs basis we have $\lambda_6 = \lambda_7 = 0$ and, therefore, $\epsilon_{31} = 0$.

\section{Scalar Couplings to the Gauge Bosons}
\label{app:scalar-gauge}

The scalar doublets couple to the gauge bosons through the covariant derivative and gauge-fixing terms:
\be
\cL_K + \sum_{i=1}^2 (D_\mu\Phi_a)^\dagger\, D^\mu\Phi_a + \cL_{\mathrm{GF}}\; =\;
\cL_{V^2} + \cL_{\phi^2} + \cL_{\phi V} + \cL_{\phi^2 V} + \cL_{\phi V^2} + \cL_{\phi^2 V^2}\, ,
\ee
where $\cL_K$ is the usual gauge-boson kinetic term and the covariant derivative is given by\footnote{The weak mixing angle $\theta_W$ is defined through the relation $ g \sin \theta_W = g^{\prime} \cos \theta_W = e$.  The operators $T_{\pm} = \frac{1}{\sqrt{2}} (T_1 \pm T_2) $ and $T_3$ can be expressed in terms of the Pauli matrices by $T_i = \dfrac{\sigma_i}{2}$ .}  \mbox{$D_{\mu}=\partial_{\mu}+ i e Q A_{\mu} + i \dfrac{g}{\cos{\theta_W}}\, Z_{\mu} (T_3 - Q \sin^2 \theta_W ) + i g  \left[ T_+ W_{\mu}^{\dag} + T_- W_{\mu}  \right]$}.   It is convenient to adopt the following $R_\xi$ gauge-fixing term ($\xi=1$),
\be
 \cL_{\mathrm{GF}}\, =\, -\frac{1}{2}\,\left(\partial_\mu A^\mu\right)^2 -
\frac{1}{2}\,\left(\partial_\mu Z^\mu + M_Z G^0\right)^2 -
\left(\partial^\mu W_\mu^\dagger + i M_W G^+\right) \left(\partial_\nu W^\nu - i M_W G^-\right)\, ,
\ee
which cancels exactly the quadratic mixing terms between the gauge and Goldstone bosons generated by
the covariant derivatives, so that $\cL_{\phi V} = 0$, and provides the Goldstone bosons with the masses
$M_{G^\pm}=M_W = gv/2$  and $M_{G^0} = M_Z = M_W/\cos{\theta_W}$. Then,
\be
\cL_{V^2}\, =\,  -\frac{1}{2}\,\left(\partial_\mu A^\mu\right)^2 -
\frac{1}{2}\,\left(\partial_\mu Z^\mu\right)^2 + \frac{1}{2}\, M_Z^2\, Z_\mu Z^\mu
- \left(\partial^\mu W_\mu^\dagger\right) \left(\partial_\nu W^\nu\right)
 + M_W^2\, W_\mu^\dagger W^\mu\, ,
\ee
while
\beqn
\cL_{\phi^2}& =&\frac{1}{2}\,\left[\partial_\mu h\, \partial^\mu h + \partial_\mu H\, \partial^\mu H +
\partial_\mu A\, \partial^\mu A \right] + \partial_\mu H^+ \partial^\mu H^-
\no\\ &+& \frac{1}{2}\,\partial_\mu G^0\, \partial^\mu G^0 -\frac{1}{2}\, M_Z^2\, (G^0)^2
+ \partial_\mu G^+\, \partial^\mu G^- - M_W^2\, G^+ G^-\, .
\eeqn

The interaction terms between the scalar and gauge bosons are given by:
\beqn
\cL_{\phi^2 V} & = & i e\, \left[A^\mu + \cot{(2\theta_W)}\, Z^\mu\right]\,
\left[ (H^+\!\lrder_\mu\! H^-) + (G^+\!\lrder_\mu\! G^-)\right]
\no\\ &+&
\frac{e}{\sin{(2\theta_W)}}\; Z^\mu\, \left[ (G^0\!\lrder_\mu\! S_1) + (S_3\!\lrder_\mu\! S_2)\right]
\no\\ &+&
\frac{g}{2}\; W^{\mu\dagger}\, \left[ (H^-\!\lrder_\mu\! S_3) -i\, (H^-\!\lrder_\mu\! S_2) +
(G^-\!\lrder_\mu\! G^0) -i\, (G^-\!\lrder_\mu\! S_1)\right]
\no\\ &+&
\frac{g}{2}\; W^\mu\, \left[ (H^+\!\lrder_\mu\! S_3) +i\, (H^+\!\lrder_\mu\! S_2) +
(G^+\!\lrder_\mu\! G^0) +i\, (G^+\!\lrder_\mu\! S_1)\right]\, ,
\eeqn
\beqn\label{eq:Lphv2}
\cL_{\phi V^2} & = &  \frac{2}{v}\; S_1\,\left[\frac{1}{2}\, M_Z^2\, Z_\mu Z^\mu + M_W^2\, W^\dagger_\mu W^\mu\right]
\no\\ &+& \left( e M_W\, A^\mu - g M_Z\sin^2{\theta_W}\, Z^\mu\right)\,\left( G^+ W_\mu + G^- W^\dagger_\mu\right)\, ,
\eeqn
\beqn
\cL_{\phi^2 V^2} & = & \frac{1}{v^2}\, \left[\frac{1}{2}\, M_Z^2\, Z_\mu Z^\mu + M_W^2\, W^\dagger_\mu W^\mu\right]
\,\left[ H^2 + h^2 + A^2 + (G^0)^2 \right]
\no\\ &+&
 \left\{e^2\,\left[A^\mu + \cot{(2\theta_W)}\, Z^\mu\right]^2 + \frac{g^2}{2}\, W_\mu^\dagger W^\mu\right\}
 \,\left( G^+ G^- + H^+ H^-\right)
 \no\\ &+&
 \frac{eg}{2}\,\left( A^\mu - \tan{\theta_W}\, Z^\mu \right)\, \Bigl[S_1 \left( G^+ W_\mu + G^- W^\dagger_\mu\right)
 + S_2 \left( H^+ W_\mu + H^- W^\dagger_\mu \right)   \no\\ &&\hskip 3.85cm\mbox{} +\;
  i \,S_3 \left( H^- W^\dagger_\mu - H^+ W_\mu \right)
 + i\, G^0 \left( G^- W^\dagger_\mu - G^+ W_\mu \right) \Bigr] \, ,\quad
\eeqn
with $S_i = \mathcal{R}_{ji}\varphi_j^0$ ($\varphi_j^0=\{h,H,A\}$) and the usual notation $ A \lrder_{\mu} B \equiv  A (\partial_{\mu} B) - (\partial_{\mu} A) B  $.

\section{Statistical treatment and data}
\label{statistical}

To obtain the preferred values for the parameters of the \athdm~we build a global $\chi^2$ function
\begin{equation}
\chi^2\; =\; \sum_k \dfrac{\left(  \mu_k - \hat \mu_k \right)^2}{\sigma_k^2}\,,
\end{equation}
where $\sigma_i$ is the experimental error extracted from the data at $1~ \sigma$.     Errors on the reported Higgs signal strengths $\hat \mu_k$ are symmetrized using
\be
\delta \hat \mu_k\; =\; \sqrt{  \dfrac{ (\delta \hat \mu_+)^2 + (\delta \hat \mu_-)^2  }{2}  }\,,
\ee
where $\delta \hat \mu_{\pm}$ are the one-sided errors given by the experimental collaborations.   We use the latest data available after the ``Hadron Collider Physics Symposium 2012 (HCP2012)", including the latest update from ATLAS of the high-resolution channels $\gamma \gamma$, $ZZ^{(*)}$~\cite{:2012gk}.  For the diphoton channels we use the data given by ATLAS and CMS at 7 and 8 TeV, provided in Refs.~\cite{:2012an,Chatrchyan:2012tx,:2012gk,:2012gu}.    For the rest of the channels we use the averages listed in Table~\ref{tab:7and8LHC}, which include the $7 \oplus 8$~TeV data reported by ATLAS and CMS together with CDF and D\O~data~\cite{:2012zzl,Aaltonen:2012if,:2012tf} at $\sqrt{s} = 1.96$~TeV.

\begin{table}[tb]\begin{center}
\caption{ \it \small Higgs signal strengths in each of the channels considered in this work.  Averages obtained from ATLAS and CMS data at $7 \oplus 8$~TeV together with CDF and D\O\ data at $\sqrt{s}= 1.96$~TeV.    (*) We do not consider non-inclusive measurements in the $\tau \tau$ channel.  Due to the large current errors associated with these measurements,  our conclusions would not be modified at this level.}   \label{tab:7and8LHC}
\vspace{0.2cm}
\begin{tabular}{|c|c|c|}
\hline
Channel &  $\hat \mu_k$&   Comment     \\
\hline
$  b \bar b V$   &   $1.1 \pm 0.44$ & ATLAS, CMS, CDF and D\O~\cite{:2012gk,:2012gu,Aaltonen:2012if,:2012tf} (our average) \\
$  W W j j $   &  $-0.2 \pm 1.56$ & ATLAS and CMS~\cite{:2012gk,:2012gu,Chatrchyan:2012tx} (our average) \\
$  WW $    &  $ 0.76 \pm 0.21$ & ATLAS, CMS, CDF and D\O~\cite{:2012gk,:2012gu,:2012an,Aaltonen:2012if,:2012zzl} (our average) \\
$  ZZ$   &   $0.96 \pm 0.26 $ & ATLAS and CMS~\cite{:2012gk,:2012gu} (our average)\\
$  \tau \tau$ (incl.) (*) &   $  0.89 \pm 0.86  $ & ATLAS and CMS~\cite{:2012gk,:2012gu} (our average) \\
$  \gamma \gamma$   &   $  1.66 \pm 0.32  $ & ATLAS and CMS~\cite{:2012gk,:2012gu} (our average) \\
$  \gamma \gamma jj$   &   $  2.18 \pm 0.84  $ & ATLAS and CMS~\cite{:2012gk,:2012gu} (our average) \\
\hline
\end{tabular}
\end{center}\end{table}

For a general channel with inclusive production we have (neglecting the subdominant production channels)
\be
\mu_k^{\varphi_i^0}\; =\; \dfrac{   \sigma_{gg} }{    \sigma_{gg}^{\sm}    } \cdot \dfrac{ \text{Br}(\varphi_i^0 \rightarrow k)}{ \text{Br}( \varphi_i^0 \rightarrow k)_{\sm}}   \,.
\ee
For the Higgs searches in the $\gamma \gamma$ channel, the ATLAS and CMS collaborations have established different categories.    To take this into account, we write the Higgs signal strength in a given $\gamma \gamma$ channel as
\be  \label{photonXS}
\mu_{\gamma \gamma}^{\varphi_i^0}\;  =\; \dfrac{  \epsilon_{ggF} \, \sigma_{ggF} +  \epsilon_{\text{VBF}}\,  \sigma_{VBF} + \epsilon_{\text{\text{VH}}}\,  \sigma_{\text{VH}} }{   \epsilon_{ggF} \, \sigma_{ggF}^{\sm}   +   \epsilon_{\text{VBF}}\, \sigma_{\text{VBF}}^{\sm}    +   \epsilon_{\text{\text{VH}}}\,  \sigma_{\text{VH}}^{\sm}   } \cdot  \dfrac{ \text{Br}(\varphi_i^0 \rightarrow \gamma \gamma)}{ \text{Br}( \varphi_i^0 \rightarrow \gamma \gamma)_{\sm}} \,,
\ee
where the coefficients $\epsilon_{(ggF, \text{VBF}, \text{VH})}$ accounting for the relative weight of each production channel have been provided by ATLAS and CMS~\cite{ATLAS:2012ad,Chatrchyan:2012twa}.    The top-quark-fusion contribution could be added in a similar way.
In Eq.~\eqref{photonXS}, the SM production cross sections and decay widths are taken from the web page of the LHC Higgs Cross Section Working Group~\cite{Dittmaier:2011ti}. For the gluon-fusion production mechanism we have
\be
\sigma( gg \rightarrow \varphi_i^0 )\;  \equiv\; \sigma_{ggF}\; = \;    C^{\varphi_i^0}_{gg}  \sigma_{ggF}^{\text{SM}}  \,,
\ee
where the scaling of the gluon-fusion cross section $C^{\varphi_i^0}_{gg}$ was defined in section~\ref{sec:SignalStrengths}.  Vector-boson fusion scales with the coefficient $\mathcal{R}_{i1}$ as
\be
\sigma(  q  q^{\prime} \rightarrow  q q^{\prime} \varphi_i^0  )\;  \equiv\;    \sigma_{\text{VBF}} \; =\; (\mathcal{R}_{i1})^2\,  \sigma_{\text{VBF}}^{\sm} \,,
\ee
and similarly for the associated production with a vector boson
\be
 \sigma(  q  \bar q \rightarrow  V \varphi_i^0  )  \;\equiv\; \sigma_{\text{VH}}\; =\; (\mathcal{R}_{i1})^2 \, \sigma_{\text{VH}}^{\sm}      \, .
\ee

\section{Perturbativity Constraints} 
\label{perturbatvity}

The charged Higgs boson contribution to $\varphi^0_i\to\gamma\gamma$
depends crucially on the value of the neutral scalar coupling to a pair of charged Higgs bosons.  To assure the validity of perturbation theory, upper bounds on the quartic Higgs self-couplings are usually imposed requiring these to be smaller than \mbox{$8 \pi$} (see \cite{Gunion:1989we,Branco:2011iw} and references therein).  The cubic Higgs self-couplings are also bounded indirectly in this way.  In this work we consider an alternative perturbativity bound on the relevant Higgs cubic coupling which is more restrictive for light charged Higgs masses.   Consider the $\varphi^0_i H^+H^-$ one-loop vertex correction given by Fig.~\ref{vertex}. The contribution of this diagram is finite and can give us an idea about the allowed magnitude of the cubic coupling in order not to spoil the perturbative convergence.
\begin{figure}[tb]
\centering
\includegraphics[scale=0.55]{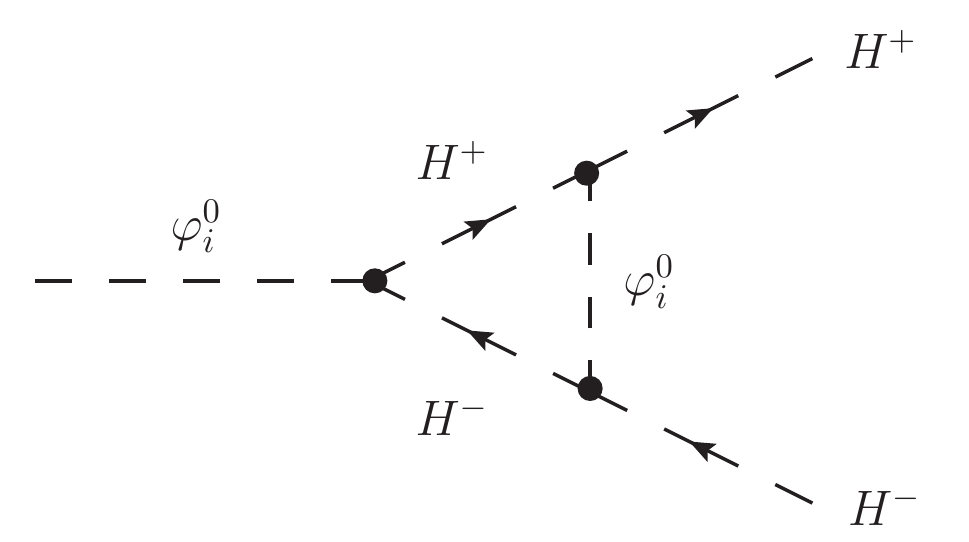}
\caption{\it \small Diagram contributing to the one-loop $\varphi^0_i H^+H^-$ vertex correction.}
\label{vertex}
\end{figure}
\noindent We obtain:
\begin{align}
(\lambda_{\varphi^0_i H^+H^-})_{\mathrm{eff}} \; =\;
\lambda_{\varphi^0_i H^+H^-}\, \left[ 1 + \frac{v^2 \lambda^2_{\varphi^0_i H^+H^-}}{16 \pi^2 M_{H^\pm}^2}\, \mathcal{Z}\left(\frac{M_{\varphi_i^0}^2}{M_{H^\pm}^2}\right) \right]\; \equiv\;\lambda_{\varphi^0_i H^+H^-} \; \left(1+\Delta\right) \, ,
\end{align}
where
\be
\mathcal{Z}(X)\; =\; \int_0^1 dy\; \int_0^{1-y}   dz  \;  \left[ (y+z)^2+ X \, (1-y-z-y z)\right]^{-1} \, .
\ee
Allowing the correction to be at most 50\% ($\Delta \leqslant 0.5$) constraints the allowed parameter space in the $(\lambda_{\varphi^0_i H^+H^-}, M_{H^\pm})$ plane to be
within the blue (hashed) region indicated in Fig.~\ref{gbounds}.

\section{Oblique Parameters} 
\label{oblique}

Possible deviations from the SM in the gauge-boson self-energies are usually characterized through the oblique parameters $S$, $T$ and $U$
\cite{Peskin:1991sw}. Taking as a reference SM Higgs mass
$M_{h,\mathrm{ref}}=126$~GeV, the most recent global fit to electroweak precision observables
quotes the values~\cite{Baak:2012kk,LEPEWWG}:
%
\begin{align}
S=0.03 \pm 0.10 \, , \qquad\qquad T = 0.05 \pm 0.12 \, , \qquad\qquad U= 0.03 \pm 0.10 \, .
\end{align}
The expressions for the oblique parameters in the CP conserving A2HDM are adapted from  Ref.~\cite{Haber:2011}:

\begin{align}
 S\; =\; \frac{1}{\pi M_Z^2}\, & \Biggl\{ \cos^2{\tilde{\alpha}}\; \biggl[
\mathcal{B}_{22}(M_Z^2;M_Z^2,M_h^2)- M_Z^2\,
\mathcal{B}_{0}(M_Z^2;M_Z^2,M_h^2) + \mathcal{B}_{22}(M_Z^2;M_H^2,M_A^2)  \biggr] \notag \\
& +\; \sin^2{\tilde{\alpha}}\; \biggl[
\mathcal{B}_{22}(M_Z^2;M_Z^2,M_H^2)- M_Z^2\,
\mathcal{B}_{0}(M_Z^2;M_Z^2,M_H^2) + \mathcal{B}_{22}(M_Z^2;M_h^2,M_A^2)  \biggr]
\notag \\
& -\; \mathcal{B}_{22}(M_Z^2;{M^2_{H^\pm}},{M^2_{H^\pm}})
-\mathcal{B}_{22}(M_Z^2;M_Z^2,M_{h,\mathrm{ref}}^2)
+ M_Z^2\,\mathcal{B}_{0}(M_Z^2;M_Z^2,M_{h,\mathrm{ref}}^2)\Biggr\}\, ,
\end{align}
\begin{align}
T\;  =\;  \frac{1}{16\pi M_W^2s_W^2}\, & \Biggl\{ \cos^2{\tilde{\alpha}}\; \biggl[
 \mathcal{F}(M_{H^\pm}^2,M_H^2) - \mathcal{F}(M_H^2,M_A^2) + 3  \!\  \mathcal{F}(M_Z^2,M_h^2) - 3 \!\ \mathcal{F}(M_W^2,M_h^2)  \biggr] \notag \\
& + \; \sin^2{\tilde{\alpha}}\; \biggl[
 \mathcal{F}(M_{H^\pm}^2,M_h^2) - \mathcal{F}(M_h^2,M_A^2) + 3  \!\  \mathcal{F}(M_Z^2,M_H^2) - 3 \!\ \mathcal{F}(M_W^2,M_H^2)  \biggr] \notag \\
& +\; \mathcal{F}(M_{H^\pm}^2,M_A^2) -  3 \!\ \mathcal{F}(M_Z^2,M_{h,\mathrm{ref}}^2) +3  \!\  \mathcal{F}(M_W^2,M_{h,\mathrm{ref}}^2) \Biggr\}\, ,
\end{align}
\begin{align}
U\; =\; \mathcal{H}(M_W^2)-\mathcal{H}(M_Z^2)+\frac{1}{\pi M_W^2}\, &\Biggl\{
  \sin^2{\tilde{\alpha}} \; \mathcal{B}_{22}(M_W^2;M^2_{H^\pm},M_h^2)+
\cos^2{\tilde{\alpha}} \; \mathcal{B}_{22}(M_W^2;M^2_{H^\pm},M_H^2) \notag \\  & +\;  \mathcal{B}_{22}(M_W^2;M^2_{H^\pm},M_A^2)  - 2  \!\ \mathcal{B}_{22}(M_W^2;M^2_{H^\pm},M^2_{H^\pm})
\Biggr\} \notag  \\
-\frac{1}{\pi M_Z^2}\, &\Biggl\{
 \sin^2{\tilde{\alpha}} \; \mathcal{B}_{22}(M_Z^2;M^2_h,M_A^2)  +
\cos^2{\tilde{\alpha}} \; \mathcal{B}_{22}(M_Z^2;M^2_H,M_A^2) \notag  \\
& -\;  \mathcal{B}_{22}(M_Z^2;M_{H^\pm}^2,M_{H^\pm}^2) \Biggr\}
\, ,
\end{align}
where
\begin{align}
\mathcal{H}(M_V^2)\;\equiv\;  \frac{1}{\pi M_V^2}\; &\Biggl\{  \cos^2{\tilde{\alpha}}\;\biggl[
\mathcal{B}_{22}(M_V^2; M_V^2,M_h^2)
-M_V^2\,\mathcal{B}_0(M_V^2;M_V^2,M_h^2)\biggr]
\notag \\
& +\; \sin^2{\tilde{\alpha}}\;\biggl[
\mathcal{B}_{22}(M_V^2; M_V^2,M_H^2)
-M_V^2\,\mathcal{B}_0(M_V^2;M_V^2,M_H^2)\biggr]
\notag \\
& -\; \mathcal{B}_{22}(M_V^2; M_V^2,M_{h,\mathrm{ref}}^2)
+M_V^2\,\mathcal{B}_0(M_V^2;M_V^2,M_{h,\mathrm{ref}}^2)\Biggr\}\, .
\end{align}
The loop functions are given by
\beqn
\label{eqnb}
B_{22}(q^2;m_1^2,m_2^2) &=& \frac{1}{4}\, (\Delta+1)\, [m_1^2+m_2^2-\frac{1}{3}\, q^2]-\frac{1}{2}\,\int^1_0 dx\; X\; \log{(X-i\epsilon)}\, ,
\\
B_{0}(q^2;m_1^2,m_2^2) &=& \Delta-\int^1_0 dx\; \log{(X-i\epsilon)}\, ,
\\
\mathcal{F}(m_1^2,m_2^2) &=& \frac{1}{2}\, (m_1^2+m_2^2)-\frac{m_1^2m_2^2}{m_1^2-m_2^2}\;
\log{\left(\frac{m_1^2}{m_2^2}\right)}\, ,
\eeqn
with
\begin{equation}
X \;\equiv\; m_1^2\, x + m_2^2\, (1-x) -q^2\, x(1-x)\, ,
\qquad\quad
\Delta \;\equiv\; \frac{2}{4-d}+\ln 4\pi-\gamma_E\, ,
\end{equation}
in $d$ space-time dimensions, where $\gamma_E$ is the Euler-Mascheroni constant,  and where we have defined:
\beqn
\mathcal{B}_{22}(q^2;m_1^2,m_2^2) &\equiv&
B_{22}(q^2;m_1^2,m_2^2)-B_{22}(0;m_1^2,m_2^2)\,,\label{b22} \\[6pt]
\mathcal{B}_{0}(q^2;m_1^2,m_2^2) &\equiv& B_{0}(q^2;m_1^2,m_2^2)-B_{0}(0;m_1^2,m_2^2)\,.\label{bzero}
\eeqn

\end{appendix}

\end{document}